\documentclass[aps,pre,preprint,onecolumn,citeautoscript,superscriptaddress,eqsecnum]{revtex4-1}
\usepackage{amsmath,amssymb,bm} 
\usepackage{graphicx}
\usepackage{color} 
\usepackage[papersize={8.5in,11in}]{geometry}
\usepackage[colorlinks=true]{hyperref}

\usepackage[caption=false]{subfig}

\hypersetup{
    bookmarks=true,         % show bookmarks bar?
    unicode=false,          % non-Latin characters 
    pdftoolbar=true,        % show Acrobat
    pdfmenubar=true,        % show Acrobat 
    pdffitwindow=false,     % window fit to page when opened
    pdfstartview={FitH},    % fits the width of the page to the window
    pdftitle={My title},    % title
    pdfauthor={Author},     % author
    pdfsubject={Subject},   % subject of the document
    pdfcreator={Creator},   % creator of the document
    pdfproducer={Producer}, % producer of the document
    pdfkeywords={keyword1} {key2} {key3}, % list of keywords
    pdfnewwindow=true,      % links in new window
    colorlinks=true,       % false: boxed links; true: colored links
    linkcolor=magenta, %red,          % color of internal links (change box color with linkbordercolor)
    citecolor=blue,        % color of links to bibliography
    filecolor=magenta,      % color of file links
    urlcolor=blue           % color of external links
} 

\usepackage{amsfonts}
\usepackage{upgreek}
\usepackage{slashed}
\usepackage{latexsym}

\usepackage{scrpage2}			
\usepackage{graphicx}					
\usepackage{amsfonts}
\usepackage{amsthm}

\usepackage{xcolor}
\usepackage{hyperref}

\newcommand{\beq}{\begin{equation}} 
\newcommand{\eeq}{\end{equation}}
\newcommand{\bqa}{\begin{eqnarray}} 
\newcommand{\eqa}{\end{eqnarray}}
\newcommand{\nn}{\nonumber \\}

\renewcommand{\vec}[1]{{\mathbf{#1}}}

\def\be{\begin{eqnarray}}
\def\ee{\end{eqnarray}}

\begin{document}

\title{UV/IR Mixing In Non-Fermi Liquids: Higher-Loop Corrections In Different Energy Ranges}

\author{Ipsita Mandal\\
{\normalsize{Perimeter Institute for Theoretical Physics,}}
{\normalsize{31 Caroline St. N., Waterloo ON N2L 2Y5, Canada}}
}

\date{\today}

\begin{abstract}
We revisit the Ising-nematic quantum critical point
with an $m$-dimensional Fermi surface by applying a dimensional regularization scheme, introduced in  Phys.~Rev.~B~92,~035141~(2015). We compute the contribution from two-loop and three-loop diagrams in the intermediate energy range controlled by a crossover scale. We find that for $m=2 $, the corrections continue to be one-loop exact for both the infrared and intermediate energy regimes.
\end{abstract}

\maketitle

\tableofcontents

\section{Introduction}

Unconventional metallic states 
lying outside the framework of Laudau Fermi liquid theory have been the subject of intensive studies \cite{holstein,reizer,leenag,HALPERIN,polchinski,ALTSHULER,YBKim,nayak,SSLee,Shouvik1,metlsach1,metlsach,chubukov1,Chubukov,mross,Jiang,Lee-Dalid,shouvik2,lawler1,ips1,ips-subir,ips-sudip1,ips-sudip2} in the recent times.
From the point of view of condensed matter systems, we want to construct minimal field theories
that can capture universal low-energy physics, thus enabling us to understand the dynamics in controlled ways.

Non-Fermi liquids arise when a gapless boson is coupled with a Fermi surface.
Depending on the system, the critical boson  can carry zero momentum or some finite momentum. In the former case,
examples include the Ising-nematic critical point \cite{metlsach1,ogankivfr,metzner,delanna,kee,lawler1,lawler2,rech,wolfle,maslov,quintanilla,yamase1,yamase2,halboth,
jakub,zacharias,YBKim,huh,ips1,ips-subir} and the Fermi surface coupled with an emergent gauge field \cite{MOTRUNICH,LEE_U1,PALEE,MotrunichFisher,ips-sudip1,ips-sudip2}, when the fermions lose
coherence across the entire Fermi surface.
The latter scenario is realised in systems like
the spin density wave (SDW) or charge density wave (CDW) critical points \cite{metlsach,chubukov1,Chubukov,shouvik2,sur16,ipsc2}, 
where electrons on hot spots (or hot lines) play a special role 
because these are the ones which remain strongly coupled with the critical boson in the low energy limit. 
The above systems are examples of critical Fermi surfaces where the 
Fermi surfaces are well-defined through weaker non-analyticities (such as power-law singularities) 
of the electron spectral function \cite{sudip,senthil}, although there is no finite jump or discontinuity in the electron occupation number as is seen in Fermi liquids.
The Fermi surface at the quantum critical point is thus identified from a non-analyticity of the spectral function. The latter
is inherited from that of the underlying Fermi liquid before the coupling with a gapless boson is turned on right at the quantum phase transition point. The effect of such coupling of the Fermi surface with critical bosons on potential pairing instability is another topic which has been examined carefully \cite{ips-sudip1,ips-sudip2,Max,ips-sc}.

In this paper, we will focus on the Ising-nematic quatum critical point. This system is worthy of
investigation because there has been considerable experimental evidence that a nodal nematic phase occurs in certain cuprate  superconductors in the underdoped regime, and probably a quantum phase transition occurs from this anisotropic state to an isotropic one. Measurements of strongly temperature-dependent transport anisotropies \cite{exp1} and neutron scattering experiments \cite{exp2} performed on such materials provide such evidence.
Let us first review the dimensional regularization scheme that has been devised to study such critical points \cite{Lee-Dalid,ips1}.
Denoting the dimension of 
Fermi surface by $m$ and the space dimensions by $d$, the number of spatial dimensions perpendicular to the Fermi surface 
is given by $(d-m) $.
while $d$ controls the strength of quantum fluctuations, the $m$ tangential directions control the extensiveness of gapless modes.
Tuning $d$, we can compute the upper critical dimension $d_c(m)$ as a function of $m$, such that theories below upper critical dimensions
flow to interacting non-Fermi liquids at low energies,
whereas systems above upper critical dimensions
are expected to be described by Fermi liquids.
In our earlier work in Ref.~\cite{ips1}, we have shown that
theories with $m = 1$ are fundamentally different 
from those with $m>1$. This
is due to an emergent locality in momentum space that
is present for $m = 1$ \cite{SSLee}.
On the other hand, for non-Fermi liquids
with $m > 1$, any naive scaling
based on the patch description is bound to break down as the size
of Fermi surface ($k_F$) qualitatively modifies the scaling. This is the result of a UV/IR
mixing, where low-energy physics is affected by gapless
modes on the entire Fermi surface in a way that their
effects cannot be incorporated within the patch description \cite{ips1}.

In Ref.~\cite{ips1}, we identified the upper critical dimension $d_c (m)$ at which
the one-loop fermion self-energy diverges logarithmically. Using $\epsilon =d_c (m) - d$ as an expansion parameter, we could perturbatively access the stable non-Fermi liquid states that arise in $d < d_c (m)$.
While computing two-loop corrections, we found that there exists a crossover scale defined by the dimensionless quantity,
%%%%
\beq
\lambda_{\text{cross}} \equiv
 \tilde{e}^{2} \, 
\left ( \frac{  k_F }  {\Lambda  } \right )^{ m-1 } \,,
\label{crossover}
\eeq
%%%%%%%%%%%
where $\tilde e$ is the effective coupling constant which remains small during perturbative expansions and $\Lambda $ is the Wilsonian cut-off for energy scales and momenta away from the Fermi surface.
For $m=1$, the $k_F$-dependence drops out from everywhere. 
Since $\tilde{e} \sim \mathcal{O} (\epsilon)$ within the perturbative window, 
one always deals with the limit
\beq
\label{lim3}
\lambda_{\text{cross}} << 1 \quad \mbox{for} \,\, m=1\,.
\eeq
The $m>1$ case is quite different. In the $\lambda_{\text{cross}} >> 1$ limit for $m>1$, the higher-loop corrections have been shown to be suppressed by
positive powers of $\tilde e$ and $\Lambda/k_F$. Due to this suppression by $1/k_F$, there is no logarithmic or higher-order divergence at the critical dimension. As a result, the critical exponents are not modified by the two-loop diagrams in the $k_F \rightarrow \infty$ limit. However, there exists a large energy window for small $\epsilon$ and $(m-1)$, before 
the theory enters into the low-energy limit controlled by $\lambda_{cross} >> 1$. In this paper, we will carry out two-loop and three-loop calculations in this intermediate energy scale characterized by $\lambda_{cross} << 1$, in order to examine whether there are non-trivial quantum corrections from higher-loop diagrams for $m>1$. We will also compute those three-loop diagrams in the $\lambda_{\text{cross}} >> 1$ limit and confirm that the $\Lambda/k_F$-suppression continues to hold as predicted in Ref.~\cite{ips1}.

The paper is organized as follows: In Sec.~\ref{model}, we revisit the action which describes the Ising-nematic quantum
critical point for a system with an $m$-dimensional Fermi surface embedded in $d$ spatial dimensions, providing a way for achieving perturbative control by dimensional regularization. 
In Sec.~\ref{rg}, we continue to review the renormalization group scheme applied for locating the infrared fixed point.
In Sec.~\ref{cscale}, we explain the crossover scale that governs the transition from infrared to intermediate energy scales. The counterterms obtained from  two-loop diagrams are discussed in Sec.~\ref{hcounter}, followed by a computation of the critical exponents in Sec.~\ref{crexp}. We conclude with a summary and some outlook in Sec.~\ref{conclusion}. 
Details on the computation of the Feynman diagrams at two-loop and three-loop orders can be found in Appendices~\ref{app:twoloop} and \ref{3loop} respectively.

%%%%%%%%%%%%%%%%%%%%%%%%
\section{Model}
\label{model}
 %%%%%%%%%%%%%%%%%%%%%%%%%%%%
 In the patch coordinate system used in Ref.~\cite{ips1}, the action for the Ising-nematic critical point involving an $m$-dimensional Fermi surface embedded in $d$ spatial dimensions, can be written as
\begin{eqnarray}
\label{act2}
S  &=&   \sum_{j} \int dk \bar \Psi_j(k)
\Bigl[ 
i \vec \Gamma \cdot \vec K  
+ i \gamma_{d-m} \, \delta_k \Bigr] \Psi_{j}(k) \, \exp \Big \lbrace \frac {{\vec{L}}_{(k)}^2}  { \mu \, {\tilde{k}}_F } \Big \rbrace +
\frac{1}{2} \int  dk ~
  {\vec{L}}_{(k)}^2\,  \phi(-k) \, \phi(k) \nonumber \\
&& +     \frac{i \, e \, \mu^{x/2} }{\sqrt{N}}  \sum_{j}  
\int dk dq  \,
\phi(q) \, \bar \Psi_{j}(k+q)\,  \gamma_{d-m} \Psi_{j}(k) \,.
\end{eqnarray}
Here, $\vec K ~\equiv ~(k_0, k_1,\ldots, k_{d-m-1})$ includes
the frequency and the first $(d-m-1)$ components 
of  the $d$-dimensional momentum vector, ${\vec L}_{(k)} ~\equiv~ (k_{d-m+1}, \ldots, k_{d})$ and $\delta_k =  k_{d-m}+ {\vec{L}}_{(k)}^2$.
In the $d$-dimensional momentum space,
$k_1,..,k_{d-m}$ (${\vec L}_{(k)}$) represent(s) the
$(d-m)$ ($m$) directions perpendicular (tangential) to the Fermi surface.
The spinor $ \Psi_j^T(k) = \left( 
\psi_{+,j}(k),
\psi_{-,j}^\dagger(-k)
\right)$ includes the right and left moving fermion fields $\psi_{+,j}(k )$ and $\left ( \psi_{-,j}(k ) \right)$
with flavour $j=1,2,..,N$.
$\vec \Gamma \equiv (\gamma_0, \gamma_1,\ldots, \gamma_{d-m-1})$ represents the gamma matrices associated with $\vec K$.
Since we are ultimately interested in the physical situations when co-dimension $1 \leq d-m \leq 2$, 
we consider only $2 \times 2$ gamma matrices with
$\gamma_0= \sigma_y , \, \gamma_{d-m} = \sigma_x$
and $\bar \Psi \equiv \Psi^\dagger \gamma_0$. 
The theory has an implicit UV cut-off for $\vec K$ and $k_{d-m}$,
which we denote as $\Lambda$ and we are interested in the limit $\Lambda \ll k_F$ corresponding to the low energy effective action.
Here the dispersion is kept parabolic, while the exponential factor effectively makes
the size of the Fermi surface finite
by damping the propagation of fermions 
with $|{\vec{L}}_{(k)}| > k_F^{1/2}$ as the bare fermion propagator is given by 
$G_0 (k) =\frac{1}{i} \, \frac{\vec \Gamma \cdot \vec K +
\gamma_{d-m} \delta_k} 
{\vec K^2  + \delta_k^2} 
%\, \times \, 
\exp \Big \lbrace - \frac {{\vec{L}}_{(k)}^2}  { \mu \, {\tilde{k}}_F } \Big \rbrace$.

%%%%%%%%%%%%%%%%%%
 \begin{figure}
 \centering
 \subfloat[][]{\includegraphics[width=0.25 \textwidth]{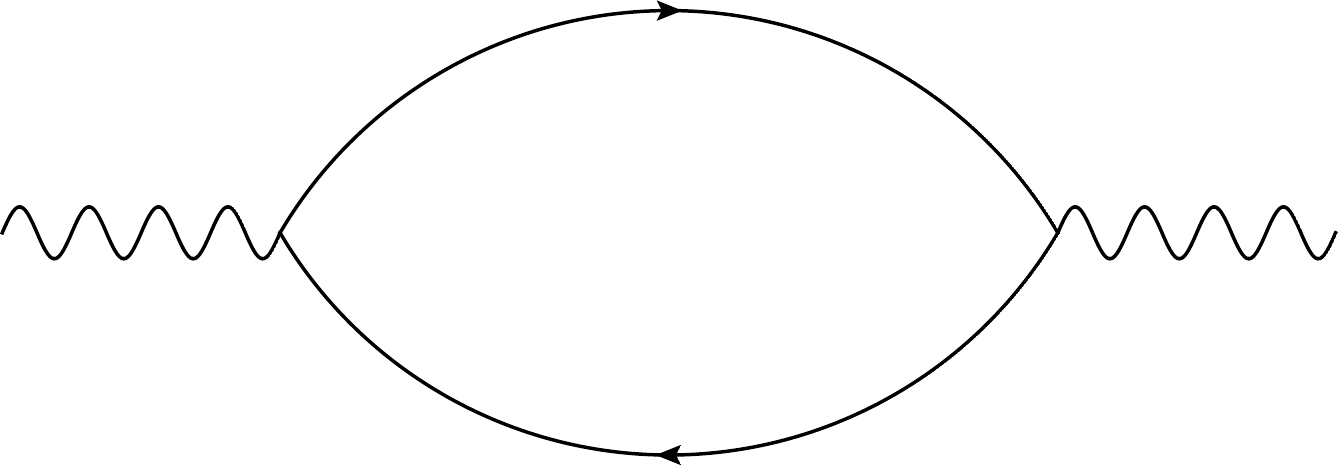}}\quad
 \subfloat[][]{\includegraphics[width=0.25\textwidth]{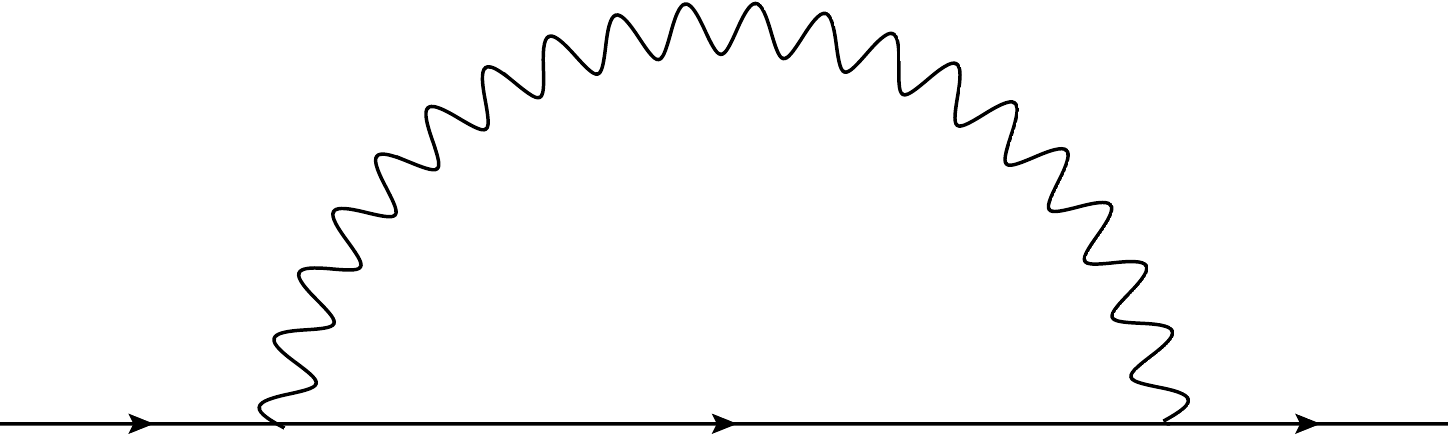}}\quad 
 \subfloat[][]{\includegraphics[width=0.25\textwidth]{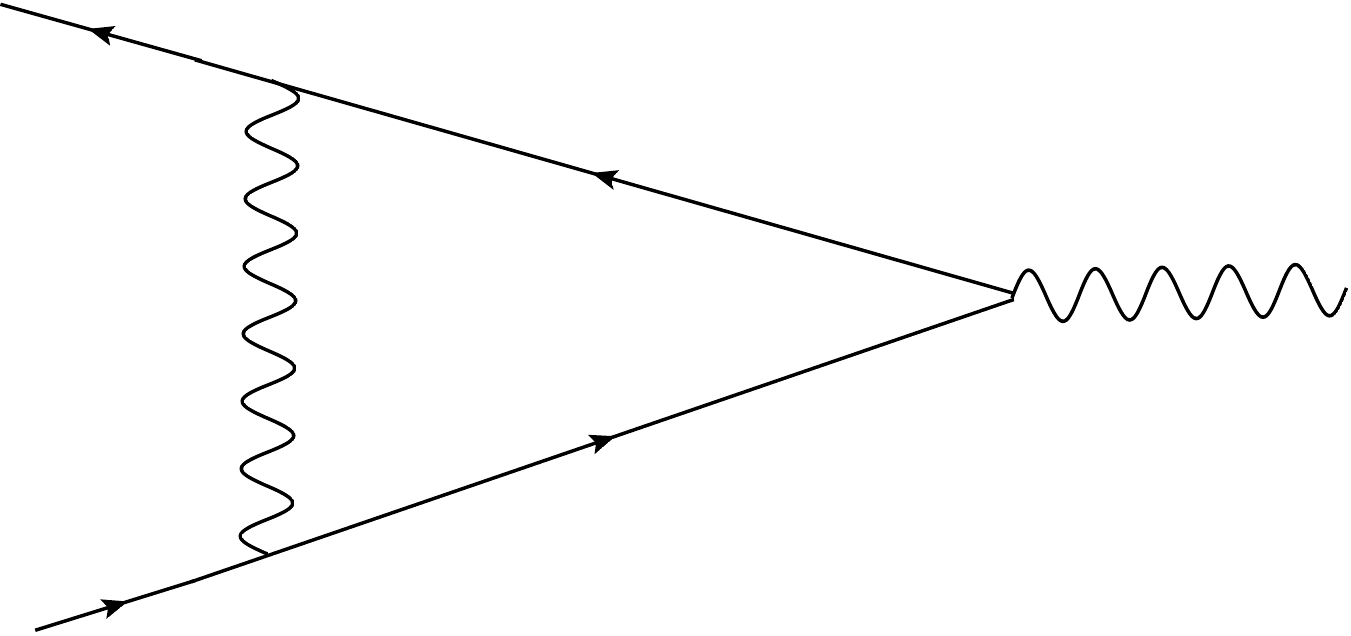}}
 \caption{\label{oneloop}
The one-loop diagrams for (a) the boson self-energy, (b) the fermion self-energy, and (c) the vertex correction (c).
Solid lines represent the bare fermion propagator, whereas wiggly lines in (b) and (c) represent the dressed boson propagator which includes the one-loop bosonic self-energy correction computed from (a).}
 \end{figure}
%%%%%%%%%%%%%%%%%%%%%%

 Let us review the results found from the one-loop diagrams in Fig.~\ref{oneloop} for the above action.
 The dressed boson propagator, which includes the one-loop self-energy is given by
\beq
\label{babos}
D_1(k) = \frac{1}{ {\vec{L}}_{(k)}^2 +\beta_{d} \, e^2 \, \mu^{x}
\displaystyle\frac{ ( \mu \, {\tilde{k}}_F )^{ \frac{m-1}{2}}  \, |\vec K|^{d-m}}{ |\vec{L}_{(k)}| } } \,,
\eeq
to the leading order in $k/k_F$,
for $|\vec K|^2/|\vec L_{(k)}|^2, ~\delta_k^2/|\vec L_{(k)}|^2 << k_F$. Here
\bqa
\beta_d = \frac{  \Gamma^2 (\frac{d-m+1} {2})}
{2^{\frac{2d+m-1}{2}}  \pi^{\frac{d-1}{2}}  | \cos \lbrace  \frac{\pi (d-m+1)} {2} \rbrace |   \Gamma(\frac{d-m}{2}) \Gamma (d-m+1)}
\nn
\eqa
is a parameter of the theory that 
depends on the shape of the Fermi surface.
The one-loop fermion self-energy $\Sigma_1 (q)$ blows up logarithmically 
in $\Lambda$ at the critical dimension
\beq
d_c(m) = m + \frac{3}{m+1} \,.
\eeq
Now we consider the space dimension $d=d_c(m) - \epsilon$.
%% for a given $m$.
In the dimensional regularization scheme, 
the logarithmic divergence in $\Lambda$
turns into a pole in $1/\epsilon$:
\beq
\label{sigma4}
\Sigma_1 (q) = 
\left( - \frac{e^{ 2 (m+1 ) /3}  } {{\tilde{k}}_F ^{ \frac{(m-1) (2-m)} {6}}N} 
\frac{u_1}{\epsilon} 
+ \mbox{ finite terms}
\right)
(i \vec \Gamma \cdot \vec Q) \,,
\eeq
to the leading order in $q/k_F$, where
\bqa
\label{u1}
u_1 &=&  \frac{ 1   }
{ \pi^{\frac{m-2}{2}}
(4 \pi)^{\frac{3}{2(m+1)}} 2^{m-1}   | \sin \lbrace (m+1) \pi/3 \rbrace |  \beta_{d}^{\frac{2-m}{3}}  (m+1) } 
\times \frac{  \Gamma (  \frac{m+4}{2(m+1)})  } 
{  \Gamma{(m/2)}  \Gamma(\frac{2-m}{2(m+1)})  \Gamma (  \frac{2m+5}{2(m+1)}) }\,.
\eqa
The one-loop vertex correction vanishes due to a Ward identity \cite{Lee-Dalid}.

\section{Renormalization Group Equations}
\label{rg}
\

To remove the UV divergences in the $\epsilon \rightarrow 0$ limit,
we add counterterms using the minimal subtraction scheme, such that
the renormalized action is given by:
\begin{widetext}
\begin{eqnarray}\label{act7}
S_{ren} & = &  \sum_{j} \int d k_B
\, \bar \Psi_{Bj}(k_B)
\Bigl[ 
i  \vec \Gamma \cdot \vec K_B + 
i \gamma_{d-m} \delta_{k_B}  \Bigr] \Psi_{Bj}(k_B) \, \exp \Big \lbrace \frac {{\vec{L}}_{(k),B}^2}  {  k_{F,B} } \Big \rbrace
+
\frac{1}{2} \int  {d k_B}  ~
 {\vec{L}}_{(k)}^2\,  \phi_B(-k_B) \, \phi_B(k_B) \nonumber \\
 && +     \frac{i \, e_B }{\sqrt{N}}  \sum_{j}  
\int d k_B\, d q_B \,
\phi_B(q_B) \, \bar \Psi_{Bj}(k_B+q_B) 
\, \gamma_{d-m} \Psi_{Bj}(k_B) \, ,
\end{eqnarray}
\end{widetext}
where
\bqa
\vec K & = & \frac{Z_2}{Z_1} \vec K_B \, , \quad k_{d-m} = k_{B, d-m} \, , \quad {\vec{L}}_{(k)} = {\vec{L}}_{(k), B} \,, \nn
\Psi(k) & = &  Z_\Psi^{-1/2} \,\Psi_B(k_B)\,, \quad \phi(k) = Z_\phi^{-1/2} \phi_B(k_B)\,, \nn
e_B & = & Z_3^{-1/2} \left( \frac{Z_2}{Z_1} \right)^{(d-m)/2} \mu^{ x/2} \, e \,,\quad   k_F =\mu \, {\tilde{k}}_F \,,
\eqa
with 
\beq
Z_\Psi =  Z_2 \left( \frac{Z_2}{Z_1} \right)^{(d-m)}\,, \quad
Z_\phi = Z_3 \left( \frac{Z_2}{Z_1} \right)^{(d-m)}\,.
\eeq
%%%%%%%%%%%%%%
The subscript ``B'' denotes the bare quantitites.

The renormalized Green's functions, defined by
\bqa
&&
\Bigl< \phi( k_{1} ) .. \phi(k_{{n_\phi} })
\Psi( k_{{n_\phi} +1} ).. \Psi(k_{{n_\phi}+n_\psi })
\bar \Psi( k_{n_{\phi}+n_\psi+1} )  .. \bar \Psi(k_{n_\phi+2 n_\psi} ) \Bigr> \nn
%%%%%%%%%%%%%%%%%
&& = G^{({n_\psi} ,{n_\psi} , {n_\phi} )}( \{ k_i \}; \tilde e, \tilde k_F , \mu )
\delta^{d+1} \left( \sum_{i=1}^{{n_\phi+n_\psi} } k_i  - \sum_{j={n_\phi} + n_\psi +1}^{2 {n_\psi} + {n_\phi}} k_j \right),\nonumber
\eqa
satisfy the RG equations
%%%%%%%%%%%%%%%%%%%
\bqa
\Bigg\{
&&
- \sum_{i=1}^{2 {n_\psi}  + {n_\phi} } \left(
z \, \vec K_i \cdot \nabla_{K_i}
+ k_{i, d-m}  \frac{\partial}{\partial k_{i,d-m}}
+ \frac{ \vec{L}_{(k_i)} } {2}   \cdot  \nabla_{{L}_{(k_i)}}
\right)  
- \frac{d \tilde k_F}{dl}  \frac{\partial} {\partial \tilde k_F}  
-  \frac{d {\tilde e}}{dl}   \frac{\partial}{\partial {\tilde e}} \nn
&&
 + 2  \, {n_\psi}  \left ( -  \frac{2 \, d_c - 2 \, \epsilon + 4 -m }{4} +  \eta_\psi \right)
+  {n_\phi} \left ( -  \frac{2 \, d_c - 2 \, \epsilon +4 -m }{4} +  \eta_\phi \right) \nn
&&
+  d_c - \epsilon+1 - \frac{m}{2}
+ (d_c - \epsilon - m) (z-1)
\Bigg\}
\, G^{({n_\psi} , {n_\psi}  ,{n_\phi})}(\{ k_i \}; {\tilde e}, \tilde k_F , \mu ) 
=0 \,. 
\label{RGeq}
\eqa
%%%%%%%%%%%%%
Here
\beq
{\tilde e} \equiv 
\frac{e^{ 2 (m+1 ) /3}  } {{\tilde{k}}_F ^{ \frac{(m-1) (2-m)} {6}}} \,,
\label{eq:eeff}
\eeq 
%%%%%%%%%%%%
$z$ is the dynamical critical
exponent,  and $n_\psi$ ($n_\phi$ ) is the anomalous
dimensions for the fermion (boson), which can be expressed as
\bqa
z=1 + \frac { \partial \ln (Z_2/Z_1)} { \partial l}  ,
\,  n_\psi =-\frac{1} {2} \frac { \partial \ln (Z_{\psi})} { \partial l} ,
\,   n_\phi =-\frac{1} {2} \frac { \partial \ln (Z_{\phi})} { \partial l}\,. \nn
\eqa
%%%%%%%%%%%%
Earlier, from the computation of one-loop beta functions \cite{ips1}, it has been established that
the higher order corrections are controlled not by $e$, 
but by the effective coupling ${\tilde e}$.
The one-loop beta function for $\tilde e$ is given by
\beq
\label{bstable}
\frac{d {\tilde e} }{dl}  = \frac{(m+1)\, \epsilon}{3} \, {\tilde e} \,
%%- \frac{ (m+1) \lbrace 3 - (m+1) \epsilon \rbrace \, u_1} {9N}  {\tilde e}^2
- \frac{ (m+1)\,  u_1} {3N} \, {\tilde e}^2 \,,
\eeq
to order ${\tilde e}^2$, which shows that that there is an IR stable fixed point at
\beq
\label{efp}
{\tilde e}^*= \frac{N \epsilon} {u_1}  + \mathcal{O}(\epsilon^2)\,.
\eeq
%%%%%%%%%%%%%%%%%%
%%%%%%%%%%%%%%%%%
\section{Crossover Scale}
\label{cscale}

The interplay between $k_F$ and $\Lambda$ plays an important role for $m>1$ in determining the magnitudes of the higher-loop corrections \cite{ips1}.
Let $k = ( {\bf K}, k_{d-m}, \vec L_{(k)} )$ be the momentum that flows through a boson propagator 
within a two-loop or higher-loop diagram.
When $|{\bf K}|$ is of the order $\Lambda$,
the typical momentum carried by a boson along the tangential direction of the Fermi surface is given by
%%%%%%%%%%%%%%
\beq
\label{cross}
 |{\vec{L}}_{(k)} |^3 \sim 
\tilde \alpha  \, \Lambda^{d-m} 
\,,
\eeq
%%%%%%%%%
where
\beq 
 \tilde{\alpha} =   \beta_{d} \, e^2 \, \mu^{x} \,( \mu \, {\tilde{k}}_F )^{ \frac{m-1}{2}},
 \label{alpha}
\eeq
as can be seen from the form of the boson propagator in Eq.~(\ref{babos}).
If $ \left( \tilde \alpha  \, \Lambda^{d-m}  \right)^{1/3} >> \Lambda^{1/2} $,
the momentum imparted from the boson to fermion is much larger than $\Lambda^{1/2}$, supressing the loop contributions by a power of $\Lambda/k_F$ at low energies.
On the contrary, no such suppression arises if $ \left( \tilde \alpha  \, \Lambda^{d-m}  \right)^{1/3} << \Lambda^{1/2} $.
The crossover is controlled by the dimensionless quantity,defined in Eq.~(\ref{crossover}),
which determines whether $ \left( \tilde \alpha  \, \Lambda^{d-m}  \right)^{1/3} >> \Lambda^{1/2} $ or 
$ \left( \tilde \alpha  \, \Lambda^{d-m}  \right)^{1/3} << \Lambda^{1/2} $.

%%%%%%%%%%%%%%%%%%%%%%%%%%%%%%%%%
\section{Counterterms at two-loop level}
\label{hcounter}

It has been demonstrated earlier \cite{ips1} that all loop corrections beyond one-loop level are expected to be suppressed by positive powers of $\tilde e$ and $\Lambda/k_F$ in the $\lambda_{\text{cross}} >> 1$ limit for $m>1$.
Here we focus on the two-loop corrections for the $\lambda_{cross} << 1$ limit, which includes the $m=1$ case. The details of the computation can be found in Appendix~\ref{app:twoloop}.
We have used $\Pi_2 (q) $ to denote the two-loop boson self-energy obtained from Fig.~\ref{fig:bos2}(a).
$\Sigma_{2a} (q)$ and $\Sigma_{2b} (q) $ are the fermion self-energy corrections computed from Fig.~\ref{fig:ferm2}(a), 
which are proportional to $\gamma_{d-m} \, \delta_q$ and $( \vec \Gamma \cdot \vec Q ) $ respectively. 
Other diagrams in Figs.~\ref{fig:bos2}(b)-(e) and \ref{fig:ferm2}(b)-(c) do not contribute \cite{Lee-Dalid}.
From the Ward identity, the vertex correction
at the two-loop level can be obtained from
the two-loop fermion self-energy correction.

Being UV-finite, the diagram in Fig. \ref{fig:bos2}(a) renormalizes $\beta_d$ in the boson propagator
by a finite amount, $\beta_d^{a}  \sim O(\tilde e /N)$,
where
\bqa
\Pi_2 (k) = \beta_d^a \,e^2 \, \mu^x\, k_F^{\frac{m-1} {2} } \,.
\eqa
The numerical factor $\beta_d^a$ can be computed for the desired values of $d$
and $m$ from the expressions in Appendix~\ref{2loopbos}.
Once this correction is fed back to the one-loop fermion self-energy in Eq. (\ref{sigma4}),
we obtain a correction to the UV-divergent fermion-self energy given by:
\bqa
%%%%%%%%%%%
\Sigma^{ba}_2(k) &=& \frac{ \beta_d^{ \frac{2-m} {3} }  }
 { \left( \beta_d - \beta_d^a  \right)^{ \frac{2-m} {3}} }
 \Sigma_1(k) -  \Sigma_1(k) 
 %%%%%%%%5
= \frac{ (2-m) \,  \beta_d^a }  {  3\,  \beta_d }  \Sigma_1(k) 
%%%%%%%%%%%%
= \left(  - \frac{ \tilde e^2  }
{N^2} \frac{ u'_2 } {\epsilon} + \mbox{finite terms}  \right)
(i \vec \Gamma \cdot \vec K ) ,\nn
\eqa
where
\bqa
u'_2 =
-\frac { (2-m) \,\beta_d^a}  {  3\,  \beta_d }  
\frac{ N^2 \,  \Sigma_1 (k)} { \tilde e ^2 \, (i \vec \Gamma \cdot \vec K ) }
\eqa
is a number independent of $\tilde e$, $k$ and $N$.

The two-loop fermion self-energy 
in Fig. \ref{fig:ferm2}(a) 
is given by 
\bqa
\label{twoloopf}
\Sigma_2 (q) 
=  \frac{(i \,e)^4 \mu^{2 \, x}}{N^2} 
\int 
{dp\, dl}  D_1 (p) \, D_1 (l) 
\gamma_{d-m} 
\,G_0 (p+q)
\,\gamma_{d-m}   G_0 (p+l+q)
\,\gamma_{d-m} \,G_0 (l+q)\, \gamma_{d-m}.\nn
\eqa
The computation described in Appendix~\ref{2loopferm} 
gives
\beq\label{twoloopf2}
\Sigma_2 (q) = 
-  \frac{ \tilde e^2 \, u_2 } 
{N^2 \, \epsilon } 
 ( i \, \vec \Gamma \cdot \vec K )
-  \frac{ \tilde e^2 \, v_2} 
{N^2  \, \epsilon}   ( i \, \gamma_{d-m} \delta_k )
+ \mbox{ finite terms},
\eeq
where
$u_2 $ and $v_2$ are obtained from the expressions there.

The counterterms that are necessary 
to cancel the UV divergences upto two-loop level are given by
\bqa
\label{actct2}
 S_{CT}^{(2loop)}  &=& \sum_{j} 
\int dk 
\, \bar \Psi_j(k)\,
[i A_1^{(2)}( \vec \Gamma \cdot \vec K )+ i A_2^{(2)}  
\gamma_{d-m} \delta_k ]\, \Psi_{j}(k) \nn
%%%%%%%%%%%%
&&
 +   \,  A_2^{(2)} \frac{i\, e \, \mu^{x /2} }{\sqrt{N}}
  \sum_{j}  
\int dk \,  dq  
\, \phi(q) \bar \Psi_{j}(k+q) \, \gamma_{d-m} \Psi_{j}(k) \,,\nn
\eqa
where
\bqa\label{Z1and2}
A_{1}^{(2)} = -  \frac{ \tilde e^2 }{N^2} ( u_2 + u'_2)\,, \quad
A_{2}^{(2)} =  -  \frac{ \tilde e^2 }{N^2} v_2 \,.
\eqa

We have also computed some relevant three-loop diagrams in Appendix~\ref{3loop}, both for the $\lambda_{cross} \gg 1$ and $\lambda_{cross} \ll 1$ limits. It is found that none of these diagrams produce a divergent contribution in either limit and the one-loop exactness for $m=2$ continues to hold even in the intermediate energy range characterized by $\lambda_{cross} \ll 1$.

%%%%%%%%%%%%%%%%%
\section{Critical exponents}
\label{crexp}

The counter terms up to the two-loop level are given by
\bqa
Z_{1,1} = - \frac{ \tilde  e  }{N} u_1 -  \frac{ \tilde e^2 }{N^2} ( u_2 + u'_2 )\, , \quad
Z_{2,1} =  -  \frac{ \tilde e^2 }{N^2} \, v_2\, , \quad
Z_{3,1} & = & 0 \,.
\eqa

The beta function for $\tilde e$ is then given by
%%%%%%%%%%
\bqa
\beta= \frac{ (m+1) \,\epsilon}{2} \tilde e
- \frac{  (m+1)^2
\left (  \frac{3 } {m+1}- \epsilon
\right )  \, u_1} 
{ 9 \, N } \tilde e^2 
- \frac{ (m+1)^3 \,
\left (  \frac{3 } {m+1}- \epsilon
\right )
\lbrace  u_1^2 
+\frac {6\, (u_2 + u_2' -v_2) } {m+1}
\rbrace} 
{ 27 \, N^2 } \tilde e^3 \,, \nn
\eqa
%%%%%%%%%%%%
which has a stable interacting fixed point at
\bqa
\frac{ \tilde e^{*} }{N} & = & 
\frac{\epsilon} {u_1}  
- \frac { u_2 +u_2' -v_2 } {u_1^3} \epsilon^2 \,.
\eqa
To the two-loop order, the dynamical critical exponent and 
the anomalous dimensions at the critical point are given by
\bqa
z = 1+ \frac{ m+1 }{3 }  \epsilon
+ \frac{ ( m+1)^2  } {9 }  \epsilon^2 \,, 
\quad
\eta_\psi = -\frac{\epsilon}{2} 
+ \frac { ( m+1 ) \, v_2 } { 3 \, u_1^2 } \epsilon^2 \, ,\quad
\eta_\phi = -\frac{\epsilon}{2} \,. \label{phical2}
\eqa
For $m=2$, we have found that $ u_2 = v_2 =u_2'=0$ for both $\lambda_{cross} \gg 1 $ and $\lambda_{cross} \ll 1 $. The answers for the $m=1$ case reduce to those found in Ref.~\cite{Lee-Dalid}.

%%%%%%%%%%%%%
\section{\label{conclusion}Conclusion}

To summarize, we have revisited the Ising-nematic quantum critical point
with an $m$-dimensional Fermi surface by applying a dimensional regularization scheme. 
We have considered the behaviour of two-loop and three-loop diagrams in the intermediate energy range controlled by a crossover scale determined by the dimensionless parameter $\lambda_{cross}$. We have found that for $m=2 $, the results continue to be one-loop exact for both the infrared and intermediate energy regimes. We have thus shown that the critical exponents at the low-energy fixed point are not modified by these higher-loop diagrams, due to the UV/IR mixing for $m>1$.
This is likely to be the case for all other higher-loop diagrams as well.

A few comments are in order. We would like to stress that UV/IR mixing is not an artifact of the chosen RG scheme, as of course no physical observable should. This is not observed in relativistic field theories where we do not have a finite-density electron-system and hence no concept of Fermi surface or $k_F$. The reason that $k_F$ becomes a ``naked" scaled for $m>1 $ is that the massless boson affects the low-energy physics by inducing strong interactions between the fermionic modes on the entire Fermi  surface. We expect such behaviour to also emerge in systems with finite-density electrons interacting with massless transverse gauge bosons.

%%%%%%%%%%%%%%%%%%%%

\begin{acknowledgments}

We thank 
Denis Dalidovich and Sung-Sik Lee for stimulating discussions.
This research was supported by NSERC, the Templeton Foundation and the Perimeter Institute for Theoretical Physics. Research at Perimeter Institute is supported by the Government of Canada through the Department of Innovation, Science and Economic Development Canada and by the Province of Ontario through the Ministry of Research, Innovation and Science.
\end{acknowledgments}

%%%%%%%%%%%%%%%%%%%%%%%%%%%%%%%%5

\appendix

%%%%%%%%%%%%%%%%%%%%%%5
\section{Computation of the Feynman diagrams at two-loop Level}
\label{app:twoloop}

\begin{figure} 
\centering
\includegraphics[width=0.7  \textwidth]{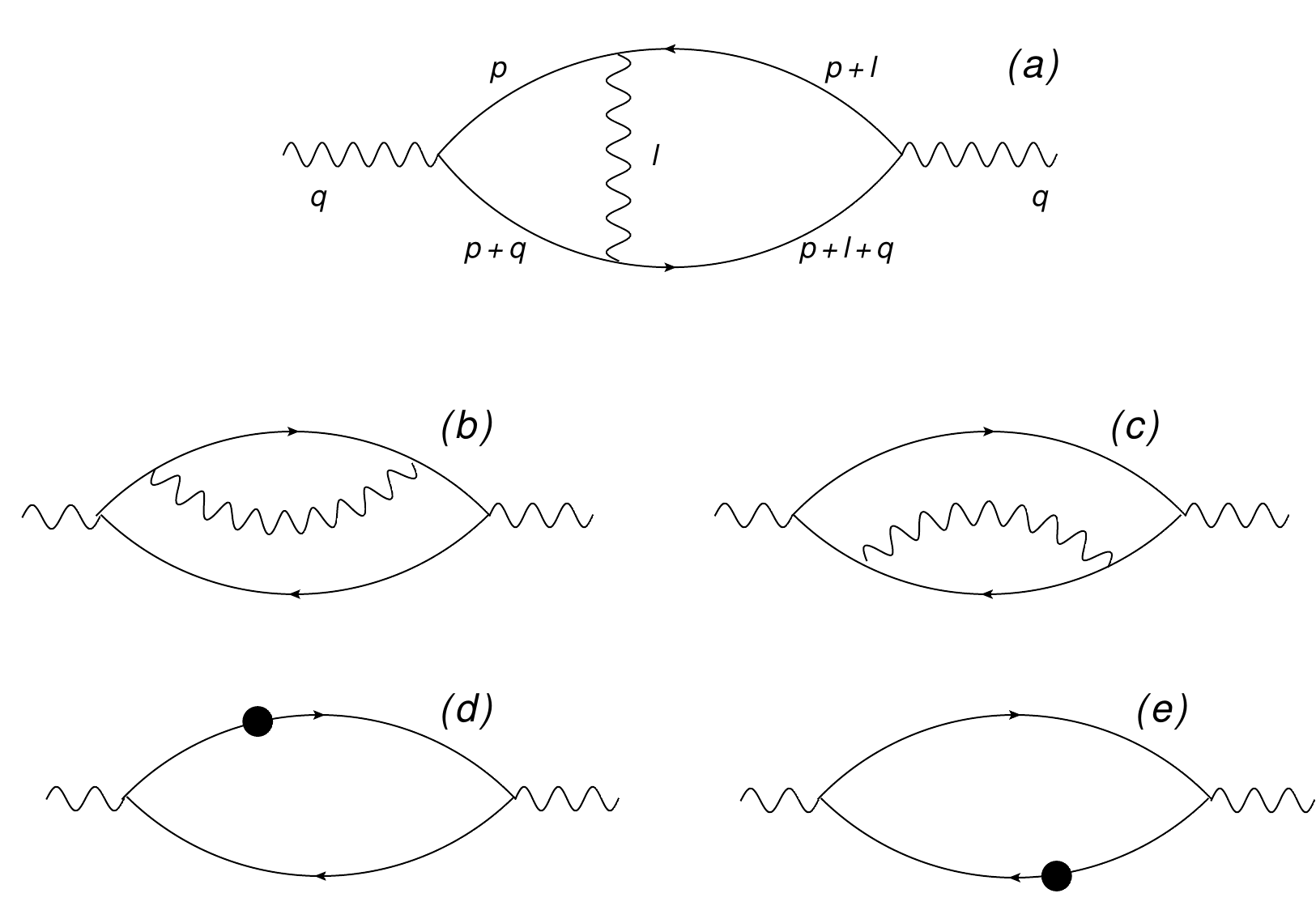}
\caption{The diagrams for two-loop boson self-energy.}
\label{fig:bos2}
\end{figure}
%%%%%%%%%%%%%%%%%%%%%%%%

%%%%%%%%%%%%%%%%%%%%%%%%%%%%
\begin{figure} 
\centering
\includegraphics[width=0.7  \textwidth]{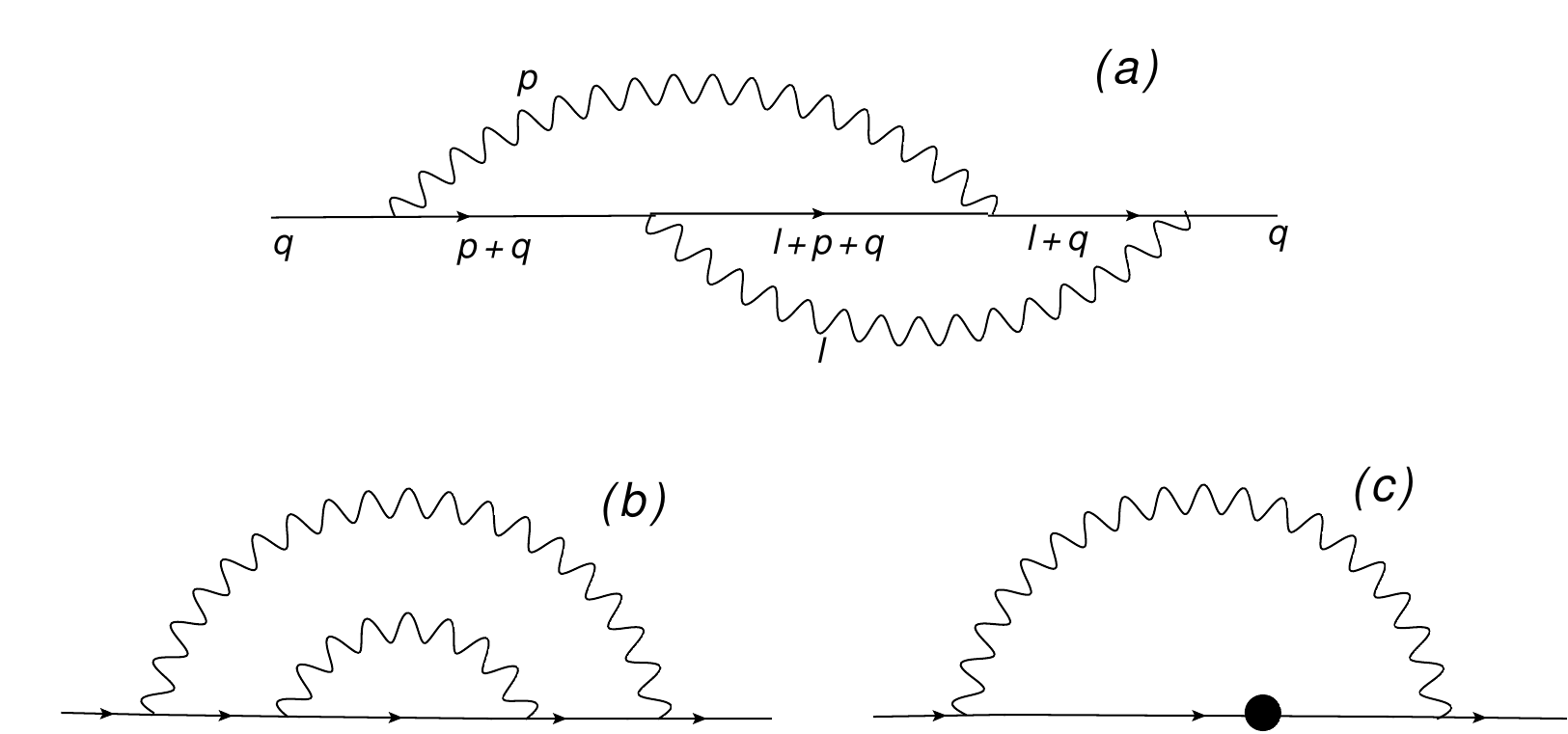}
\caption{The diagrams for two-loop fermion self-energy.}
\label{fig:ferm2}
\end{figure}
%%%%%%%%%%%%%%%%%%%%%%%%%%%

%%%%%%%%%%%%%
\begin{figure}
\centering
\includegraphics[width=0.9  \textwidth]{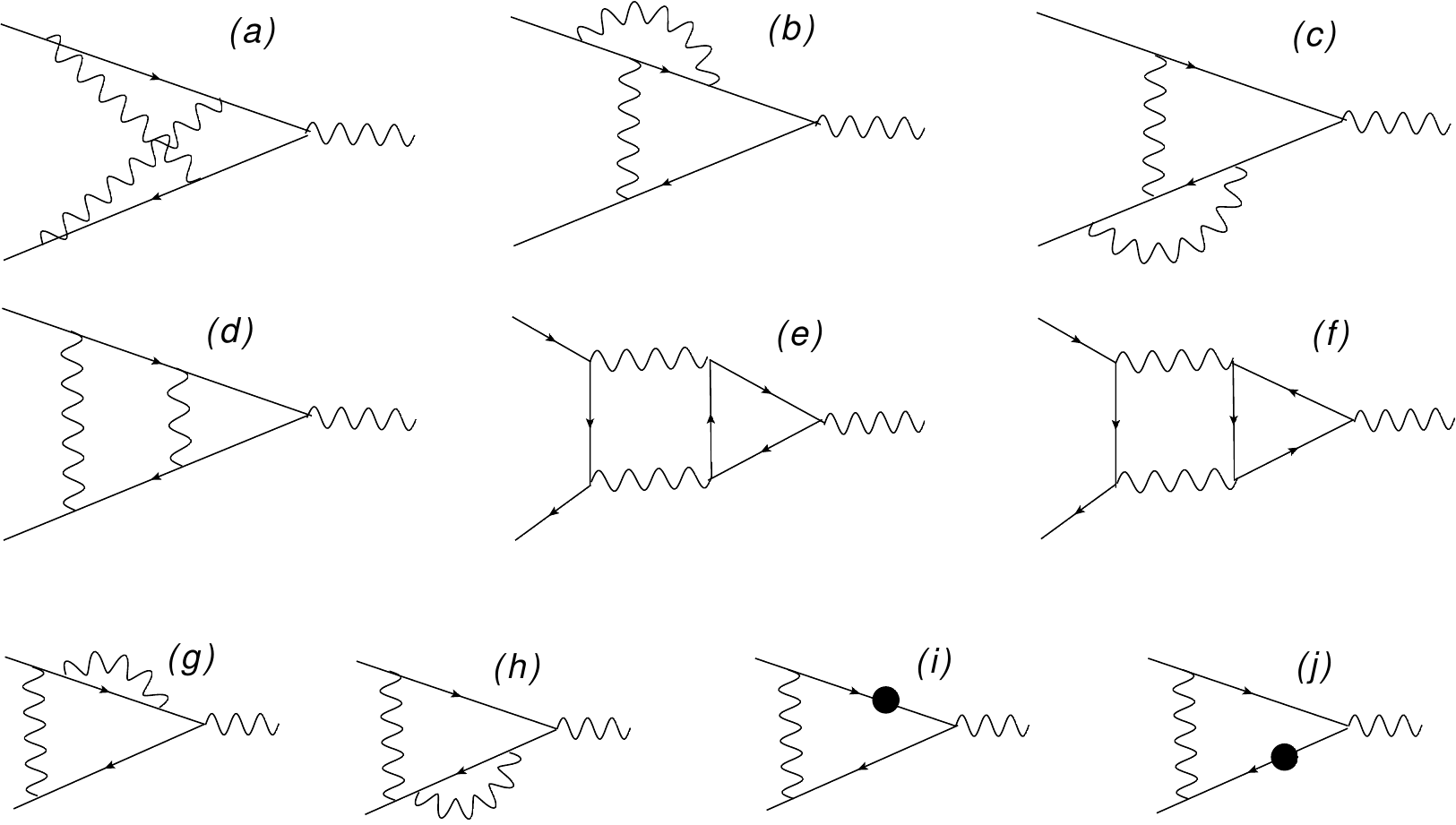}
\caption{The diagrams for two-loop vertex corrections.}
\label{fig:vert2}
\end{figure} 
%%%%%%%%%%%%%%%%%%%

All the two-loop diagrams are shown in 
Figs.~\ref{fig:bos2}, \ref{fig:ferm2} and \ref{fig:vert2}.  
The black circles in  
Figs.~\ref{fig:bos2} (d)-(e), \ref{fig:ferm2}(c) and 
\ref{fig:vert2}(i)-(j) denote the one-loop
counterterm for the fermion self-energy,
\beq
\label{2legv}
i \, A_1^{(1)} \bar \Psi (\vec \Gamma \cdot \vec Q ) \Psi.
\eeq 
%%%%%%%%%%
Among the self-energy diagrams, only Figs.~\ref{fig:bos2}(a) and \ref{fig:ferm2}(a) contribute \cite{Lee-Dalid}.
The vertex correction can be obtained from the  fermion self-energy correction through the Ward identity.

Here we consider the energy limit $ \lambda_{\text{cross}} << 1 $, which includes the case with $m=1$.

%%%%%%%%%%%%%%%%%
%%%%%
%%%%%%%%

\subsection{Two-loop contribution to boson self-energy}
\label{2loopbos}

We compute the two-loop boson self-energy shown 
in Fig.~\ref{fig:bos2} (a):
\beq
\label{prbos2a}
\Pi_2 (q) = -\frac{e^4 \mu^{2\, x }  \, N } {N^2}
 \int {dl\, dp} 
\, D_1 (l) \, {\rm Tr} \{ \gamma_{d-m}\, G_0 (p)\, \gamma_{d-m}\, G_0 (p+l)\, 
\gamma_{d-m} \,G_0 (p+l+q) \,\gamma_{d-m} \,G_0 (p+q) \}\,.
\eeq
%%%%%%%%%%%%%%%%%%%%%%%%%%
Taking the trace, we obtain
\beq\label{prbos2b}
\Pi_2 (q) = -\frac{e^4 \mu^{2\, x}}{N}  \,
\int {dl\, dp} \,D_1(l) \,
\frac{{\cal B}_{1}}{{\cal D}_{1}} \,
\exp\left( {-\frac{{\vec{L}}_{(p)}^2 + {\vec{L}}_{(p+q)}^2 +  {\vec{L}}_{(p+l)}^2+  {\vec{L}}_{(p+l+q)}^2} { k_F }} \right)\,,
\eeq
where
%%%%%%%%%%%%%%%%%%%%%%%%
\bqa
{\cal B}_{1}& =& 2 \,[ \,\delta_{p+l} \, \delta_{p+q+l} -(\vec P +\vec L)
\cdot (\vec P +\vec L +\vec Q)\,]\, 
 \, [ \, \delta_{p+q} \, \delta_{p} -(\vec P +\vec Q)
\cdot \vec P \,] \nn
&& -\,  2 \, [\, (\vec P +\vec L)\cdot (\vec P +\vec Q)\, ]\,
[ \,(\vec P +\vec L +\vec Q)\cdot \vec P] 
\nonumber \\
&& 
+\,  2 \, [ \,(\vec P +\vec L)\cdot \vec P] \,
[ \,(\vec P +\vec L +\vec Q)\cdot (\vec P +\vec Q)\, ] \nn
&&
- \, 2 \, [\, \delta_{p+l} \,  (\vec P +\vec L +\vec Q) + \delta_{p+l+q}  \, (\vec P +\vec L)\, ]
\cdot [\, \delta_{p+q} \, \vec P + \delta_{p} \, (\vec P +\vec Q)\, ] \,, 
\label{B1} \nonumber \\
%%%%%%%%%%%%%%%%%%%%%
{\cal D}_{1} &=& [\, \delta_{p}^2 +\vec P^2] \,  [ \, \delta_{p+q}^2 +(\vec P +\vec Q)^2\,] \,
[ \, \delta_{p+l}^2 +(\vec P +\vec L)^2]\,
[ \, \delta_{p+l+q}^2 +(\vec P +\vec L +\vec Q)^2]\,. \label{D1}
\eqa
%%%%%%%%%%%%%%%%%%%%%%%%%%%%
Shifting the variables as
\beq
p_{d-m } \rightarrow p_{d-m } - \vec L_{(p)}^2\,, \qquad 
l_{d-m } \rightarrow l_{d-m } - p_{d-m } -  \vec L_{(p+l)}^2\,,\nonumber
\eeq
%%%%%%%%
we can substitute
\beq
\delta_{p} \rightarrow p_{d-m} \,, \quad \delta_{p+q} \rightarrow p_{d-m}
+2 \,  \vec L_{(p)} \cdot  \vec L_{(q)} +\delta_q \,, \quad
\delta_{l+p} \rightarrow l_{d-m } \,, \quad \delta_{p+l+q} \rightarrow l_{d-m }
+2 \, \vec L_{(p+l)} \cdot  \vec L_{(q)}  +\delta_q \,. \nonumber
\eeq
%%%%%%%%%%%%%%%%%%%5
Integration over $p_{d-m }$ and $l_{d-m }$ gives us
%%%%%%%%%%%%%%%%%%%%5
\beq
\Pi_2 (q) = -\frac{e^4 \mu^{2\, x }}{N} 
\int \frac{d \vec L_{(l)} \, d\vec L\, d\vec L_{(p)}\, d\vec P } {(2\pi)^{2\,d}}\, D_1 (l)\,
\frac{{\cal B}_{2}}{{\cal D}_{2}}\,
\exp\left( {-\frac{{\vec{L}}_{(p)}^2 + {\vec{L}}_{(p+q)}^2 +  {\vec{L}}_{(p+l)}^2+  {\vec{L}}_{(p+l+q)}^2} { k_F }} \right)\,,
\eeq
where
\bqa
{\cal B}_{2} &=& 2\, \left(\, |\vec P +\vec L|+ |\vec P +\vec L +\vec Q|\, \right)
\left( \,|\vec P +\vec Q|+ |\vec P|\, \right)\nonumber\\
&& \quad \times\, \left. \Bigl\{ 
\left[ \,|\vec P +\vec L| \,|\vec P +\vec L +\vec Q|-
(\vec P +\vec L) \cdot (\vec P +\vec L +\vec Q)\, \right]
\left[ \,|\vec P +\vec Q|\, |\vec P|-
(\vec P +\vec Q) \cdot \vec P  \,  \right] \right.\nonumber\\
&&  \quad \quad \quad - \left. [\, (\vec P +\vec L)\cdot (\vec P +\vec Q)\, ]\,
[ \, (\vec P +\vec L+\vec Q)\cdot \vec P \,]
+ [\, (\vec P +\vec L)\cdot \vec P\, ]\,
[ \, (\vec P +\vec L+\vec Q)\cdot (\vec P +\vec Q)\, ] \right. \Bigr\} \nonumber\\
&&  - 2\, \left( \,2 \,  \vec L_{(p+l)} \cdot  \vec L_{(q)}+\delta_q \right) 
\left( \, 2 \,  \vec L_{(p)} \cdot  \vec L_{(q)} +\delta_q \,\right) \nn
&&  \quad \times\,   \left[\, |\vec P +\vec L +\vec Q |\, (\vec P +\vec L)- 
|\vec P +\vec L| (\vec P +\vec L +\vec Q )\right]\cdot
\left[ \,|\vec P +\vec Q |\, \vec P - 
|\vec P|\, (\vec P +\vec Q )\,\right], \label{B2} \\
%%%%%%%%%%%
{\cal D}_{2} &=& 4 \,  | \vec P|\,|\vec P +\vec Q |
  \,|\vec P +\vec L| \, |\vec P +\vec Q +\vec L| \,
 \left[ \left( \, 2 \, \vec L_{(p+l)} \cdot  \vec L_{(q)}  +\delta_q \,\right)^2 + 
\left(|\vec P +\vec L|+ |\vec P +\vec Q +\vec L| \right)^2  \,
\right] \nonumber\\
&&  \times \, \left[ \, \left(\, 2 \, \vec L_{(p)} \cdot  \vec L_{(q)}  +\delta_q \right)^2 + 
\left(|\vec P |+ |\vec P +\vec Q | \right)^2  \right]. \label{D2}
\eqa
%%%%
Without loss of generality, we can choose the coordinate system such that ${\vec{L}}_{(q)} = ( q_{d-m+1},0, 0,\ldots,0) $ with $q_{d-m+1}>0$.
After making a further change of variables as
\bqa
\vec L  \rightarrow   \vec L -\vec P, \quad
\vec P \rightarrow   \vec P -\frac{\vec Q}{2}\,, \quad
2 \, | \vec L_{(q)} |\, p_{d-m+1} + \, \delta_q   \rightarrow  p_{d-m+1}  \,,
\eqa
and integrating over $p_{d-m+1} $ (neglecting the corresponding exponential damping part), we obtain:
%%%%%%%%%%%%%%%%%%%%%%%%%5
\bqa
\label{prbos2d}
\Pi_2 (q) &\simeq&  -\frac{e^4 \mu^{2\, x}}{N}  \,
\int \frac{d \vec L_{(l)} \, d\vec L}{(2\pi)^{d}} \frac{{d \vec{u}}_{(p)} \, d\vec P}{(2\pi)^{d-1}} \,
D_1(   \vec L_{(l)} ,|\vec L -\vec P| )\,
\frac{{\cal B}_{3}(\vec L, \vec P, \vec Q)}
{{\cal D}_{3} (l, \vec P, q)} 
\, \exp\left( {-\frac{   3\, {\vec{u}}_{(p)}^2 } { k_F }} \right)\nn
%%%%%%%%%%%%%%%%%
%%%
&\simeq &-\frac{e^4 \mu^{2\, x}}{N}  \,\left(\frac{k_F}{12 \, \pi} \right)^{\frac {m-1} {2}}
\int \frac{d \vec L_{(l)} \, d\vec L}{(2\pi)^{d}} \frac{ d\vec P}{(2\pi)^{d-m}} \,
D_1(   \vec L_{(l)} ,|\vec L -\vec P| )\,
\frac{{\cal B}_{3}(\vec L, \vec P, \vec Q)}
{{\cal D}_{3} (l, \vec P, q)} \,,\nn
\eqa
%%%
where
%%%%%%%%%%%%%%%%%%%
\bqa
 {\vec{u}}_{(k)} &=&(k_{d-m+2},\ldots,k_d)\,,\nn
{\cal B}_{3} (\vec L, \vec P, \vec Q) &=&
{\cal B}_{4} (\vec L, \vec P, \vec Q)
\,  \bar {\cal D } (\vec L, \vec P, \vec Q)\,,
 \label{B3} \\
%%%%
{\cal D}_{3} (l, \vec P, q) & = &
8\,| \vec L_{(q)} | \,
{\cal D}_{4} (\vec L, \vec P, \vec Q)
\,  \Bigl\{ \bar {\cal D }^2 (\vec L, \vec P, \vec Q) + 4  ( \vec L_{(q)} \cdot \vec L_{(l)}\,)^2  \Bigr\} \,,
\label{D3}  \nn
%%%%
{\cal B}_{4} (\vec L, \vec P, \vec Q) &=&
 \left( \,|\vec L -\vec Q/2|\,|\vec L +\vec Q/2| 
-\vec L^2 +\vec Q^2/4 \, \right)
\left( \,|\vec P -\vec Q/2| \, |\vec P +\vec Q/2| 
-\vec P^2 +\vec Q^2/4 \, \right) \nonumber\\
&& \quad -   \left. \left[ \, (\vec L -\vec Q/2)\cdot (\vec P +\vec Q/2)\, \right]
\left[\,(\vec L +\vec Q/2)\cdot (\vec P -\vec Q/2)\, \right]\right.\nonumber\\ 
&& \quad +    \left. \left[\, (\vec L -\vec Q/2)\cdot (\vec P -\vec Q/2)\, \right]
\left[ \,(\vec L +\vec Q/2)\cdot (\vec P +\vec Q/2)\, \right]
\right. \nonumber\\
&&  \quad  -  \left. 
 |\vec L +\vec Q/2|\, |\vec P +\vec Q/2|\,
[ \,(\vec L -\vec Q/2 )\cdot (\vec P -\vec Q/2 )\, ] \right.\nonumber\\
&&  \quad +  \left. |\vec L +\vec Q/2|\, |\vec P -\vec Q/2|\,
[\, (\vec L -\vec Q/2 )\cdot (\vec P +\vec Q/2 )\, ] \right.\nonumber\\
&&  \quad  + \left. |\vec L -\vec Q/2|\, |\vec P +\vec Q/2|\,
[ (\vec L +\vec Q/2 )\cdot (\vec P -\vec Q/2 )\,]\right.\nonumber\\
&&   \quad - |\vec L -\vec Q/2|\, |\vec P -\vec Q/2|\,
[ \,(\vec L +\vec Q/2 )\cdot (\vec P +\vec Q/2 )\,] \,, \label{B4} \\
%%%%%%
{\cal D}_{4} (\vec L, \vec P, \vec Q) & = &
 |\vec L-\vec Q/2|\, |\vec L +\vec Q/2| \,
|\vec P-\vec Q/2|\, |\vec P +\vec Q/2 |\,, 
\label{D4} \\
%%%%%%%%
\bar {\cal D } (\vec L, \vec P, \vec Q)  &=&
|\vec L-\vec Q/2|+ |\vec L+\vec Q/2| + 
|\vec P-\vec Q/2|+ |\vec P+\vec Q/2| \,.
\eqa
%%%
%%%
Note that we can ignore the exponential damping part for $\vec L_{(l)}$.

For $ \lambda_{\text{cross}} << 1$, the angular integrals along the Fermi surface directions give a factor proportional to
\bqa
\label{ang1}
 \int d \Omega_{m-1}  \int_0^{\pi} d \theta \,
\frac {  {\bar {\cal D }} (\vec L, \vec P, \vec Q) \, \sin^{m-2} \theta}
{  {\bar {\cal D}}^2(\vec L, \vec P, \vec Q) + 4 (  |\vec L_{(l)}| \, |\vec L_{( q )}| \, \cos \theta )^2   } 
\simeq 
 \int_0^{\pi} \frac { d \theta \,  \sin^{m-2} \theta}
{  {\bar {\cal D}}(\vec L, \vec P, \vec Q)    } 
= 
\frac{ 2\, \pi^{m/2}  }
{ {\bar {\cal D}} (\vec L, \vec P, \vec Q)  \, \Gamma \left ( \frac{m} {2} \right)  } \nn
\eqa
%%%%%%%%%%%%
in the limit $ \frac{ {\bar {\cal D }} (\vec L, \vec P, \vec Q)  }  { 2 \, |\vec L_{(l)}|  \, |\vec L_{( q )}| }  >>  1 $, which is valid when $ |\vec L_{( q )}|^2 <<  \frac{ \Lambda } 
{ \left ( \lambda_{\text{cross}} \right )^{\frac{1} { m+1} }  }  $.
This follows from the fact that
the main contribution to the integral over $|\vec L_{(l)}|$ comes from 
$|\vec L_{(l)}| \sim \tilde \alpha^{\frac{ 1} { 3 } } \, |\vec L-\vec P|^{\frac{d-m}{3}}$ (see also Eqs.~(\ref{crossover}) and (\ref{cross}) ).

Integrating $|\vec L_{(l)}|$, we get
\bqa
%%%%%
\label{p2in}
\Pi_2 (q) &\sim& 
-\frac{e^4 \mu^{2\, x} \, \pi
\,\left(\frac{k_F}{12 \, \pi} \right)^{\frac {m-1} {2}}
} 
{ 24 \, N \, | \vec L_{(q)} |
 \,  \tilde \alpha^{\frac{ (2-m)}{3}} 
\sin \left( \frac{ (m+1) \, \pi } {3}  \right)  }  
%%%%%%%%%%%%%%%%%%%%%%%
\int  \frac{ d\vec L\,d\vec P}{(2\pi)^{2d-2m}} \,
\frac{{\cal B}_{4}(\vec L, \vec P, \vec Q) }
{ {\bar {\cal D}}(\vec L, \vec P, \vec Q) \,
{\cal D}_{4} (\vec L, \vec P, \vec Q) \,|\vec L - \vec P|^{\frac{(d-m)\, ( 2-m)}{3} } 
}  \,.\nn
%\,, \,\, \text{for}\, m<2 \,.\nn
\eqa
%Note that we have ignored the $\frac{ \sqrt{ \pi }  \, \Gamma \left (\frac{m-1} {2} \right) } { \left (\frac{m} {2} \right)} $, since they just contribute non-singular $m$-dependent numbers when we have angular integrals for $m>1$. For  $m=1$, there is no angular integral, and this factor is replaced by unity.
If we work in the the $(d-m)$-dimensional spherical coordinate system
such that
\bqa
\vec P \cdot \vec Q  = |\vec P | \, |\vec Q | \, \cos \theta_p\,,
\quad \vec L \cdot \vec Q = |\vec P |\, |\vec Q | \, \cos \theta_l \,,
\quad \vec P \cdot \vec L = |\vec P | \, |\vec L | ( \cos \theta_p \cos \theta_l + \sin \theta_p \sin \theta_l \cos \phi_l ) \,,\nn
\eqa
the integration measures are given by
\bqa
&& d \vec P = \frac{ 2 \pi^{\frac{d-m-1 } {2}} } { \Gamma \left( \frac{d-m-1 } {2} \right) }
 |\vec P |^{d-m-1} \,  \sin^{d-m-2} \theta_p \, d  |\vec P | \, ~d \theta_p \,,\nn
&&  
d \vec L = \frac{ 2 \pi^{\frac{d-m-2 } {2}} }{\Gamma \left( \frac{d-m-2} {2} \right) } 
 |\vec L  |^{d-m-1} \sin^{d-m-2 } \theta_l \sin^{d-m -3} \phi_l \, d |\vec L  | \, d \theta_l \, d \phi_l \, .\nn
\eqa
The factor $\frac {1} { \Gamma \left( \frac{d-m-1 } {2} \right)}$ from these integration measures then clearly cancels out the apparent divergence from the $ \frac{1} {\sin \left( \frac{ (m+1) \, \pi } {3}  \right) }$ factor in Eq.~(\ref{p2in}) for $m=2$.

In order to extract the leading order dependence on $|\vec Q|$, 
we write $\vec Q = |\vec Q| \, \vec n$, where $\vec n$ is the unit vector along $\vec Q$,
and redefine variables as
\beq
\vec L = \tilde {\vec  L} \,|\vec Q| \,, \quad \vec P  = \tilde {\vec  P} \,|\vec Q|\,.
\eeq
For $d=d_c-\epsilon$, the total powers of $e$ come out out to be $2+\frac{2 \, (m+1) }{3}$. Hence we find that
%%%%%%%%%%%%%%%
\beq
\label{prbos2e}
\Pi_2 (q) \sim 
- \frac{e^2 \, k_F^{\frac{m-1}{2} } \, |\vec Q|^{\frac{3} {m+1} } }
{N \, | \vec L_{(q)} | }
 \tilde e
\int  \frac{ d  \tilde {\vec  L}  \,d \tilde {\vec  P }}
{(2\pi)^{2d-2m}} \,
\frac{{\cal B}_{4}(\tilde {\vec  L} , \tilde {\vec  P }, \vec n ) }
{ {\bar {\cal D }} ( \tilde {\vec  L}, \tilde {\vec  P}, \vec n ) \,
{\cal D}_{4} ( \tilde {\vec  L}, \tilde {\vec  P}, \vec n )
\,| \tilde {\vec  L}  - \tilde {\vec  P} |^{\frac{(d-m)\, ( 2-m)}{3} } }  \,.
\eeq 
%%%%%%%%%%%%
The UV-divergent behaviour will be dictated by the form of the integrand for  
$ |\tilde {\vec L }|\gg 1$ and $  |\tilde {\vec P }| \gg 1$. 
In this limit,
\beq
| \tilde {\vec L } \pm \vec n/2|
\approx | \tilde {\vec L } | \pm \frac{ \tilde {\vec L } \cdot \vec n}
{2\, |\tilde {\vec L } |}
+ \frac{1}{8\,  | \tilde {\vec L } | } - \frac{(\vec n \cdot \tilde {\vec L })^2}{8 \,    | \tilde {\vec L } |^3} \,, 
%%%%%%%%%%%%%%%
\quad 
| \tilde {\vec P } \pm \vec n/2|
\approx | \tilde {\vec P } | \pm \frac{ \tilde {\vec P } \cdot \vec n}
{2\, |\tilde {\vec P } |}
+ \frac{1}{8\,  | \tilde {\vec P } | } - \frac{(\vec n \cdot \tilde {\vec L })^2}{8 \,    | \tilde {\vec P } |^3} \,, 
\eeq
%%%%%%%%%%%%
so that 
\bqa
\label{looppi2}
%%%%%%%%%
&& \frac{{\cal B}_{4}(\tilde {\vec  L} , \tilde {\vec  P }, \vec n ) }
{ {\bar {\cal D }} ( \tilde {\vec  L}, \tilde {\vec  P}, \vec n ) \,
{\cal D}_{4} ( \tilde {\vec  L}, \tilde {\vec  P}, \vec n )
 } \nn
 &\approx& 
 \frac{
 - \tilde {\vec L } ^2 \,  \tilde {\vec P } ^2
 +    (  \tilde {\vec L } \cdot  \tilde {\vec P }  )\, | \tilde {\vec L } |  \, | \tilde {\vec P } |
 +   (  \tilde {\vec L } \cdot   {\vec n }  )^2 \,  \tilde {\vec P }^2
  +  (  \tilde {\vec P } \cdot   {\vec n }  )^2 \,  \tilde {\vec L }^2
  -| \tilde {\vec L } |\, | \tilde {\vec P } | \,  (  \tilde {\vec L } \cdot   {\vec n }  )\,  (  \tilde {\vec P } \cdot   {\vec n }  )
   -(  \tilde {\vec L } \cdot  \tilde {\vec P } )\,  (  \tilde {\vec L } \cdot   {\vec n }  )\,  (  \tilde {\vec P } \cdot   {\vec n }  )
 } 
 { 2 \, | \tilde {\vec L } |^3 \, | \tilde {\vec P } |^3 \left ( | \tilde {\vec L } |+ | \tilde {\vec P } |\right)  } \,,\nn
%%%%%%%%%%
\eqa
which shows that the degree of divergence for the $ \tilde {\vec L }$ and $ \tilde {\vec P }$ integrals is $\frac{1-2 \, m} {m+1}$ at $d=d_c$. This means that
the integrals are convergent and there is no UV divergence. We get a finite correction
\beq
\Pi_2 (q) \sim
 \frac{ \tilde e } {N} \, \Pi_1(q) \,,
\eeq
which is suppressed by $ \frac{  \tilde e } {N}$ compared to the one-loop result. However, the overall coefficient of this correction vanishes at $d-m=1$, as can be clearly seen from Eq.~(\ref{looppi2}).
%%%%%%%%

%%%%%%
%%%%%%%%%%%%%%%%%%%%%%%%%%%%%%%%%%%%%%%%%%%%%5

\subsection{Two-loop contribution to fermion self-energy}
\label{2loopferm}

The two-loop fermion self-energy 
in Fig.~\ref{fig:ferm2}(a) 
is given by
%%%%%%%%%%%%%%%%%%%%%%%% 
\beq\label{twoloopff}
\Sigma_2 (q) = \frac{(ie)^4 \mu^{2\,x}}{N^2} 
\int {dp \,dl} \, D_1 (p)\, D_1 (l) \,
\gamma_{d-m} \, G_0 (p+q) \, \gamma_{d-m} \, G_0 (p+l+q) \, \gamma_{d-m} \, G_0 (l+q) \, \gamma_{d-m} \,.\nonumber
\eeq
%%%%%%%%%%%%%%%%%%%%%%
Using the gamma matrix algebra, we find that 
the self-energy can be divided into two parts:
%%%%%%%%%%%%%%%%%%%%%%%%%%%%%%%%
\beq
\Sigma_2 (q) = \Sigma_{2a} (q) + \Sigma_{2b} (q) \,,
\eeq
%%%%%%%%%%%%%%%%%%%
where
\bqa
\Sigma_{2a,2b} (q) &=& \frac{i \, e^4 \, \mu^{2\,x}}{N^2} 
\int {dp \,dl} \,
 D_1 (p) \,D_1 (l) \frac{{\cal C}_{a,b}}{[(\vec P +\vec Q)^2 +\delta_{p+q}^2]\,
[(\vec P +\vec L+ \vec Q)^2 +\delta_{p+l+q}^2]\,
[(\vec L +\vec Q)^2 +\delta_{l+q}^2] } \,, \nn
\eqa
with
%%%%%%%%%%%%%%%%%%%%%%%%%
\bqa
{\cal C}_{a} =\,  & \gamma_{d-m} & \, \Big [ \delta_{p+q} \,\delta_{p+l+q} \, \delta_{l+q}
-   \delta_{p+l+q} 
\, \lbrace \,  \vec \Gamma \cdot (\vec P +\vec Q) \rbrace
\, \lbrace \,  \vec \Gamma \cdot (\vec L +\vec Q) \,  \rbrace  
\nn
&& - \,
\delta_{l+q} \, \lbrace \, \vec \Gamma \cdot (\vec P +\vec Q)\rbrace
\, \lbrace \,  \vec \Gamma \cdot (\vec P +\vec L+ \vec Q) \rbrace 
-  \delta_{p+q} \, \lbrace \,  \vec  \Gamma \cdot (\vec P +\vec L+ \vec Q)  \rbrace
\, \lbrace \,  \vec \Gamma \cdot (\vec L +\vec Q) 
 \rbrace  \Big ] \,, \nn
%%%%%%%%
{\cal C}_{b} & = & [\, \vec \Gamma \cdot (\vec P +\vec Q) \, ] \,
[ \, \vec \Gamma \cdot (\vec P +\vec L+ \vec Q) \, ] \,
[ \, \vec \Gamma \cdot (\vec L +\vec Q) \, ]\, 
- \delta_{p+q} \,\delta_{l+q}\,  [\vec \Gamma \cdot (\vec P +\vec L+ \vec Q)] 
\nn &&- \delta_{p+l+q} \,\delta_{l+q} 
\, [ \, \vec \Gamma \cdot (\vec P +\vec Q)\, ]
 \,   \,  - \delta_{p+q} \,\delta_{p+l+q} \, [ \, \vec \Gamma \cdot (\vec L +\vec Q) \, ] \,.
\eqa
%%%%%%%%%%%
Shifting the variables as
\beq
p_{d-m} \rightarrow  p_{d-m} -\delta_q -2\,  \vec L_{(p)} \cdot \vec L_{(q)} -  \vec L_{(p)}^2
\,, \quad
l_{d-m} \rightarrow  l_{d-m} -\delta_q -2 \, \vec L_{(l)} \cdot \vec L_{(q)} -  \vec L_{(l)}^2 \,, \nonumber
\eeq
%%%%%%
and integrating over $p_{d-m}$ and $l_{d-m}$, we obtain
%%%%%%%%%%%%%%%%%%%%%%%%
\bqa
\Sigma_{2a} (q) & = & \frac{ i \, e^4 \, \mu^{2\,x}}  {4 \,N^2} 
\int \frac{d\vec P d\vec L}{(2\pi)^{2d-2m}} \frac{d \vec L_{(p)} \,  d \vec L_{(l)}}{(2\pi)^{2m}} \,
\frac{\gamma_{d-m} \,
(\delta_q -2\,  \vec L_{(l)} \cdot \vec L_{( p )} )  \, 
{\bar {\cal C}}_a (\vec L, \vec P, \vec Q)   \,D_1 (p)\, D_1 (l)  }
{ (\delta_q -2\, \vec L_{(l)} \cdot \vec L_{(p)} )^2 + {\bar {\cal C}} (\vec L, \vec P, \vec Q) ^2   } \,,  \nn
\Sigma_{2b} (q) & = & \frac{ i \, e^4 \, \mu^{2\,x}}  {4 \,N^2}  
\int \frac{d\vec P d\vec L}{(2\pi)^{2d-2m}} \frac{d \vec L_{(p)} \,  d \vec L_{(l)}}{(2\pi)^{2m}} \,
 \frac{  {\bar {\cal C}} (\vec L, \vec P, \vec Q) \,
{\bar {\cal C}}_b  (\vec L, \vec P, \vec Q)   \,D_1 (p) \,D_1 (l) }
{(\delta_q -2\, \vec L_{(l)} \cdot \vec L_{(p)} )^2 +  {\bar {\cal C}} (\vec L, \vec P, \vec Q) ^2   } \,, \label{2loosigb}
\eqa
%%%%%%%%%%%%%%%%%%%%%%5
where 
%%%%%
\bqa
%%%%%%%%%%%%%%%%
{\bar {\cal C}} (\vec L, \vec P, \vec Q) &=& |\vec P +\vec Q| + |\vec L +\vec Q|
+|\vec P +\vec L +\vec Q | \,,\nn
%%%%%%%%%%%%%%%%%%%%%%%%%%%%%
{\bar {\cal C}}_a (\vec L, \vec P, \vec Q) & = &
1-\frac{ [\vec \Gamma \cdot (\vec P +\vec Q)] \,
[\vec \Gamma \cdot (\vec  P +\vec L +\vec Q)]}{|\vec P +\vec Q|\,
|\vec P +\vec L +\vec Q|}  - \frac{ [\vec \Gamma \cdot (\vec P +\vec L +\vec Q)] \,
[\vec \Gamma \cdot (\vec L +\vec Q)]}{|\vec P +\vec L + \vec Q|\,
|\vec L +\vec Q|} 
\nn &&
+ \,
\frac{ [\vec \Gamma \cdot (\vec P +\vec Q)] \,
[\vec \Gamma \cdot (\vec L +\vec Q)]}{|\vec P +\vec Q|\,
|\vec L +\vec Q|} \,, \nn
{\bar {\cal C}}_b (\vec L, \vec P, \vec Q) & = &
\frac{ [\vec \Gamma \cdot (\vec P +\vec Q)] \,
[\vec \Gamma \cdot (\vec  P +\vec L +\vec Q)] \,
[\vec \Gamma \cdot (\vec L +\vec Q)]}
{|\vec P +\vec Q|\, |\vec P +\vec L +\vec Q| \, |\vec L +\vec Q|} 
- \frac{[\vec \Gamma \cdot (\vec L +\vec Q)]}{|\vec L +\vec Q|}
+ \frac{[\vec \Gamma \cdot (\vec L + \vec P+ \vec Q)]}
{|\vec L +\vec P+ \vec Q|}
\nn &&
- \, \frac{[\vec \Gamma \cdot (\vec P +\vec Q)]}{|\vec P +\vec Q|} \,.
\label{cab}\nn
%%%%%%%%%%%%
\eqa
%%%

\subsubsection{\underline{For $2-m $ away from zero}}
\label{mneq2}

The angular integrals along the Fermi surface directions give a factor proportional to
\bqa
\label{angcase}
&& \int d\Omega_{m } \, d\Omega_{m-1}
 \int_0^{\pi} d \tilde \theta \,
\frac {  \delta_q -2\,  |\vec L_{(l)}| \,| \vec L_{(p)}| \, \cos  \tilde \theta }
{
\Big \lbrace  {\bar {\cal C}} (\vec L, \vec P, \vec Q) ^2 
+ (\delta_q -2\,  |\vec L_{(l)}| \,| \vec L_{(p)}| \, \cos  \tilde \theta  )^2 
\Big \rbrace^2   }
\sin^{m-2}  \tilde \theta \nn
&&
 \simeq 
%%%%%%%%%%%
 \frac{  4 \, \pi^{m } \, \delta_q }
 {  \Big \lbrace  \delta_q ^2
 + {\bar {\cal C}} (\vec L, \vec P, \vec Q) ^2  
\Big \rbrace
 \, \Gamma^2 \left( \frac{m}{2} \right ) 
} 
 \,, 
\eqa
%%%
for $\Sigma_{2a}$;
and
%%%%%%%%%%%%%%%%%%%%%%%
\bqa
&& \int  d\Omega_{m } \, d\Omega_{m-1}
 \int_0^{\pi}  d  \tilde \theta \,
\frac {   \sin^{m-2}  \tilde \theta}
{
\Big \lbrace  {\bar {\cal C}} (\vec L, \vec P, \vec Q) ^2 
+ (\delta_q -2\,  |\vec L_{(l)}| \,| \vec L_{(p)}| \, \cos  \tilde \theta  )^2 
\Big \rbrace^2   }\nn
& \simeq &
%%%%%%%%%%%%
 \frac{ 4 \, \pi^{m}  }
 {  \Big \lbrace  \delta_q ^2
 + {\bar {\cal C}} (\vec L, \vec P, \vec Q) ^2  
\Big \rbrace
 \, \Gamma^2 \left( \frac{m}{2} \right ) } \,, 
\eqa
%%%%%%%%%%%%%%%%%%
for $\Sigma_{2b }$, when $\frac{ {\bar {\cal C}} (\vec L, \vec P, \vec Q)} {2\, |\vec L_{(l)}|  \, |\vec L_{(p)}| } >>1$ is satisfied. 
For $  \lambda_{ \text{cross}} << 1 $,  the terms only from the limit $\frac{ {\bar {\cal C}} (\vec L, \vec P, \vec Q)} {2\, |\vec L_{(l)}|  \, |\vec L_{(p)}| }  >> 1 $ are important.  This follows from the fact that the main contribution to the integrals over $|\vec L_{(l)}|$ and $|\vec L_{(p)}|$ come from 
$|\vec L_{(l)}| \sim \tilde \alpha^{\frac{ 1} { 3 } } \,  |\vec L|^{\frac{d-m}{3}}$ and $|\vec L_{(p)}| \sim \tilde \alpha^{\frac{ 1} { 3 } } \,  |\vec P|^{\frac{d-m}{3}}$ respectively. 

We can extract the UV divergent pieces 
by setting $\vec Q =0$ for $\Sigma_{2a}(q)$ and
expanding the integrand for small $ |\vec Q |$ for $\Sigma_{2b}(q)$.
Integrating over $|\vec L_{(l)}|$ and $|\vec L_{( p )}|$, we get
\bqa
\Sigma_{2a} (q) & \sim & \frac{ i \, e^4 \, \mu^{2\,x}  \, \gamma_{d-m}\,\delta_q}  
{ \tilde \alpha^{\frac{ 2\, (2-m)}{3}} \, N^2 \, 
\sin^2 \left ( \frac { ( m+1) \, \pi} { 3 } \right ) } 
\int \frac{d\vec P\, d\vec L}{(2\pi)^{2d-2m}}\,
\frac{  
{\bar {\cal C}}_a (\vec L, \vec P, 0)   }
{  \left(\,|\vec L|\, |\vec P|\, \right)^{\frac{(d-m)\, (2-m)}{3}} \,
\lbrace
\delta_q^2 +  \left( \,P+ L
+|\vec P +\vec L | \, \right)^2  \rbrace   } \,,\nn
&& \\
\Sigma_{2b} (q) & \sim & \frac{ i   \left( \vec \Gamma \cdot \vec Q \right)
 e^4 \, \mu^{2\,x}
}  
{  \tilde \alpha^{\frac{ 2\, (2-m)}{3}} \,N^2
\, 
\sin^2 \left ( \frac { ( m+1) \, \pi} { 3 } \right ) }  
\int \frac{d\vec P\, d\vec L}{(2\pi)^{2d-2m}} \,
 \frac{ {\cal C}'_b  (\vec L, \vec P, \delta_q )  }
{ \left(\,|\vec L|\, |\vec P|\, \right)^{\frac{(d-m)\, (2-m)}{3}}   
\lbrace
\delta_q^2 +  \left( \,P+ L
+|\vec P +\vec L | \, \right)^2  \rbrace} \, , \label {2b} \nn
\eqa
where
%%%
\bqa
\label{tcb}
&& {\cal C}_b^\prime  (\vec L, \vec P, \delta_q ) \nn 
&=&
\frac{P+L+|\vec P +\vec L|}{PL\,|\vec P +\vec L| \, (d-m) }
\nn && \times \, 
 \Big [  (d-m-1)  \, 
\big \lbrace \, L^2 +P^2 +(\vec P\cdot \vec L) +PL-
(P+L)|\vec P +\vec L| \, \big \rbrace
+\,
\frac{2P^2 L^2 -2(\vec P\cdot \vec L)^2} {|\vec P +\vec L|^2}
\Big ] \nn
&& + \left[ \frac{\delta_q^2 - \left( \,P + L+|\vec P +\vec L| \, \right )^2 } 
{\delta_q^2 + \left ( \, P + L+|\vec P +\vec L| \, \right)^2 } \right]
\left( 1+\frac{(\vec P\cdot \vec L)}{PL} \right)
\frac{(\, P+L-|\vec P + \vec L|\,) \,(\,P+L+2|\vec P + \vec L| \,)}
{|\vec P + \vec L|^2}.\nn
\eqa
In Eq.~(\ref{tcb}),
we have used the equality 
$(\vec P \cdot \vec Q) (\vec \Gamma \cdot \vec L) =
(\vec P \cdot \vec L) (\vec \Gamma \cdot \vec Q)/(d-m)$. This holds inside the integration because
the denominator in Eq. (\ref{2b}) is 
invariant under $(d-m)$-dimensional rotation,
and the transformations 
$P_\nu \rightarrow -P_\nu$,
$L_\nu \rightarrow -L_\nu$ for each $\nu$.
We then perform the rescaling
\beq
P_\nu \rightarrow P_\nu \, |\delta_q| \,, \qquad 
L_\nu \rightarrow L_\nu \, |\delta_q | \,,
\eeq 
and for $d-m>1$, introduce the 
spherical coordinate in $(d-m)$ dimensions to
integrate over $\vec L$ and $\vec P$. 
Let $\theta$ be the angle 
between $\vec L$ and $\vec P$. 
Making a change of variables 
%%%%%%%%%%%%%%%%%%
\beq
L\rightarrow P \,  l \quad (0<l<\infty), \qquad P\rightarrow P , 
\eeq 
for $ d_c -d = \epsilon \,,$ we obtain
%%%%
\bqa
\Sigma_{2a} (q) &\sim&  \frac { i \, \tilde{e}^2 \,   \, \gamma_{d-m}\,\delta_q}  
{  N^2 \,|\delta_q|^{2\, (m+1)\epsilon/3} }
%%%%%%%%%%%%
 \int d\Omega_{d-m} \,  \int d\Omega_{d-m -1}
 \nn
&& \times \,
%%%
\int_0^{\infty } \int_0^{\infty } dl \, dP \,
P^{ 1 - \frac{2\,(m+1)\, \epsilon} {3}  }  \,
l^{- \frac{ m+1 }  {3} \epsilon }
 \int_0^{\pi}d\theta \,
\frac{  \left (\, \sin \theta \right)^{\frac{1-2 \, m}{m+1} - \epsilon} \, \left(\, 1+\cos\theta \, \right) }{1+P^2 \, [\,1+l+\eta\,]^2}  \left(
1-\frac{1+l}{\eta} \right), \label{2loosiga2} \\
\Sigma_{2b} (q) &\sim& 
\frac {  i \, \left( \vec \Gamma \cdot \vec Q \right) \, \tilde{e}^2}  
{  N^2 \,|\delta_q|^{2\, (m+1)\epsilon/3}\, (d-m) }
%%%%%%%%%%%%
 \int d\Omega_{d-m} \,  \int d\Omega_{d-m -1}
\int_0^{\infty } \int_0^{\infty } dl \, dP \,
 P^{ 1 - \frac{2\, (m+1)\,\epsilon} {3}  } \,
 l^{ - \frac{ m+1 }  {3} \epsilon  } 
%%%%
\nn
&& \times \,
\int_0^{\pi}  d\theta \, \left (\, \sin \theta \right)^{\frac{1-2 \, m}{m+1} - \epsilon} \,
 \Big [
\frac{ 1-P^2 (1+l+\eta)^2}   { \lbrace \,1+P^2 (1+l+\eta)^2 \, \rbrace^2} \,
\frac{(1+l-\eta) (1+\cos\theta) (1+l+2\eta)} {\eta^2} \nn
&&
\quad +\,
\frac{1+l+\eta}{1+P^2 \, (1+l+\eta)^2}  \frac{1} {l \,\eta} 
\Big \lbrace \left ( \, \frac{2-m}{m+1} -\epsilon\, \right) \left( 1+l^2 + l \,(1+\cos\theta) -(1+l)\, \eta \right) +
\frac{2\,l^2 \sin^2\theta}{\eta^2} \Big \rbrace
\Big ] ,\nn
 \label{2loosigb2}
\eqa
%%%
where 
$
\eta \equiv \eta (l,\theta) \equiv \sqrt{1+l^2 + 2l \cos\theta} \,.
$
%%%%%%%%%%%%
In order to extract the leading $1/\epsilon$ contribution  
in Eqs.~(\ref{2loosiga2})-(\ref{2loosigb2}), 
we use  
\bqa
&& \int_0^{\infty} \frac{dP\,P^{1 - \frac{2\, (m+1)\,\epsilon}{3} } } {1+P^2 \,(1+l+\eta)^2}=
 \frac{\pi} {2 \, \sin \left( \frac{(m+1)\,\pi \, \epsilon}{3} \right) } 
\frac{1}{(1+l+\eta)^{ 2 - \frac{2\, (m+1)\,\epsilon}{3} } } \,,
\nn
&&
\int_0^{\infty} \frac{dP\,P^{1 - \frac{2\,(m+1)\, \epsilon}{3} } \,
 [\,1-P^2\, (1+l+\eta)^2\,]}
{[  \, 1+P^2 \, (1+l+\eta)^2 \,]^2}=
-\frac{(1 - \frac{2 \, (m+1) }{3} \epsilon \,) \,\pi}
{2 \, \sin \left( \frac{ (m+1)\,\pi \, \epsilon}{3} \right) } 
\frac{1}{(1+l+\eta)^{ 2 - \frac{2\, (m+1)\,\epsilon}{3} } } 
\,.
\eqa

Let us also compute the residue when $d-m$ is away from $1$. In that case,
%%%%
\bqa
\Sigma_{2a} (q) &\sim &  \frac { 3\,i \, \tilde{e}^2 \,   \, \gamma_{d-m}\,\delta_q}  
{ 2\, (m+1)\, \epsilon \, N^2 }
\frac{ 4 \, \pi^{d-m-1/2}}
{ 
\sin^2 \left ( \frac { ( m+1) \, \pi} { 3 } \right )
\, \Gamma \left( \frac{ d-m} {2} \right ) \,
 \Gamma \left( \frac{ d-m-1} {2} \right ) } \nn
 && \times \, 
\int_0^{\infty } dl
 \int_0^{\pi} d\theta \,
\frac{  \left (\, \sin \theta \right)^{\frac{1-2 \, m}{m+1}} \, \left(\, 1+\cos\theta \, \right) }{ [\,1+l+\eta\,]^2}  
\left( 1-\frac{1+l}{\eta} \right)
\nn
%%%%%%%%%
&=& \frac { 3\,i \, \tilde{e}^2 \,   \, \gamma_{d-m}\,\delta_q}  
{ 2\, (m+1)\, \epsilon \, N^2  } 
%%%%%%%%%%%
\frac{ 4 \, \pi^{d-m-1/2}}
{ 
\sin^2 \left ( \frac { ( 2-m) \, \pi} { 3 } \right )
\, \Gamma \left( \frac{ 3 } {2 \,(m+1)}  - \frac{\epsilon} {2} \right ) \,
 \Gamma \left( \frac{ 2-m } {2 \,(m+1)}  - \frac{\epsilon} {2} \right ) }
%%%%%%%%
\nn &&
\times \,
\int_0^{\pi} d\theta \,
\frac{  \left (\, \sin \theta \right)^{\frac{1-2 \, m}{m+1}} \, \left(\, 1+\cos\theta \, \right) }
{ 8\, \sin^4\left( \frac{\theta}{2}\right ) }  
\Big [
4-\cos \theta \,
\Big \lbrace  4 - 2\,\ln \left( \cos^2 (\theta /2) \right) \Big \rbrace 
+ 6 \ln \left( \cos^2 (\theta /2) \right)
\Big] \nn
%%%%%%%%%
&=& \frac { 3\,i \, \tilde{e}^2 \,   \, \gamma_{d-m}\,\delta_q}  
{ (m+1)\, \epsilon \, N^2  } 
\frac{  \pi^{d-m-1/2}}
{ 
\sin^2 \left ( \frac { ( 2-m) \, \pi} { 3 } \right )
\, \Gamma \left( \frac{ 3 } {2 \,(m+1)}  \right ) \,
 \Gamma \left( \frac{ 2-m } {2 \,(m+1)}  - \frac{\epsilon} {2} \right ) }
 %%%%%%%%%%%%%%
 \nn &&
\times \,
 \int_0^{\pi}d\theta \,
\frac{  \left (\, \sin \theta \right)^{\frac{1-2 \, m}{m+1}} \, \left(  1+\cos\theta \right) }
{  \left( 1-\cos \theta \right )^2 } 
\Big [
4-\cos \theta \, \Big \lbrace  4 - 2\,\ln \left( \frac{1+\cos \theta} {2} \right) \Big \rbrace 
+ 6 \ln \left( \frac{1+\cos \theta} {2} \right)
\Big] \nn
%%%%%%%%%
&=& \frac { 3\,i \, \tilde{e}^2 \,   \, \gamma_{d-m}\,\delta_q}  
{  (m+1)\, \epsilon \, N^2  } 
\frac{  \pi^{d-m-1/2}}
{ 
\sin^2 \left ( \frac { ( 2-m) \, \pi} { 3 } \right )
\, \Gamma \left( \frac{ 3 } {2 \,(m+1)}  \right ) \,
 \Gamma \left( \frac{ 2-m } {2 \,(m+1)}  - \frac{\epsilon} {2} \right ) }
\,\times \,I_{ua} \,,\nn
%%%%%%%%%%%%
I_{ua} &=&
\int_{-1}^{1} du \,
\frac{  \left( \sqrt { 1-u^2 } \right )^{ \frac{-3\,m } {m+1}  } \, \left(\, 1+ u \, \right) }
{  \left( 1- u \right )^2 }  
\Big [
4-u \, \Big \lbrace  4 - 2\,\ln \left( \frac{1+ u } {2} \right) \Big \rbrace
+ 6 \ln \left( \frac{1+ u} {2} \right)
\Big] \,.\nn
\eqa 
%%%%%%%%%%%%%%%%
%%%%%%%%%%%%%
%%%%%%%%%%%%%%
Also,
\bqa
\Sigma_{2b} (q) &\sim& 
\frac { 3\, i \, \left( \vec \Gamma \cdot \vec Q \right) \, \tilde{e}^2}  
{  (m+1)\,(d-m)\,\epsilon \, N^2  } 
\frac{ 2 \, \pi^{d-m-1/2}}
{ 
\sin^2 \left ( \frac { ( 2-m) \, \pi} { 3 } \right )
\, \Gamma \left( \frac{ 3 } {2 \,(m+1)}  \right ) \,
 \Gamma \left( \frac{ 2-m } {2 \,(m+1)}  - \frac{\epsilon} {2} \right ) } 
\, \times \, I_{\theta b} \,,\nn
%%%%%%%%%%%%%%
I_{\theta b} &=&
\int_0^{\infty } dl 
\int_0^{\pi} d\theta 
\frac{  \left (\, \sin \theta \right)^{\frac{1-2 \, m}{m+1}} }
{ [\,1+l+\eta\,]^2}  
\Big [
-\frac{(1+l-\eta) (1+\cos\theta) (1+l+2\eta)} {\eta^2} \nn
&&
+\,
\frac{1+l+\eta} {l \,\eta} \,
\Big \lbrace  \, \frac{2-m}{m+1}  \left( 1+l^2 + l \,(1+\cos\theta) -(1+l)\, \eta \, \right) +
\frac{2l^2 \sin^2\theta}{\eta^2} \Big \rbrace
\Big ]  \,.
\eqa

%%%%%%%%%%
\begin{figure}
        \centering
        \subfloat[][]{\includegraphics[width=0.45 \textwidth]{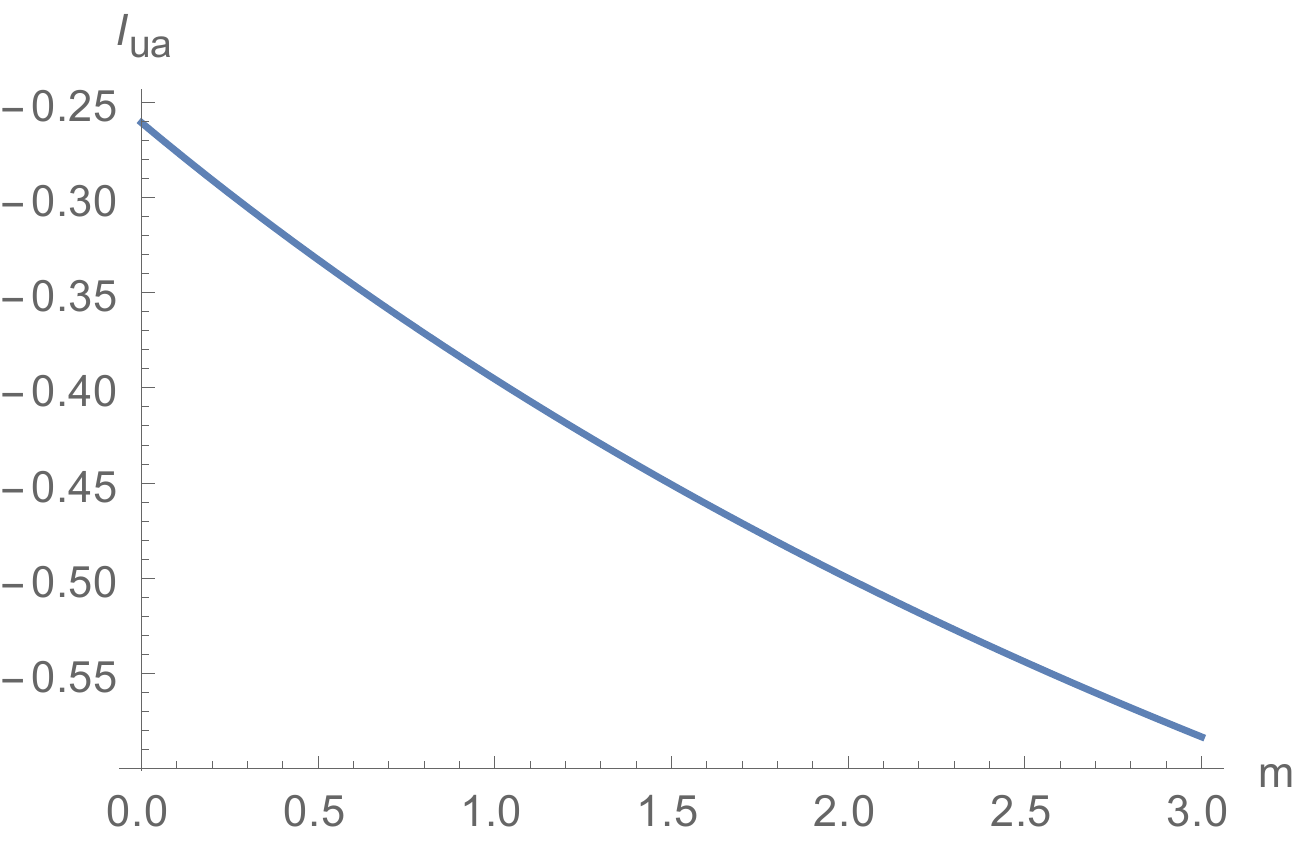}
              { \label{fig:iua}}} \quad
          \subfloat[][]
                {\includegraphics[width=0.42 \textwidth]{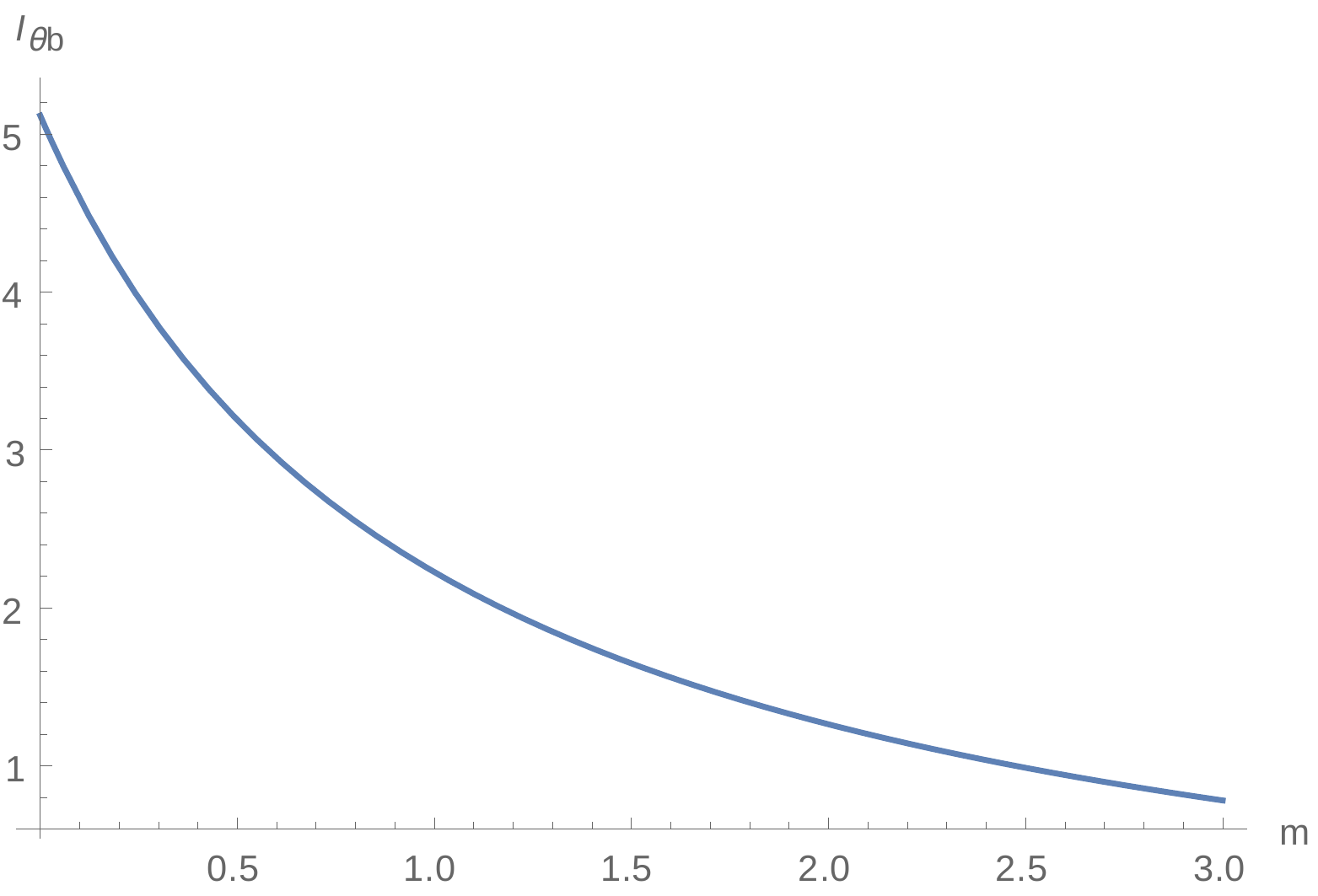}
		{\label{fig:itb}}}
        \caption{\label{fig:iut}(a) Plot of $I_{u a}$ versus $m$. 
                 (b) Plot of $I_{ \theta a}$ versus $m$.}
\end{figure}
%%%%%%%%%%%%
Fig.~\ref{fig:iut} shows the plots of the integrals $ I_{u a}$ and $ I_{ t b} $ as functions of $m$. Clearly, they are perfectly well-behaved functions in our range of interest for $m$. The residues thus can be read off from these functions at the desired value of $m$. We also note that overall coefficients vanish at $d-m=1$, again indicating that there is no fermion self-energy correction at two-loop order for $d=3$ and $m=2$.

%%%%%%%%%%%%%%%%%%%%%%%%%%%%%%%%%%%%%%%%%%%%%%%%%%
\section{Computation of the Feynman diagrams at three-loop level}
\label{3loop}

Since the number of diagrams increases dramatically at higher loops,
it is extremely hard to go beyond the two-loop level systematically.
Nevertheless, we will consider some three-loop diagrams which
can potentially contribute to the anomalous dimension of the boson through
a non-trivial correction to $Z_3$, given that $Z_3=1$ up to the two-loop order. Here we will consider
both the $ \lambda_{\text{cross}} >> 1 $ and $ \lambda_{\text{cross}} << 1 $ limits.

%%%%%%%%%%%%%%%%%%%

Let us first evaluate the function
\beq
\label{f0}
f_t(l,q) = -\frac{(i \, e)^3 \mu^{3 x /2}}{N^{3/2}} \, N 
\int  {dp} \, 
{\rm Tr} \{ \gamma_{d-m} \, G_0 (p+l) \, \gamma_{d-m} \, G_0 (p+q) \, \gamma_{d-m} \, G_0 (p)\} \,,
\eeq
which is formed by a fermion loop with three external boson propagators. This will be useful for all our three-loop calculations.
Taking the trace in Eq.~(\ref{f0}), we obtain
\begin{eqnarray}
\label{f1}
f_t(l,q) &=& - \frac{2e^3 \mu^{3 x /2}}{\sqrt{N}}
\int \frac{d \vec P \, dp_{d-m+2} \ldots dp_d}{(2\pi)^{d-2}} \, \sum_{i=1}^{4} \kappa_i \,,
\end{eqnarray}
where
\begin{eqnarray}
\kappa_1 &=& \int \frac{dp_{d-m} \, dp_{d-m+1}} {(2\pi)^2}
\frac{ \delta_{p} \, \delta_{p+q} \, \delta_{p+l}}
{den_{\kappa}} 
\exp\left( {-\frac{{\vec{L}}_{(p)}^2 + {\vec{L}}_{(p+q)}^2 +  {\vec{L}}_{(p+l)}^2} { \mu \, {\tilde{k}}_F }} \right)
\,,\nn
\kappa_2 &=& -\int \frac{dp_{d-m} \, dp_{d-m+1}} {(2\pi)^2}
\frac{\delta_{p} \,(\vec P +\vec Q)\cdot (\vec P +\vec L)}
 {den_{\kappa}} 
\exp\left( {-\frac{{\vec{L}}_{(p)}^2 + {\vec{L}}_{(p+q)}^2 +  {\vec{L}}_{(p+l)}^2} { \mu \, {\tilde{k}}_F }} \right)
\,,\nn
\kappa_3 &=& -\int \frac{dp_{d-m} \, dp_{d-m+1}} {(2\pi)^2}
\frac{\delta_{p+q} \,(\vec P +\vec L)\cdot \vec P}
{den_{\kappa} }
\exp\left( {-\frac{{\vec{L}}_{(p)}^2 + {\vec{L}}_{(p+q)}^2 + {\vec{L}}_{(p+l)}^2} { \mu \, {\tilde{k}}_F }} \right)
 \,,\nn
\kappa_4 &=& -\int \frac{dp_{d-m} \, dp_{d-m+1}} {(2\pi)^2}
\frac{\delta_{p+l} \,(\vec P +\vec Q)\cdot \vec P}
 {den_{\kappa}}
\exp\left( {-\frac{{\vec{L}}_{(p)}^2 + {\vec{L}}_{(p+q)}^2 +  {\vec{L}}_{(p+l)}^2} { \mu \, {\tilde{k}}_F }} \right)
\,,\\
den_{\kappa} &=& [ \, \delta_{p}^2 +P^2 \,] \,
 [\, \delta_{p+q}^2 + (\vec P +\vec Q)^2 \,]\,
[\, \delta_{p+l}^2 +(\vec P +\vec L)^2 \, ]\,.
\end{eqnarray}

%%%%%%%%%%%%%%%%%%
%\subsubsection{\underline{For  $\frac{ |q_{d-m}| }  { |\vec{L}_{(q)}| \sqrt{2 k_F} } , \frac{ |\vec P| + |\vec P + \vec Q| }  { |\vec{L}_{(q)} | \sqrt{2 k_F}}, \frac{ |\vec P | + |\vec P + \vec L| }  {  |\vec{L}_{(q)}| \sqrt{2 k_F} } << 1 :$}}

We assume that we are in the region $\frac{ |q_{d-m}| }  { |\vec{L}_{(q)}| \sqrt{2 k_F} } , \frac{ |\vec P| + |\vec P + \vec Q| }  { |\vec{L}_{(q)} | \sqrt{2 k_F}}, \frac{ |\vec P | + |\vec P + \vec L| }  {  |\vec{L}_{(q)}| \sqrt{2 k_F} } << 1$ and choose the coordinate system such that ${\vec{L}}_{(q)} = ( q_{d-m+1},0, 0,\ldots,0) $,  without any loss of generality.
We then redefine some variables as:
\beq
x_1 = p_{d-m} +{\vec{L}}_{(p)}^2 \,, \qquad
x_2= \delta_{q} + 2 \, p_{d-m+1} \,|{\vec{L}}_{(q)}| \,,
\quad    d p_{d-m}  \, d p_{d-m +1 } = \frac{  dx_1 \, d x_2} {  |q_{d-m+1} | } \,, \nonumber
\eeq
so that 
\begin{equation}
\delta_p = x_1\,, \quad
\delta_{p+q} =x_1+x_2 \,, \quad
\delta_{p+l} = \delta_p + \delta_l +2 \, p_{d-m+1} \,l_{d-m+1} + 2 \, {\vec{u}}_{(p)} \cdot {\vec{u}}_{(l)}
= x_1 + \frac{l_{d-m+1}} { q_{d-m+1} }  x_2 + \Delta_t (p,l, q)\,,
\nonumber
\end{equation}
with 
\begin{equation}
\Delta_t (p,l, q) = \delta_l -  \frac{l_{d-m+1}} { q_{d-m+1} } \, \delta_q 
+ 2\, {\vec{u}}_{(p)} \cdot {\vec{u}}_{(l)} \,, 
\quad {\vec{u}}_{(k)} =(k_{d-m+2},\ldots,k_d)\,.
\end{equation}
Here the vector
${\vec{u}}_{(k)} $
consists of the last $(m-1)$ components of ${\vec{L}}_{(k)}$.
Neglecting the exponential damping factors for $x_2$, we get
%%%%%%%%%%%%%%%%%%%%%%%%%
\begin{eqnarray}
\label{f4}
&& \tilde \kappa_1 \nn
&& \equiv 
\frac{1}{2 \, | q_{d-m+1} | }
\int \frac{dx_1 \, dx_2}{(2\pi)^2 }
\frac{x_1 \, (x_1 + x_2) \, 
\left ( x_1 + \frac{l_{d-m+1}} { q_{d-m+1} }  x_2 + \Delta_t (p,l, q) \right)   
}
{ 
\Big \lbrace  x_1^2  +\vec P ^2 \Big \rbrace   
\Big \lbrace  \left( x_1 + x_2 \right)^2  +(\vec P +\vec Q)^2 \Big \rbrace   
\Big \lbrace  \left( x_1 + \frac{l_{d-m+1}} { q_{d-m+1} }  x_2 + \Delta_t (p,l, q) \right)^2  +(\vec P +\vec L)^2 \Big \rbrace } \nn
%%%%%%%%%%%%%%%%%%%%%%%%%%%%%%%
&&=
\frac{1} {4   }
\int \frac{dx_1}  {2\pi }
\frac{x_1 \,
\Big [ | l_{d-m+1} | \, |\vec P +\vec Q| 
+  | q_{d-m+1}|   |\vec P +\vec L|
\Big ]  
\, sgn \left (  q_{d-m+1} \right) 
}
{ 
\Big \lbrace  x_1^2  +\vec P ^2 \Big \rbrace   
\Big \lbrace  \left( 
l_{d-m+1} \, x_1  
-  q_{d-m+1}  (x_1 + \Delta_t)
\right)^2  
+( \,| l_{d-m+1} | \, |\vec P +\vec Q| 
+  | q_{d-m+1}|   |\vec P +\vec L|\, )^2 \Big \rbrace } \nn
%%%%%%%%%%%%%%%%%%%%%%%%%%
&&=
\frac{    \Delta_t
} 
{8 \, |q_{d-m+1}|   }
%%%%%%%%%%%%%%
\frac{  sgn \left(  l_{d-m+1} -  q_{d-m+1} \right ) \,
 sgn \left (  l_{d-m+1} \right)   }
{  \Delta_t^2
+
\frac{ \Big [
|l_{d-m+1}  - q_{d-m+1}  |
\,|\vec P|
+ 
|l_{d-m+1}| \, |\vec P +\vec Q| 
+| q_{d-m+1} | \, |\vec P +\vec L| 
\Big]^2} 
{ q_{d-m+1}^2 }
} \,.
%%%%%%%%%%%%%%%%%%%%%%%%
\end{eqnarray}

%%%%%%%%%%%%%%%%%%%%%%%%%%%%%%%%%%%%%%%%%%%
%%%%%%%%%%%%%%%%%%%%%%%%%
\begin{eqnarray}
\label{kappatilde2}
&& \tilde \kappa_2 \nn
&& \equiv 
\frac{ 1 }
{2 \,  |q_{d-m+1} | }
\int \frac{dx_1 \, dx_2}{(2\pi)^2 }
\frac{  - x_1 \,(\vec P + \vec Q) \cdot (\vec P + \vec L)
}
{ 
\Big \lbrace  x_1^2  +\vec P ^2 \Big \rbrace   
\Big \lbrace  \left( x_1 + x_2 \right)^2  +(\vec P +\vec Q)^2 \Big \rbrace   
\Big \lbrace  \left( x_1 + \frac{l_{d-m+1}} { q_{d-m+1} }  x_2 + \Delta_t (p,l, q) \right)^2  +(\vec P +\vec L)^2 \Big \rbrace } \nn
%%%%%%%%%%%%%%%%%%%%%%%%%%%%%%%
&&=
\frac{ 1 } 
{4
 \left( l_{d-m+1}- q_{d-m+1} \right )^2 }
\int \frac{dx_1}  {2\pi }
\frac{- x_1
\, \frac{ (\vec P + \vec Q) \cdot (\vec P + \vec L) }
{|\vec P +\vec Q| 
\, |\vec P +\vec L|} 
\left ( \,| l_{d-m+1}  | \, |\vec P +\vec Q| 
+ |q_{d-m+1}  | \, |\vec P +\vec L| \,  \right )
}
{ 
\Big \lbrace  x_1^2  +\vec P ^2 \Big \rbrace   
\Big [  \Big \lbrace x_1  
 -\frac{  q_{d-m+1} \, \Delta_t }
 {
 \left( l_{d-m+1}- q_{d-m+1}  \right )
 }
\Big \rbrace ^2  
+\left ( 
\frac 
{   | l_{d-m+1}  | \, |\vec P +\vec Q| 
+ |q_{d-m+1}  | \, |\vec P +\vec L| 
}
{
|   l_{d-m+1}- q_{d-m+1}   |
}
\right )^2 \Big ] } \nn
%%%%%%%%%%%%%%%%%%%%%%%%%%
%%%%%%%%%%%%%%%%%%%%%%%%%%
&& \Rightarrow
 \kappa_2 =
- \frac{ (\vec P + \vec Q) \cdot (\vec P + \vec L) }
{|\vec P +\vec Q|  \, |\vec P +\vec L|} 
\,\frac{ sgn \left( q_{d-m+1}   \right ) }
{  sgn \left( l_{d-m+1}   \right ) }  \,  \kappa_1 \,.
%%%%%%%%%%%%%%%%%%%%%%%%
\end{eqnarray}
%%%%%%%%%%%%%%%%%%%%%%%%%%%%%%%%%%%%%%%%%%%

%%%%%%%%%%%%%%%%%%%%%%%%%
\begin{eqnarray}
\label{kappatilde3}
&& \tilde \kappa_3  \nn
&& \equiv 
\frac{1 }
{2 \, | q_{d-m+1} | }
\int \frac{dx_1 \, dx_2}{(2\pi)^2 }
\frac{  -(x_2 +  x_1 ) \,(\vec P + \vec L) \cdot  \vec P
}
{ 
\Big \lbrace  x_1^2  +\vec P ^2 \Big \rbrace   
\Big \lbrace  \left( x_1 + x_2 \right)^2  +(\vec P +\vec Q)^2 \Big \rbrace   
\Big \lbrace  \left( x_1 + \frac{l_{d-m+1}} { q_{d-m+1}  }  x_2 + \Delta_t (p,l, q) \right)^2  +(\vec P +\vec L)^2 \Big \rbrace } \nn
%%%%%%%%%%%%%%%%%%%%%%%%%%%%%%%
 &&=
\frac{ -\frac{ (\vec P + \vec L) \cdot  \vec P }
{ |\vec P +\vec L|}
\, sgn \left( l_{d-m+1}  \right ) } 
{8 \,   \left( l_{d-m+1} - q_{d-m+1}  \right )^2 }
%%%%%%%%%%%%%
\times \, 
\frac{   -   q_{d-m+1} \,  \Delta_t
}
{|\vec P| }
%%%%%%%%%%%
\times \, \frac{  1  }
{ 
\frac{ q_{d-m+1}^ 2 
 }
 {
 \left(  l_{d-m+1} -  q_{d-m+1} \right )^2 
}
\Delta_t^2
+
\Big [ 
|\vec P|
+ 
\frac 
{ |  l_{d-m+1} | \, |\vec P +\vec Q| 
+| q_{d-m+1}  | \, |\vec P +\vec L| 
}
{ |  l_{d-m+1} -  q_{d-m+1} |
}
\Big ]^2
}\nn
%%%%%%%%%%%%%%%%%%%%%%%%
%%%%%%%%%%%%%%%%%%%%%%%%%%
&& \Rightarrow
 \kappa_3 =
 \frac{ (\vec P + \vec L ) \cdot  \vec P   }
{|\vec P +\vec L |  \, |\vec P  |} 
\,  
\frac{ sgn \left( q_{d-m+1}  \right )  }
{ sgn \left(  l_{ d-m+1}-q_{ d-m+1} \right ) }
\, \kappa_1 \,.
%%%%%%%%%%%%%%%%%%%%%%%%
\end{eqnarray}
%%%%%%%%%%%%%%%%%%%%%%%%%%%%%%%%%%%%%%%%%%%
%%%%%%%%%%%%%%%%%%%%%%%%%%%%%%%%%%%%%%%%%%%

%%%%%%%%%%%%%%%%%%%%%%%%%%%%%%%%%%%%%%%%%%%
%%%%%%%%%%%%%%%%%%%%%%%%%
\begin{eqnarray}
\label{kappatilde4}
&& \tilde \kappa_4  \nn
&& \equiv 
\frac{1 }
{2 \, |q_{d-m+1} | }
\int \frac{dx_1 \, dx_2}{(2\pi)^2 }
\frac{  - \left ( x_1 + \frac{l_{d-m+1}} { q_{d-m+1} }  x_2 + \Delta_t (p,l, q) \right)    \,(\vec P + \vec Q) \cdot  \vec P
}
{ 
\Big \lbrace  x_1^2  +\vec P ^2 \Big \rbrace   
\Big \lbrace  \left( x_1 + x_2 \right)^2  +(\vec P +\vec Q)^2 \Big \rbrace   
\Big \lbrace  \left( x_1 + \frac{l_{d-m+1}} { q_{d-m+1}  }  x_2 + \Delta_t (p,l, q) \right)^2  +(\vec P +\vec L)^2 \Big \rbrace } \nn
%%%%%%%%%%%%%%%%%%%%%%%%%%%%%%%
&&=
\frac{  sgn \left ( q_{d-m+1} \right ) \,
\frac{ (\vec P + \vec  Q ) \cdot  \vec P }
{ |\vec P +\vec Q|}
} 
{8  \left(  l_{d-m+1}- q_{d-m+1} \right )^2 }
%%%%%%%%%%%%%
\frac{   -   q_{d-m+1}\,  \Delta_t
}
{|\vec P| }
%%%%%%%%%%%
\frac{  1  }
{ 
\frac{  q_{d-m+1}^2  }
{  \left(  l_{d-m+1}-  q_{d-m+1} \right )^2 }
 \Delta_t^2
+
\Big [
|\vec P|
+ 
\frac 
{  | l_{d-m+1} | \, |\vec P +\vec Q| 
+| q_{d-m+1} | \, |\vec P +\vec L| 
}
{|  l_{d-m+1} - q_{d-m+1} |
}
\Big]^2
}\nn
%%%%%%%%%%%%%%%%%%%%%%%%%%%%%%%%%%%%%%%%%%%%%%%%%
&& \Rightarrow
 \kappa_4 =
-  \frac{ (\vec P + \vec  Q ) \cdot  \vec P   }
{|\vec P +\vec  Q |  \, |\vec P  |} 
\,  \frac { \kappa_1 }
{ sgn \left(   l_{d-m+1} -   q_{d-m+1}   \right ) \, sgn \left ( l_{d-m+1} \right ) \,} \,.
\end{eqnarray}
%%%%%%%%%%%%%%%%%%%%%%%%%%%%%%%%%%%%%%%%%%%

For $\vec Q = 0 $, 
%%%%%%%%%%%%%%%%%%%%%%%%%%%%%%%
\bqa
\sum_{i=1}^{4} \kappa_i \Big |_{\vec Q =0 }
= 
\frac{    \Delta_t
} 
{4 \, q_{d-m+1}   }
\frac { \vec P \cdot (\vec P + \vec L   ) 
- | \vec P | \, | \vec P + \vec L | }
{  | \vec P | \, | \vec P + \vec L | } 
%%%%%%%%%%%%%%
\frac{ 
 \Theta \left (  l_{d-m+1} \right) -\Theta \left(  l_{d-m+1} -  q_{d-m+1} \right )  }
{  \Delta_t^2
+
 \Big [ |\vec P|
+  |\vec P +\vec L| 
\Big]^2
} \,.
\eqa

We now choose ${\vec{u}}_{(l)} = ( l_{d-m+2},0, 0,\ldots,0) $, with $l_{d-m+2}>0$, since $f_t(l,q)$ can depend only on $|{\vec{u}}_{(l)} |$.
Define $ \quad  x_3 = 2 \,|{\vec{u}}_{(l)} | \, p_{d-m+2} \,,
 \, {\vec{v}}_{(k)} =(k_{d-m+3},\ldots,k_d), \,
 \tilde \Delta_t (l, q) = \delta_l -  \frac{l_{d-m+1}} {  q_{d-m+1}  } \, \delta_q  $
 we get
%%%%%%%%%%%%%%%
 \begin{eqnarray}
 && 
 \int \frac{   d p_{d-m+2}}  {2 \pi}
 \tilde \kappa_1
%%%
\,\exp \left( {-\frac{ 3 \,p_{d-m+2}^2 +2\, |{\vec{u}}_{(l)} |\, p_{d-m+2} } { k_F }} \right) \nn
%%%%%%%%%%%
&& = \frac{
\, sgn \left( l_{d-m+1}- q_{d-m+1}\right ) \, sgn \left ( l_{d-m+1} \right )
\,
 \exp \Big [ \frac{ 2 \, \vec u_{(l)}^ 2 } { 3 \, k_F} \Big ]
} 
{ 16 \, |\vec{u}_{ (l) } |
\, | q_{d-m+1} |
}
\int_{-\infty}^{\infty} \frac{ d z_3}{2 \pi}
 \frac{ \left( z_3 + u_3  \right)
\,\exp
\left( - \frac{3  z_3 ^2 } {4 } \right)
}
{
\left( z_3 + u_3  \right)^2
+ y_3^2
} \,,\nn
%%%%%%%%%%%
\end{eqnarray}
where
%%%%%%%%%%%%%%%
\begin{equation}
u_3=\frac{ - \frac{2}{3} |{\vec{u}}_{(l)} |^2 + \tilde \Delta_t} { |{\vec{u}}_{(l)} | \, \sqrt{k_F}} \, , \quad
y_3 = \frac{1}  { |{\vec{u}}_{(l)} | \, \sqrt{k_F}}  \,
\frac{ |l_{d-m+1}  - q_{d-m+1}  |
\,|\vec P|
+ 
|l_{d-m+1}| \, |\vec P +\vec Q| 
+| q_{d-m+1} | \, |\vec P +\vec L|  } 
{   | q_{d-m+1} | } \,.
\end{equation}
%%%%%%%%%%%%%%%%%%%%%%%

\begin{enumerate}
\item{
In the limit $  u_3,\, y_3 << 1  $, we have
\beq
 \int_{-\infty}^{\infty} \frac{ d z_3}{2 \pi}
 \frac{ \left( z_3 + u_3  \right)
\,\exp
\left( - \frac{3  z_3 ^2 } {4 } \right)
} {
\left( z_3 + u_3  \right)^2
+ y_3^2}\simeq
\sqrt{ \frac{3}{4 \, \pi} } \, u_3
%%%%%%%%%%%
=
 \frac{1} { 2 \, |\vec{u}_{ (l) } |^2 }
\frac{  \tilde \Delta_t -\frac { 2 \, \vec u_{(l)}^ 2 } {3} }
{\sqrt{\pi \, k_F /3}} \,.
%%%%%%%%%%%%%%%%%%%%%%%%%%
\eeq
%%%%%%%%%%
Integrating the above over ${\vec{v}}_{(p)} $, we get
%%%%%%%%%%%%%%%
 \begin{eqnarray}
 && \int  \frac {d{\vec{v}}_{(p)} \,dp_{d-m+2}}  {(2 \pi)^{m-1} } \, \kappa_1 \nn
 %%%%%%%%%%%%%%
 &&
 =\int \frac{ d \vec v_{(p)} \, d p_{d-m+2}  }  {  (2 \pi) ^{ m-1 }}
\,  \tilde \kappa_1
 %%%
\,\exp \left( {-\frac{ 3 \, \vec v_{(p)}^2 + 3 \,p_{d-m+2}^2 +2\, |{\vec{u}}_{(l)} |\, p_{d-m+2} } { k_F }} \right) \nn
%%%%%%%%%%%
&& =
sgn \left( l_{d-m+1}- q_{d-m+1}\right ) 
\, sgn \left( l_{d-m+1} \right ) 
\frac{ \delta_l -  \frac{l_{d-m+1}} { q_{d-m+1} } \, \delta_q
-\frac { 2 \, \vec u_{(l)}^ 2 } {3} } 
 { 2^{m + 3 }  \, \pi \, |\vec{u}_{ (l) } |^2 
\,  | q_{d-m+1} |
 }
 \left(  \sqrt{ \frac{ k_F} {3 \pi} } \right)^{m-3} 
 + \mathcal{O} \left (\frac{1}{k_F^{\frac{4-m}{2}} } \right)\,.\nn
\end{eqnarray}

%%%%%%%%%%%%%%%%%%%
Hence, as long as $m>1$,
\begin{eqnarray}
\label{c1}
f_t(l,q) &\propto& e^3 \, k_F^{\frac{m-3}{2}}  , \quad \mbox{for} \quad u_3,\, y_3 << 1 \, \mbox{and} \, m>1.
\end{eqnarray}

}
%%%%%%%%%%%%%%%%%%%%%%%%%%%%%%%%%%%%%%
\vspace{2 cm}
\item{
In the limit $  u_3,\, y_3 >> 1  $, we have
\beq
 \int_{-\infty}^{\infty} \frac{ d z_3}{2 \pi}
 \frac{ \left( z_3 + u_3  \right)
\,\exp
\left( - \frac{3  z_3 ^2 } {4 } \right)
} {
\left( z_3 + u_3  \right)^2
+ y_3^2}
\simeq
\frac{ u_3} 
{  u_3 ^2 + y_3^2}
\int_{-\infty}^{\infty} \frac{ d z_3}{2 \pi}
 \exp
\left( - \frac{3  z_3 ^2 } {4 } \right)
= \frac{ u_3} 
{  u_3 ^2 + y_3^2} \, \frac{1}{\sqrt{3 \, \pi}}
\eeq
Hence we get
%%%%%%%%%%%
\begin{eqnarray}
&&  \int \frac {d{\vec{v}}_{(p)} \,dp_{d-m+2}}  {(2 \pi)^{m-1} } \, \kappa_1 \nn
%%%%%%%%%%
&& = 
\frac{  \left(  \sqrt{ \frac{ k_F} {3 \pi} } \right)^{m- 1 }
\, sgn \left( l_{d-m+1}- q_{d-m+1} \right ) 
\, sgn \left( l_{d-m+1} \right )  } 
 { 2^{m + 2 } \,
  |q_{d-m+1} |
 } \nn
 && \times \, 
\frac{ \delta_l -  \frac{l_{d-m+1}} {  q_{d-m+1} } \, \delta_q
-\frac { 2 \, \vec u_{(l)}^ 2 } {3} } 
{   
  \left( \delta_l -  \frac{l_{d-m+1}} {  q_{d-m+1} } \, \delta_q
-\frac { 2 \, \vec u_{(l)}^ 2 } {3}  \right)^2   
  +
  %%%%
\frac{ 
\Big [
|l_{d-m+1}  - q_{d-m+1}  |
\,|\vec P|
+ 
|l_{d-m+1}| \, |\vec P +\vec Q| 
+| q_{d-m+1} | \, |\vec P +\vec L| 
\Big]^2 }
{   q_{d-m+1}^2 }
} 
 +  \, \mathcal{O} \left (\frac{1}{k_F^{\frac{2-m}{2}} } \right) \,.
\label{c2}
 \nn
\end{eqnarray}
%%%%%%%%%%%%%%%%
%%%%%%%%%%%%%%%%
Hence,
\begin{equation}
f_t(l,q) \propto e^3 \, k_F^{\frac{m-1}{2}}  , \quad \mbox{for} \,  u_3,\, y_3 >> 1 \,.
\end{equation}
This also corresponds to the case of $m=1$.
}
%%%%%%%%%%%%%%%%%%%%%%%%%
The case of $m=1$ has of course been discussed thoroughly in \cite{Lee-Dalid}. 

\end{enumerate}
%%%%%%
For simplicity, we have shown the final expressions for $\kappa_1$ only in the appropriate limits.

%%%%%%%%

%%%%%%%%%%%%%%%%%%%%%%%%%%%%
\subsection{Three-loop fermion self-energy diagrams with one fermion loop }

\begin{figure}[ht]   
  \centering
       \subfloat[][]{\includegraphics[width=0.32 \textwidth]{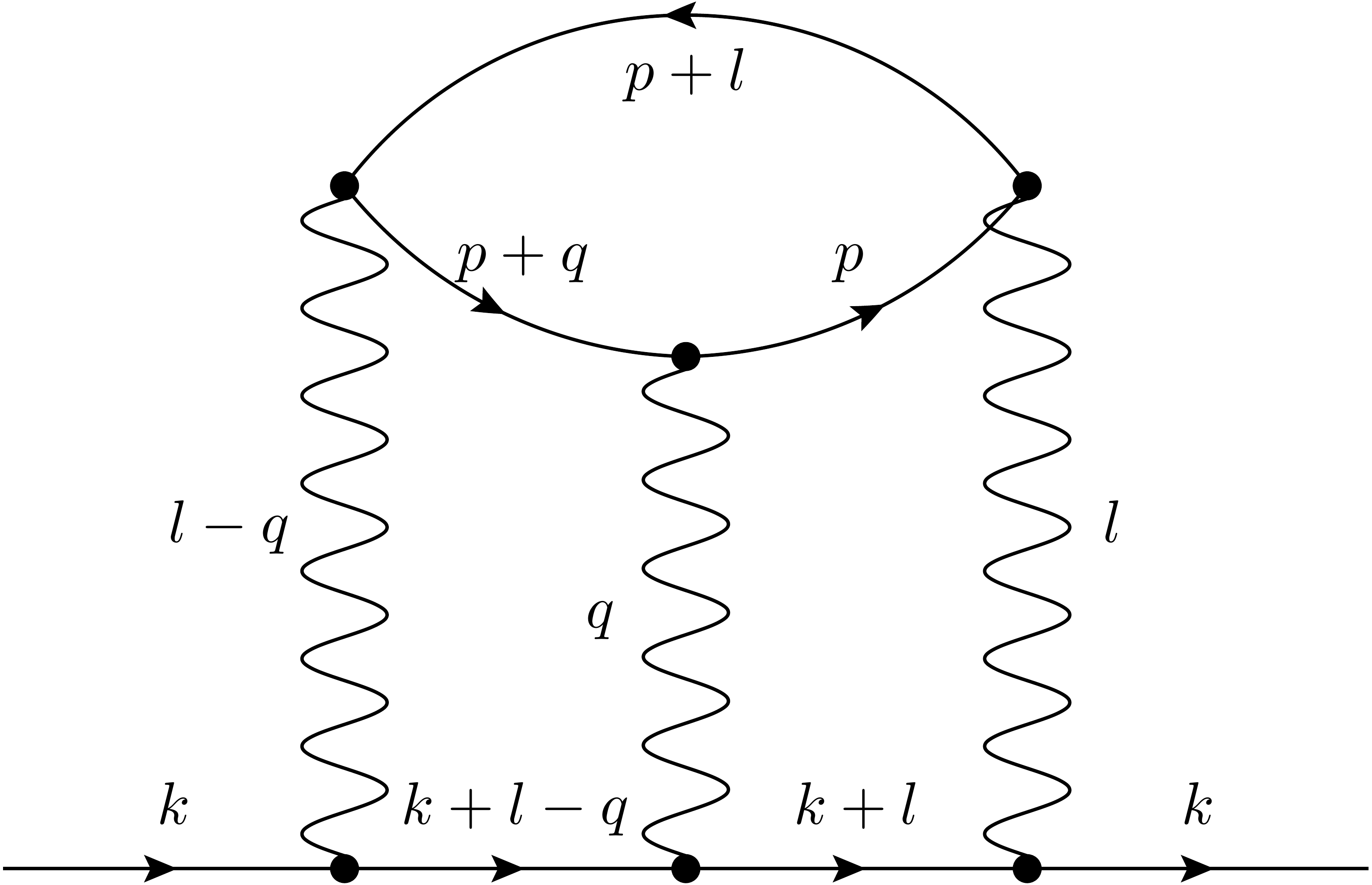}
 \label{fig:3f1}} 
 \subfloat[][]{\includegraphics[width=0.32 \textwidth]{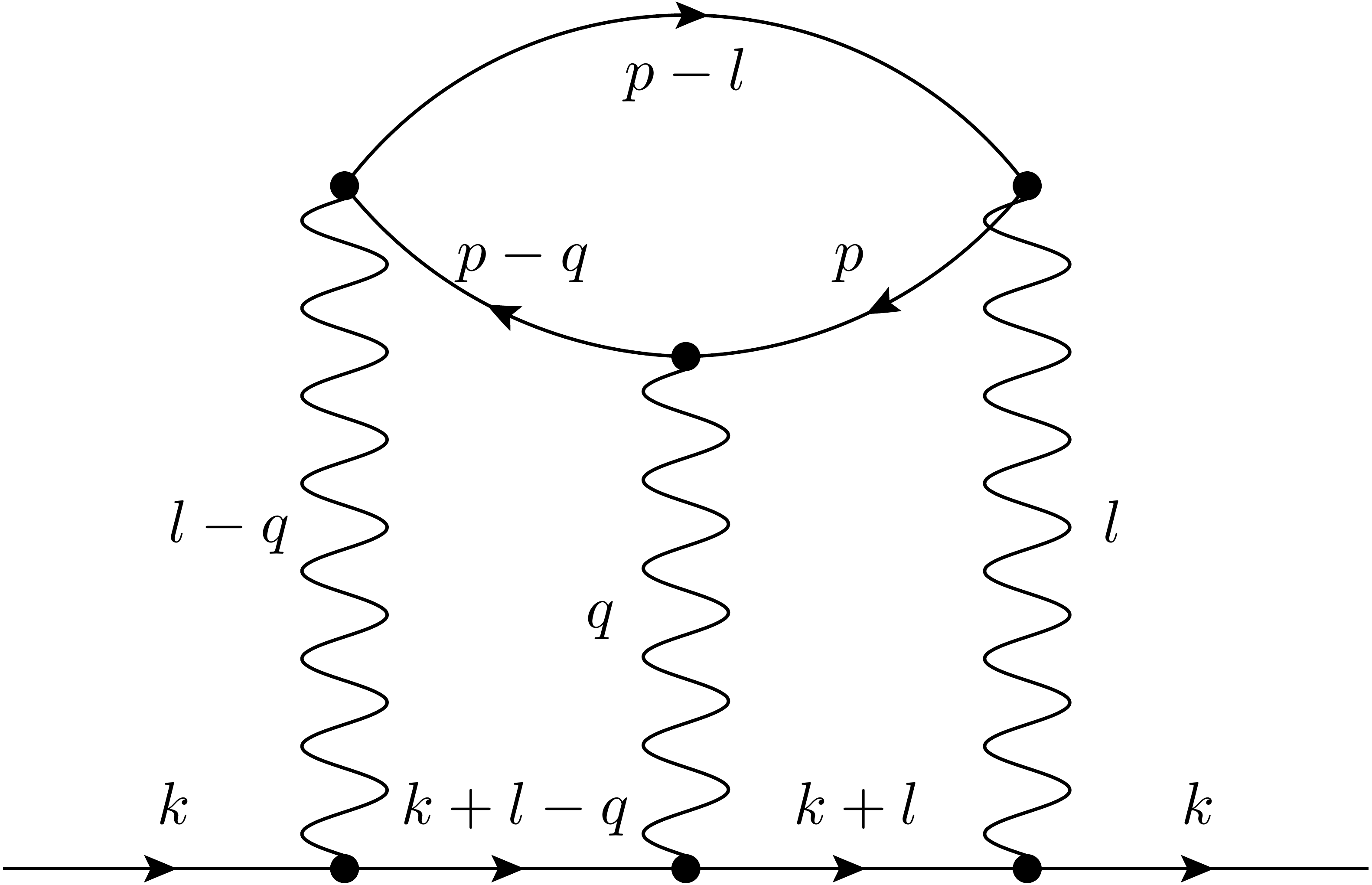}
\label{fig:3f2}} 
 \subfloat[][]{\includegraphics[width=0.32 \textwidth]{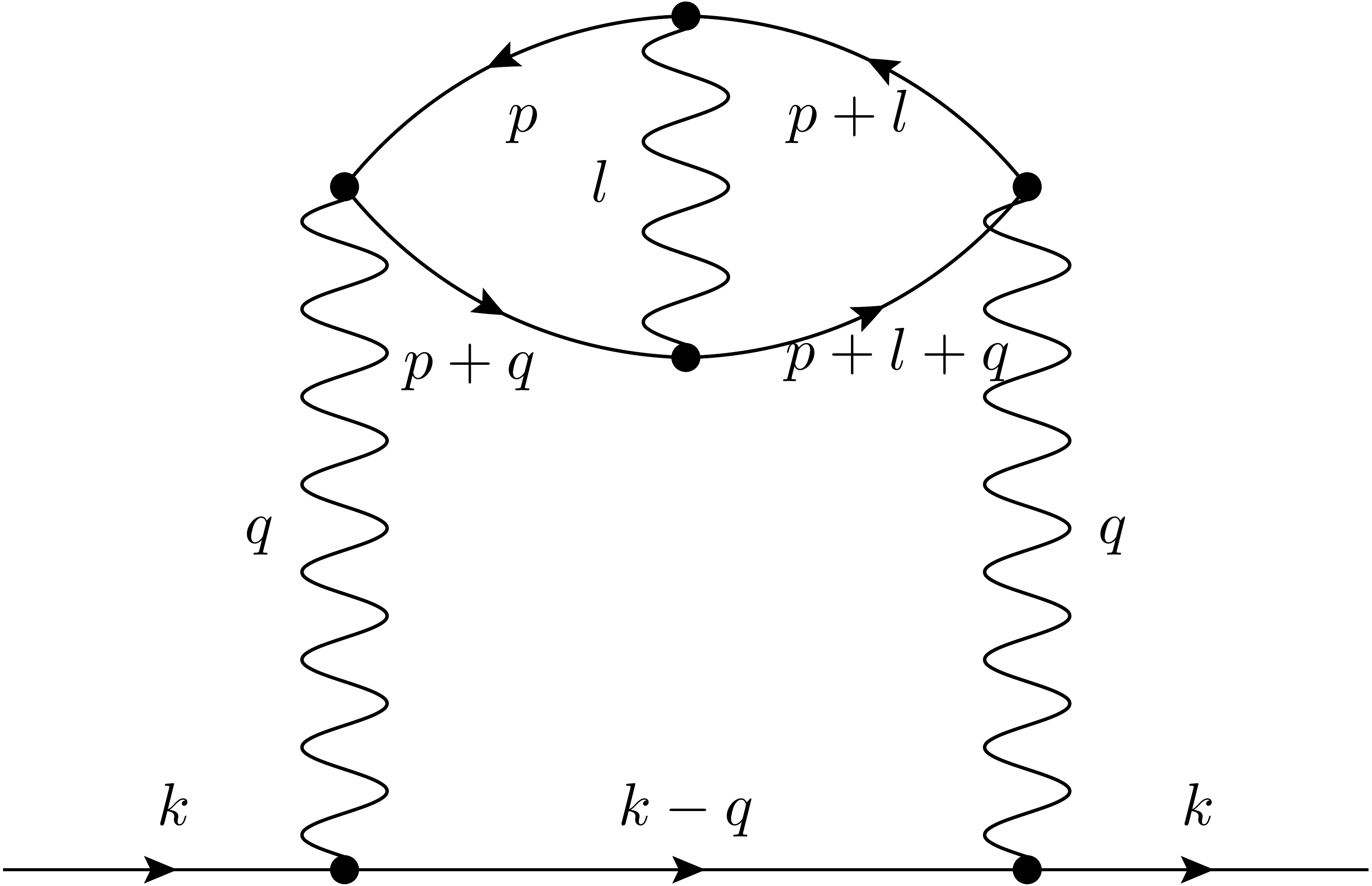}
\label{fig:3f3}} 
      \caption{\label{fig:3ferm}Three-loop fermion self-energy diagrams each with one fermion loop.}
\end{figure}

Fig.~\ref{fig:3ferm} shows three-loop fermion self-energy diagrams each containing one fermion loop.
From the computation of Fig.~\ref{fig:bos2}(a), it is clear that Fig.~\ref{fig:3ferm}(c)
does not contribute for $m>1$.
Hence we calculate the contribution coming from the diagrams in Figs.~\ref{fig:3ferm}(a) and ~\ref{fig:3ferm}(b). The integrals involve the function $f_t(l,q)$ coming from the fermion loop.  Their total contribution can be written as
%%%%%%%%%%%%%
\bqa
\label{ferm3}
\Sigma_{3 }(k) & \sim & 
\frac{ (i \, e)^3 \, \mu^{3 x /2}}{N^{3/2}}
\int  {  dq \, dl} 
\, \lbrace f_{t} (l,q) +  f_{t} (q-l,q) \rbrace \nn
&& \qquad \qquad \qquad  \times \,
 \, D_1(l-q) \, D_1(q) \, D_1(l) \,
 \gamma_{d-m} \, G_0 (k+l-q) \, \gamma_{d-m} \, G_0 (k+l) \, \gamma_{d-m}   \nn
 %%%%%%%%%%%%
 & = &
 \frac{  i \, e^3 \,\mu^{3 x /2}}{N^{3/2}} 
\int   dq \, dl
\, \big  \lbrace f_{t} (l,q) +  f_{t} (q-l, q) \big  \rbrace  \,
D_1(l-q) \, D_1(q) \, D_1(l) \,
\frac{num_{30} } {den_{30} } \,,
\eqa
%%%%
where
\begin{eqnarray}
\label{nf}
num_{30} &=&
\Big[
\lbrace \,\vec \Gamma \cdot \vec Q \, \rbrace \, 
\lbrace \,\vec \Gamma \cdot \left( \,\vec K \,+\, \vec L \, \right) \, \rbrace \, -\,
\left( \,\vec K \,+\, \vec L \, \right)^2 \,+\,  \delta_{k + l}  \,  \delta_{k + l -\,q} \,
\Big] \, \gamma_{d-m} \nn
&& +\,
\Big[
\lbrace \,\vec \Gamma \cdot \vec Q \, \rbrace \,
-\,  \lbrace \,\vec \Gamma \cdot \left( \,\vec K \,+\, \vec L \, \right) \, \rbrace \, \Big]\,   \delta_{k + l}
-\,  \lbrace \,\vec \Gamma \cdot \left( \,\vec K \,+\, \vec L \, \right) \, \rbrace \,    \delta_{k + l-q}
\,, \nn
%%%
den_{30} &=& 
 \Big [ \,  \delta_{k + l}^2 + \left  (  \,  \vec K \,+\, \vec L  \, \right)^2 
 \Big] \, \Big [ \,  \delta_{k + l-q}^2 + \left  (  \,  \vec K \,+\, \vec L \,-\,\vec Q  \, \right)^2 
 \Big] \,,
\end{eqnarray}
%%%%
and $f_t(l,q)$ is obtained by using Eq.~(\ref{fc2}) for $m> 1$ or Eq.~(\ref{c2}) for $m=1$. However, we must use these formulas with $\vec  u_l^2 =  \vec {L}_{(l)} ^2- \frac{ \left( \vec {L}_{(q)} \cdot \vec {L}_{(l)} \right)^2 } {  \vec {L}_{(q)}  ^2  }$ and $l_{d-m+1} = \frac{  \vec {L}_{(q)} \cdot \vec {L}_{(l)}  } {  |\vec {L}_{(q)}  |   }$.
%%%
Let $\theta_{ql} $ be the angle between $ \vec {L}_{(q)}$ and $ \vec {L}_{(l)}$. Then we can write $|\vec {L}_{(q)}|$, $|\vec {L}_{(l)}| \, \cos \theta_{ql}$ and $|\vec {L}_{(l)}| \, \sin \theta_{ql}$ in place of $q_{d-m+1}$, $ l_{d-m+1}$ and $ |\vec u_{(l)}|$ respectively.

We redefine the variables as:
\beq
y_1 = \delta_{k + l}  \,,\quad y_2 = \delta_{-q} - 2 \, \vec {L}_{(q)} \cdot \vec {L}_{(k+l)} \,,
\eeq
so that 
\begin{equation}
\delta_{k+l-q} = y_1 +  y_2\,, \quad
\delta_q = 2 \, \vec{L}_{(q)}^2 - y_2  - 2 \, \vec {L}_{(q)} \cdot \vec {L}_{(k+l)} \,, \quad
\delta_{l} =  y_1 - \delta_k - 2 \, \vec {L}_{(k)} \cdot \vec {L}_{(l)} \,.
\end{equation}
%%%%%%%%%%%%%%
Using Eq.~(\ref{fc2}), which is possible for $m>1$, we have:
\bqa
&& \int_{-\infty}^{\infty} \frac {d{\vec{v}}_{(p)} \,dp_{d-m+2}}  {(2 \pi)^{m-1} } \, 
\lbrace \kappa_1 (l,q) +  \kappa_1 ( q-l,  q)  \rbrace \nn
&&
 \simeq
\frac{ 
\frac{1} { \vec{u}_{ ( l) } ^2  } 
\left (\delta_l -  \frac{  |\vec {L}_{(l)}| \, \cos \theta_{ql}  } { |\vec {L}_{(q)}|  } \, \delta_q
-\frac { 2 \, \vec u_{(l)}^ 2 } {3}  \right )
+
\frac{1} { \vec{u}_{ ( q-l) } ^2  } 
\left (\delta_{ q-l} 
-  \frac{   |\vec {L}_{(q)}|  - |\vec {L}_{(l)}| \, \cos \theta_{ql}  } { |\vec {L}_{(q)}|  } \, \delta_{q}
-\frac { 2 \, \vec u_{(l)}^ 2 } {3}  \right )
}
%%%%
 { 2^{m + 3 }  \, \pi \,
 | \vec {L}_{(q)} |
 } \nn
 && \quad  \times \,
 \left(  \sqrt{ \frac{ k_F} {3 \pi} } \right)^{m-3}
 sgn \left(  |\vec {L}_{(l)}| \, \cos \theta_{ql} \right )
 \, sgn \left(  |\vec {L}_{(q)}|  -  |\vec {L}_{(l)}| \, \cos \theta_{ql} \right )   \nn
 %%%%%%%%%%%%%%%%%%%%%%
 &=&
 \frac{   \left( \frac{k_F}{3 \pi} \right)^{ \frac{m-3} {2} }  }  
{2^{m+3} \,   \pi  }
sgn \left(  |\vec {L}_{(l)}| \, \cos \theta_{ql} \right )
 \, sgn \left(  |\vec {L}_{(q)}|  -  |\vec {L}_{(l)}| \, \cos \theta_{ql} \right )
%%%%%%%%%%%%%%%%%%%%%%%
\frac{ 
\frac{ \tilde t_1} { \vec{u}_{ ( l) } ^2  } + \frac{ \tilde t_2} { \vec{u}_{ ( q-l) } ^2  } }
{
 | \vec {L}_{(q)} | } \,, \nn
\eqa
%%%%%%%%%%%%%%%%%%%%%%%%%%%%%%%%%%%%%%%%%%%%%%%%%%%%%%%%%
where
%%%%%%%%
\bqa
\label{tildet}
&& \tilde t_1 = y_1 +  \frac{  \vec{L}_{(q)} \cdot \vec{L}_{(l)} } { \vec{L}_{(q)}^2} \,y_2  -\delta_k 
-2\, \vec{L}_{(k)} \cdot \vec{L}_{(l)}
- \frac{2 \,\vec u_{(l)} ^2} {3  }   
+ 2 \left (  \vec{L}_{(q)} \cdot \vec{L}_{(l)} \right ) 
\left( \frac{  \vec{L}_{(q)} \cdot \vec{L}_{(k+l)}} { \vec{L}_{(q)}^2} -1
\right) \,,\nn
%%%%%%
&& \tilde t_2 =-  \left (y_1 +  \frac{  \vec{L}_{(q)} \cdot \vec{L}_{(l)} } { \vec{L}_{(q)}^2} \,y_2  -\delta_k
 \right  )
- \frac{2 \,\vec u_{(l)} ^2 } {3  }
+ 2 \, \vec L_{(l)} ^2 \,.
\eqa
%%%%%%%
There will be similar terms for the other $\kappa_i $'s.

One can find out the $e$ and $k_F$ dependence of the final answer by solving the following integrals, which appear for the various terms of the complete integrand:
\beq
I_{1 \Sigma} = \int {dy_1 \, dy_2} \, \frac{1} { \Big [ \,  y_1^2 + A^2 
 \Big] \, \Big [   \left(  y_1 + y_2 \right )^2 + B^2 
 \Big]  }
 = \frac{\pi^2}   {| A | \, |B|} \,.
 \eeq
 %%%
 \beq
I_{2 \Sigma} = \int {dy_1 \, dy_2} \, \frac{ y_1} { \Big [ \,  y_1^2 + A^2 
 \Big] \, \Big [   \left(  y_1 + y_2 \right )^2 + B^2 
 \Big]  }
 = 0 \,.
 \eeq
 %%%
 \beq
I_{3 \Sigma} = \int {dy_1 \, dy_2} \, \frac{ y_1 + y_2 } { \Big [ \,  y_1^2 + A^2 
 \Big] \, \Big [   \left(  y_1 + y_2 \right )^2 + B^2 
 \Big]  }
 = 0 \,.
 \eeq
 %%%
 \beq
I_{4 \Sigma} = \int {dy_1 \, dy_2} \, \frac{ y_1 \, \left ( y_1 + y_2 \right ) } { \Big [ \,  y_1^2 + A^2 
 \Big] \, \Big [   \left(  y_1 + y_2 \right )^2 + B^2 
 \Big]  }
 = \pi^2 \,.
 \eeq
 %%%
\beq
\label{i11sig}
I_{1 1\Sigma} = \int {dy_1 \, dy_2} \, \frac{ y_1 +  \frac{ | \vec{L}_{(q)} \cdot \vec{L}_{(l)}|} { \vec{L}_{(q)}^2} \,y_2} { \Big [ \,  y_1^2 + A^2 
 \Big] \, \Big [   \left(  y_1 + y_2 \right )^2 + B^2 
 \Big]  }
 = 0 \,.
 \eeq
 %%%
%%%
\beq
\label{i21sig}
I_{2 1\Sigma} = \int {dy_1 \, dy_2} \, \frac{ y_1\, \left( y_1 +  \frac{ | \vec{L}_{(q)} \cdot \vec{L}_{(l)}|} { \vec{L}_{(q)}^2} \,y_2 \right)} { \Big [ \,  y_1^2 + A^2 
 \Big] \, \Big [   \left(  y_1 + y_2 \right )^2 + B^2 
 \Big]  }
 = g_1 \left ( A, B, \frac{ | \vec{L}_{(q)} \cdot \vec{L}_{(l)}|} { \vec{L}_{(q)}^2} \right ) \,.
 \eeq
 %%%
\beq
\label{i31sig}
I_{3 1\Sigma} = \int {dy_1 \, dy_2} \, \frac{ \left( y_1 + y_2 \right) \, \left( y_1 +  \frac{ | \vec{L}_{(q)} \cdot \vec{L}_{(l)}|} { \vec{L}_{(q)}^2} \,y_2 \right)} { \Big [ \,  y_1^2 + A^2 
 \Big] \, \Big [   \left(  y_1 + y_2 \right )^2 + B^2 
 \Big]  }
 = g_2 \left ( A, B, \frac{ | \vec{L}_{(q)} \cdot \vec{L}_{(l)}|} { \vec{L}_{(q)}^2} \right ) \,.
 \eeq
 %%%
\beq
\label{i41sig}
I_{ 4 1\Sigma} = \int {dy_1 \, dy_2} \, \frac{ y_1 \,  \left( y_1 + y_2 \right) \, \left( y_1 +  \frac{ | \vec{L}_{(q)} \cdot \vec{L}_{(l)}|} { \vec{L}_{(q)}^2} \,y_2 \right)} { \Big [ \,  y_1^2 + A^2 
 \Big] \, \Big [   \left(  y_1 + y_2 \right )^2 + B^2 
 \Big]  }
 = 0 \,.
 \eeq
 %%%%

To calculate the overall powers of $ \tilde e $, $k_F$ and $\Lambda $, we scale out $\tilde \alpha $ appearing in the boson propagators by redefining variables as:
\beq
\vec L_{(q)} =  \left( \tilde \alpha \, |\vec P |^{ d-m } \right)^{ \frac{1}{3} } \, \tilde{ \vec L }_{(q)}\,,
 \quad \vec{  L }_{(l)} =  \left( \tilde \alpha \, |\vec P |^{ d-m } \right)^{ \frac{1}{3} } \, \, \tilde{ \vec L}_{(l)}\,.
\eeq
Then we have terms proportional to:
\bqa
&&
\left( \frac{ \tilde e \, \Lambda}
{k_F} \right)^{\frac{2\,m }  { m+1} }   \delta_k 
%%%
=
\left(  \frac{ \tilde e } { \lambda_{ \text{cross}} ^{ \frac{1} {m+1}  } } 
\right)^{\frac{2\,m }  { m- 1} }   \delta_k \,,
%%%%%%%%%%%%%%%%%%
%%%%%%%%%%%
\quad \left( \frac{ \tilde e  \, \Lambda}
{k_F} \right)^{\frac{2\,m }  { m+1} } i\,\left( \vec \Gamma \cdot \vec K \right) 
\sim \frac{\Lambda} { k_F } \, \Sigma_{2 b }
%%%
=
\left(  \frac{ \tilde e^2 } 
{ \lambda_{ \text{cross}} } 
\right)^{\frac{ 1 }  { m- 1} } \, \Sigma_{2 b }
\,,
\nn && 
%%%%%%%%%%%%
%%%%%%%%%%
\frac{ \tilde e^2 \, \Lambda}
{k_F} \,  \delta_k 
%%%
=
\left(  \frac{ \tilde e^{2\, m } } 
{ \lambda_{ \text{cross}}  } 
\right)^{\frac{ 1 }  { m- 1} }   \delta_k \,,
%%%%%%%%%
%%%%%%%%%%%%%\
\quad \frac{ \tilde e^2 \, \Lambda}
{k_F} \, i\,\left( \vec \Gamma \cdot \vec K \right) 
\sim
\left( \frac{\tilde e \, \Lambda}
{k_F} \right)^{\frac{2\,m }  { m+1} }  \Sigma_{2 b }
%%%
=
\left(  \frac{ \tilde e } { \lambda_{ \text{cross}} ^{ \frac{1} {m+1}  } } 
\right)^{\frac{2\,m }  { m- 1} }   \Sigma_{2 b } \,,
\eqa
%%%%%%%%%%%%%%%%%%%%%%%%%%%%
to leading order in $k$, for $m>1$. 
There will be similar terms for the other $\kappa_i $'s.
Hence we conclude that for $m>1$, the three-loop terms are suppressed compared to the the one-loop terms for $\lambda_{ \text{cross} } >> 1 $.

For $\lambda_{ \text{cross} } << 1 $, which includes the case of $m=1$, we have:
%%%%%%%%%%%
\begin{eqnarray}
&&  \int \frac {d{\vec{v}}_{(p)} \,dp_{d-m+2}}  {(2 \pi)^{m-1} } \, 
\lbrace \kappa_1 (l,q) +  \kappa_1 (-l, -q)  \rbrace \nn
%%%%%%%%%%
&& = 
\frac{  \left(  \sqrt{ \frac{ k_F} {3 \pi} } \right)^{m- 1 }
\, sgn \left( |\vec {L}_{(l)}| \, \cos \theta_{ql} - |\vec {L}_{(q)}|  \right ) 
\, sgn \left( |\vec {L}_{(l)}| \, \cos \theta_{ql} \right )  } 
 { 2^{m + 2 } \,
  | \vec {L}_{(q)}  |
 } \,
\left (
\frac{ \tilde t_1 } 
{   
  \tilde t_1^2   
  +
  %%%%
\tilde p^2
} +
\frac{ \tilde t_2 } 
{   
  \tilde t_2^2   
  +
  %%%%
\tilde p^2
}  \right )\,,
% \label{c2}
 \nn
\end{eqnarray}
%%%%%%%%%%%%%%%%
where $\tilde t_{1,2} $ has been defined in Eq.~(\ref{tildet}) and
\bqa
\tilde p =
\frac{ 
 | \, |\vec {L}_{(l)}| \, \cos \theta_{ql} - |\vec {L}_{(q)}| \, | \times
\,\, |\vec P|
+ 
|\vec {L}_{(l)}| \, | \cos \theta_{ql} | \, |\vec P +\vec Q| 
+| \vec {L}_{(q)} | \, |\vec P +\vec L| 
 }
{  | \vec {L}_{(q)} | } \,.
\label{tildep}
\eqa
%%%%%%%%%%%%%%

For the term proportional to $\gamma_{d-m}$, we need integrals of the following form:
\bqa
\label{i5sig}
I_{5 \Sigma} 
&=& 
\int \frac{dy_1 \, dy_2} {(2 \, \pi) ^2 }
\frac{ 1 }
{ \Big [ \,  y_1^2 + |\vec K +\vec L|^2 
 \Big] \, \Big [   \left(  y_1 + y_2 \right )^2 +|\vec K +\vec L-\vec Q |^2 
 \Big]} \, 
\frac{ y_1+  \frac{ l_{d-m+1}} { q_{d-m+1}} \,y_2 - \tilde a  } 
{   
  \left( y_1+  \frac{ l_{d-m+1}} { q_{d-m+1}} \,y_2 - \tilde a  \right)^2   
  +
 \tilde p^2} \nn
 %%%%%%%%%%%%%%%%%%
&=& \frac{q_{d-m+1}} {l_{d-m+1}}
\int \frac{dy_1 \, dy_2} {(2 \, \pi) ^2 }
\frac{ 1 }
{ \Big [ \,  y_1^2 + |\vec K +\vec L|^2 
 \Big] \, \Big [   \left(  y_2 + y_1 \right )^2 +|\vec K +\vec L-\vec Q |^2 
 \Big]} \, 
\frac{ y_2+  \frac{q_{d-m+1}} {l_{d-m+1}} \left( y_1 -\tilde a \right )  } 
{   
  \lbrace  y_2+  \frac{q_{d-m+1}} {l_{d-m+1}} \left( y_1 -\tilde a  \right) \rbrace ^2   
  +
 \frac{q_{d-m+1}^2} {l_{d-m+1}^2} \, \tilde p^2 } \nn
  %%%%%%%%%%%%%%%%%%
&=& \frac{q_{d-m+1}} { 2 \, l_{d-m+1} \, |\vec K +\vec L-\vec Q |} \nn
&&
\times \, \int \frac{dy_1 } { 2 \, \pi }
\frac{  \frac{q_{d-m+1}} {l_{d-m+1}} \left( y_1 -\tilde a  \right) -y_1 }
{ \Big [ \,  y_1^2 + |\vec K +\vec L|^2 
 \Big] \, \Big [   \lbrace \,
  \frac{q_{d-m+1}} {l_{d-m+1}} \left( y_1 -\tilde a  \right) -y_1 \rbrace^2
   +
   \lbrace \,
  |  \frac{q_{d-m+1}} {l_{d-m+1} } | \, \tilde p + |\vec K +\vec L-\vec Q | \, \rbrace^2 
 \Big]  }
 \nn
 %%%%%%%%%%%%%%%%%%
&=& \frac{q_{d-m+1}} { 2 \,\left (q_{d-m+1}-  l_{d-m+1} \right )
\, |\vec K +\vec L-\vec Q | } \nn
&& \times \,
\int \frac{dy_1 } { 2 \, \pi }
\frac{ y_1 - \frac{q_{d-m+1}} {   q_{d-m+1}-  l_{d-m+1}  } \, \tilde a }
{ \Big [ \,  y_1^2 + |\vec K +\vec L|^2 
 \Big] \, \Big [   \lbrace \,
  y_1 - \frac{q_{d-m+1}} {  q_{d-m+1}-  l_{d-m+1}   } \, \tilde a \,  \rbrace^2
   +
   \lbrace \,
 \frac{ |  q_{d-m+1} | \, \tilde p +  | l_{d-m+1} |\, |\vec K +\vec L-\vec Q |
 }
 { | q_{d-m+1} -l_{d-m+1} | }    \, \rbrace^2 
 \Big]  }
 \nn
 %%%%%%%%%%%%%%%%%%
 &=&
 \frac{- \tilde a }   
 { 4\, |\vec K +\vec L| \,  |\vec K +\vec L-\vec Q | \,
\Big [ \tilde a^2 +\,
\lbrace \, 
|1-\frac{  l_{d-m+1} } {  q_{d-m+1}}| \, |\vec K +\vec L|
+ | \frac{  l_{d-m+1} } {  q_{d-m+1}}| \, |\vec K +\vec L-\vec Q |
+ \tilde p
\,  \rbrace^2
 \Big ]
 } \,,
 \eqa
%%%%%%%%%%%%% 
and
\bqa
\label{i6sig}
I_{6 \Sigma} 
&=& 
\int \frac{dy_1 \, dy_2} {(2 \, \pi) ^2 }
\frac{ y_1  \left(  y_1 + y_2 \right )  }
{ \Big [ \,  y_1^2 + |\vec K +\vec L|^2 
 \Big] \, \Big [   \left(  y_1 + y_2 \right )^2 +|\vec K +\vec L-\vec Q |^2 
 \Big]} \, 
\frac{ y_1+  \frac{ l_{d-m+1}} { q_{d-m+1}} \,y_2 - \tilde a  } 
{   
  \left( y_1+  \frac{ l_{d-m+1}} { q_{d-m+1}} \,y_2 - \tilde a  \right)^2   
  +
 \tilde p^2} \nn
 %%%%%%%%%%%%%%%%%%
&=& \frac{q_{d-m+1}} {l_{d-m+1}}
\int \frac{dy_1 \, dy_2} {(2 \, \pi) ^2 }
\frac{ y_1  \left(  y_2 + y_1 \right ) }
{ \Big [ \,  y_1^2 + |\vec K +\vec L|^2 
 \Big] \, \Big [   \left(  y_2 + y_1 \right )^2 +|\vec K +\vec L-\vec Q |^2 
 \Big]} \, 
\frac{ y_2+  \frac{q_{d-m+1}} {l_{d-m+1}} \left( y_1 -\tilde a \right )  } 
{   
  \lbrace  y_2+  \frac{q_{d-m+1}} {l_{d-m+1}} \left( y_1 -\tilde a  \right) \rbrace ^2   
  +
 \frac{q_{d-m+1}^2} {l_{d-m+1}^2} \, \tilde p^2 } \nn
  %%%%%%%%%%%%%%%%%%
&=& \frac{q_{d-m+1}} { 2 \, l_{d-m+1}}
\int \frac{dy_1 } { 2 \, \pi }
\, y_1
\frac{  |  \frac{q_{d-m+1}} {l_{d-m+1} } | \, \tilde p 
+    |\vec K +\vec L-\vec Q | }
{ \Big [ \,  y_1^2 + |\vec K +\vec L|^2 
 \Big] \, \Big [   \lbrace \,
  \frac{q_{d-m+1}} {l_{d-m+1}} \left( y_1 -\tilde a  \right) -y_1 \rbrace^2
   +
   \lbrace \,
  |  \frac{q_{d-m+1}} {l_{d-m+1} } | \, \tilde p + |\vec K +\vec L-\vec Q | \, \rbrace^2 
 \Big]  }
 \nn
 %%%%%%%%%%
&=& \frac{q_{d-m+1} \,  l_{d-m+1} \, 
\left (  |  \frac{q_{d-m+1}} {l_{d-m+1} } | \, \tilde p 
+    |\vec K +\vec L-\vec Q | \right )} 
{ 2 \,\left (q_{d-m+1}-  l_{d-m+1} \right )^2
 } \nn
&& \times \,
\int \frac{dy_1 } { 2 \, \pi }
\frac{ y_1 }
{ \Big [ \,  y_1^2 + |\vec K +\vec L|^2 
 \Big] \, \Big [   \lbrace \,
  y_1 - \frac{q_{d-m+1}} {  q_{d-m+1}-  l_{d-m+1}   } \, \tilde a \,  \rbrace^2
   +
   \lbrace \,
 \frac{ |  q_{d-m+1} | \, \tilde p +  | l_{d-m+1} |\, |\vec K +\vec L-\vec Q |
 }
 { | q_{d-m+1} -l_{d-m+1} | }    \, \rbrace^2 
 \Big]  }
 \nn
 %%%%%%%%%%%%%%%%%%
 &=&
 \frac{ -  \tilde a
 \, sgn \left( l_{d-m+1}- q_{d-m+1} \right ) 
\, sgn \left( l_{d-m+1} \right ) }   
 { 4\,
\Big [ \tilde a^2 +\,
\lbrace \, 
|1-\frac{  l_{d-m+1} } {  q_{d-m+1}}| \, |\vec K +\vec L|
+ | \frac{  l_{d-m+1} } {  q_{d-m+1}}| \, |\vec K +\vec L-\vec Q |
+ \tilde p
\,  \rbrace^2
 \Big ]
 } \,.
 \eqa
%%%%%%%%%%%%% 
Setting $\vec K =\vec L_{ (k)} = 0$, we have then terms as:
\bqa
%%%%%%
&&  \frac{ \delta_k + 
\sum \limits_{\substack{
   s_1, s_2 = q,l \\
   j=1,\ldots,m
  }}  c_{s_1 s_2\,  j } \, L_{ (s_1)}^{j} \, L_{ (s_2)}^{j}
  }
%\sum _{ s_1, s_2 = q,l ; j=1,\ldots,m}  
{  \left (
\delta_k + 
\sum \limits_{\substack{
   s_1, s_2 = q,l \\
   j=1,\ldots,m
  }}  c_{ s_1 s_2\,  j } \, L_{ (s_1)}^{j} \, L_{ ( s_2 )}^{j}
\right )^2 
+\lbrace \, 
|1-\frac{  | \vec L_{l}| \, \cos \theta_{ql} } { | \vec L_{q}| }| \, | \vec L|
+ \frac{   | \vec L_{l}| \, |\cos \theta_{ql} |  } {  | \vec L_{q}| } \, | \vec L-\vec Q |
+ \tilde p
\,  \rbrace^2 } \nn
%%%%%%%%%%%%
&&+
\frac{  - \delta_k + 
\sum \limits_{\substack{
   s_1, s_2 = q,l \\
   j=1,\ldots,m
  }} \tilde  c_{s_1 s_2\,  j } \, L_{ (s_1)}^{j} \, L_{ (s_2 )}^{j}
  }
{  \left (
- \delta_k + 
\sum \limits_{\substack{
   s_1, s_2 = q,l \\
   j=1,\ldots,m
  }}  \tilde c_{s_1 s_2\,  j } \, L_{ (s_1)}^{j} \, L_{ ( s_2 )}^{j}
\right )^2 
+\lbrace \, 
|1-\frac{  | \vec L_{l}| \, \cos \theta_{ql} } { | \vec L_{q}| }| \, | \vec L|
+ \frac{   | \vec L_{l}| \, |\cos \theta_{ql} |  } {  | \vec L_{q}| } \, | \vec L-\vec Q |
+ \tilde p
\,  \rbrace^2 } \,.
\eqa
%%%%%%%%%%%
We can expand to leading order in $\delta_k$. Furthermore, in the limit $\lambda_{\text{cross}} << 1$, the main contribution to the integral over $\vec L_{ (q) }$ and $\vec L_{ (l) }$ will come from $ |  L_{ (q)}^{j} |, |  L_{ (l)}^{j'} | \sim \tilde \alpha^{1/3} \Lambda^{\frac{d-m} {3} }  << \Lambda$. So, we can also expand in small $ c_{s_1 s_2\,  j } \, L_{ (q)}^{j} \, L_{ (l)}^{j}$ and $ \tilde  c_{s_1 s_2\,  j } \, L_{ (q)}^{j} \, L_{ (l)}^{j}$, such that the leading order term proportional to $\delta_k$ can be extracted, which is:
%%%%%%%%%%%
\bqa
\frac{   
\delta_k  \,
\left ( \sum \limits_{\substack{
   s_1, s_2 = q,l \\
   j=1,\ldots,m
  }} \tilde  d_{ab j} \,  L_{ (s_1)}^{j} \, L_{ (s_2)}^{j}
  \right )^2
  }
{ \lbrace \, 
|1-\frac{  | \vec L_{l}| \, \cos \theta_{ql} } { | \vec L_{q}| }| \, | \vec L|
+ \frac{   | \vec L_{l}| \, |\cos \theta_{ql} |  } {  | \vec L_{q}| } \, | \vec L-\vec Q |
+ \tilde p
\,  \rbrace^4 } \,.
\eqa

For the term proportional to $\vec \Gamma \cdot \vec K$, we need the following integrals:
\bqa
\label{i7sig}
I_{7 \Sigma} 
&=& 
\int \frac{dy_1 \, dy_2} {(2 \, \pi) ^2 }
\frac{ y_1 }
{ \Big [ \,  y_1^2 + |\vec K +\vec L|^2 
 \Big] \, \Big [   \left(  y_1 + y_2 \right )^2 +|\vec K +\vec L-\vec Q |^2 
 \Big]} \, 
\frac{ y_1+  \frac{ l_{d-m+1}} { q_{d-m+1}} \,y_2 - \tilde a  } 
{   
  \left( y_1+  \frac{ l_{d-m+1}} { q_{d-m+1}} \,y_2 - \tilde a  \right)^2   
  +
 \tilde p^2} \nn
%%%%%%%%%%%%% 
 &=&
 \frac{
  - \, sgn \left( l_{d-m+1}- q_{d-m+1} \right ) 
\, sgn \left( l_{d-m+1} \right )
\,\left( 
 |1-\frac{  l_{d-m+1} } {  q_{d-m+1}}| \, |\vec K +\vec L|
+ | \frac{  l_{d-m+1} } {  q_{d-m+1}}| \, |\vec K +\vec L-\vec Q |
+ \tilde p \right ) }   
 { 4 \,  |\vec K +\vec L-\vec Q |\,
\Big [ \tilde a^2 +\,
\lbrace \, 
|1-\frac{  l_{d-m+1} } {  q_{d-m+1}}| \, |\vec K +\vec L|
+ | \frac{  l_{d-m+1} } {  q_{d-m+1}}| \, |\vec K +\vec L-\vec Q |
+ \tilde p
\,  \rbrace^2
 \Big ]
 } \,,\nn
 \eqa
%%%%%%%%%%%%% 
and
\bqa
\label{i8sig}
I_{8 \Sigma} 
&=& 
\int \frac{dy_1 \, dy_2} {(2 \, \pi) ^2 }
\frac{   \left(  y_1 + y_2 \right )  }
{ \Big [ \,  y_1^2 + |\vec K +\vec L|^2 
 \Big] \, \Big [   \left(  y_1 + y_2 \right )^2 +|\vec K +\vec L-\vec Q |^2 
 \Big]} \, 
\frac{ y_1+  \frac{ l_{d-m+1}} { q_{d-m+1}} \,y_2 - \tilde a  } 
{   
  \left( y_1+  \frac{ l_{d-m+1}} { q_{d-m+1}} \,y_2 - \tilde a  \right)^2   
  +
 \tilde p^2} \nn
 %%%%%%%%%%%%%%%%%%
 &=&
 \frac{ 
 \, sgn \left(  q_{d-m+1} \right ) 
\, sgn \left( l_{d-m+1} \right )
\, \lbrace \, 
|1-\frac{  l_{d-m+1} } {  q_{d-m+1}}| \, |\vec K +\vec L|
+ | \frac{  l_{d-m+1} } {  q_{d-m+1}}| \, |\vec K +\vec L-\vec Q |
+ \tilde p
\,  \rbrace
 }   
 { 4\, |\vec K +\vec L|\,
\Big [ \tilde a^2 +\,
\lbrace \, 
|1-\frac{  l_{d-m+1} } {  q_{d-m+1}}| \, |\vec K +\vec L|
+ | \frac{  l_{d-m+1} } {  q_{d-m+1}}| \, |\vec K +\vec L-\vec Q |
+ \tilde p
\,  \rbrace^2
 \Big ]
 } \,.
 \eqa
%%%%%%%%%%%%% 
%%%%%%%%%%%%% 
Setting $\delta_k =\vec L_{ (k)} = 0$, now we have terms as:
\bqa
%%%%%%
&&  \frac{1  }
%\sum _{ s_1, s_2 = q,l ; j=1,\ldots,m}  
{  \left (
\sum \limits_{\substack{
   s_1, s_2 = q,l \\
   j=1,\ldots,m
  }}  c_{ s_1 s_2\,  j } \, L_{ (s_1)}^{j} \, L_{ ( s_2 )}^{j}
\right )^2 
+\lbrace \, 
|1-\frac{  | \vec L_{l}| \, \cos \theta_{ql} } { | \vec L_{q}| }| \, | \vec L|
+ \frac{   | \vec L_{l}| \, |\cos \theta_{ql} |  } {  | \vec L_{q}| } \, | \vec L-\vec Q |
+ \tilde p
\,  \rbrace^2 } \nn
%%%%%%%%%%%%
&&+
\frac{ 1  }
{  \left (
\sum \limits_{\substack{
   s_1, s_2 = q,l \\
   j=1,\ldots,m
  }}  \tilde c_{ s_1 s_2\,  j } \, L_{ (s_1)}^{j} \, L_{ ( s_2 )}^{j}
\right )^2 
+\lbrace \, 
|1-\frac{  | \vec L_{l}| \, \cos \theta_{ql} } { | \vec L_{q}| }| \, | \vec L|
+ \frac{   | \vec L_{l}| \, |\cos \theta_{ql} |  } {  | \vec L_{q}| } \, | \vec L-\vec Q |
+ \tilde p
\,  \rbrace^2 } \,,
\eqa
%%%%%%%%%%%
which can be expanded to leading order in small $ c_{ s_1 s_2\,  j } \, L_{ (q)}^{j} \, L_{ (l)}^{j}$ and $ \tilde  c_{ s_1 s_2\,  j } \, L_{ (q)}^{j} \, L_{ (l)}^{j}$. The leading order term proportional to $\vec \Gamma \cdot \vec K$ can now be extracted, which is:
%%%%%%%%%%%
\bqa
\frac{   
\vec \Gamma \cdot \vec K  \,
\left ( \sum \limits_{\substack{
   s_1, s_2 = q,l \\
   j=1,\ldots,m
  }} \tilde  g_{  s_1 s_2\,  j } \,  L_{ (s_1)}^{j} \, L_{ (s_2)}^{j}
  \right )^2
  }
{ \lbrace \, 
|1-\frac{  | \vec L_{l}| \, \cos \theta_{ql} } { | \vec L_{q}| }| \, | \vec L|
+ \frac{   | \vec L_{l}| \, |\cos \theta_{ql} |  } {  | \vec L_{q}| } \, | \vec L-\vec Q |
+ \tilde p
\,  \rbrace^4 } \,.
\eqa

Again, to calculate the overall powers of $ {\tilde{e}}$, $k_F$ and $\Lambda$, we scale out $\tilde \alpha $ appearing in the boson propagators by redefining variables as:
\beq
\vec L_{(q)} = \left( \tilde \alpha \, |\vec P |^{d-m} \right)^{ \frac{1}{3} } \, \tilde{ \vec L }_{(q)}\,,
 \quad \vec{  L }_{(l)} =  \left( \tilde \alpha \, |\vec P |^{ d-m } \right)^{ \frac{1}{3} } \, \, \tilde{ \vec L}_{(l)}\,.
\eeq
Then the overall dependence is 
%%%%%%%%%%%%%%%%%%%%%%%%
\bqa
\Sigma_{3a} (q) & \sim &  {\tilde{e}} ^{ \frac { 2\,(m+3 ) } { m+1} }
 \left ( \frac{ k_F}  { \Lambda }  \right ) ^{ \frac {2\,( m-1) } { m+1 } }     \gamma_{ d-m } \,  \delta_q 
 %%%%%%%%%%%%%%
 = { \lambda_{\text{cross}} } ^{ \frac { m+3  } { m+1} }
 \left ( \frac{  \Lambda }  {k_F }  \right ) ^{ m-1 }    \gamma_{ d-m } \,  \delta_q \,,\nn
 %%%%%%%%%%%%%%%%%%%%
 \Sigma_{3b} (q) & \sim &   {\tilde{e}} ^{ \frac { 2\,(m+3 ) } { m+1} }
 \left ( \frac{ k_F}  { \Lambda }  \right ) ^{ \frac{ 2\, (m-1) } { m+1 } }   \left (\vec \Gamma \cdot \vec Q   \right )
 = { \lambda_{\text{cross}} } ^{ \frac { m+3  } { m+1} }
 \left ( \frac{  \Lambda }  {k_F }  \right ) ^{ m-1 } 
 \left (\vec \Gamma \cdot \vec Q   \right ).
\eqa
 This shows that there is a logarithmic divergence at $m=1$. However, for $m>1$, in the limit  $ \lambda_{\text{cross}} \ll 1 $, the integral is not divergent, a behaviour which is also seen for the $\lambda_{\text{cross}} \gg 1 $ limit.

%%%%%%%%%%%%%%%%%%%%%%%%%%%
%%%%%%%%%%%%%%%%%%%%%%%%%%%%%%%%%%%%%%%%%%%%%%%%%%%%%%%%%%
 \subsection{Three-loop Aslamazov-Larkin-type contribution to boson self-energy}
 \label{3loopAL}

\begin{figure}
\centering
\includegraphics[width= 0.6 \textwidth]{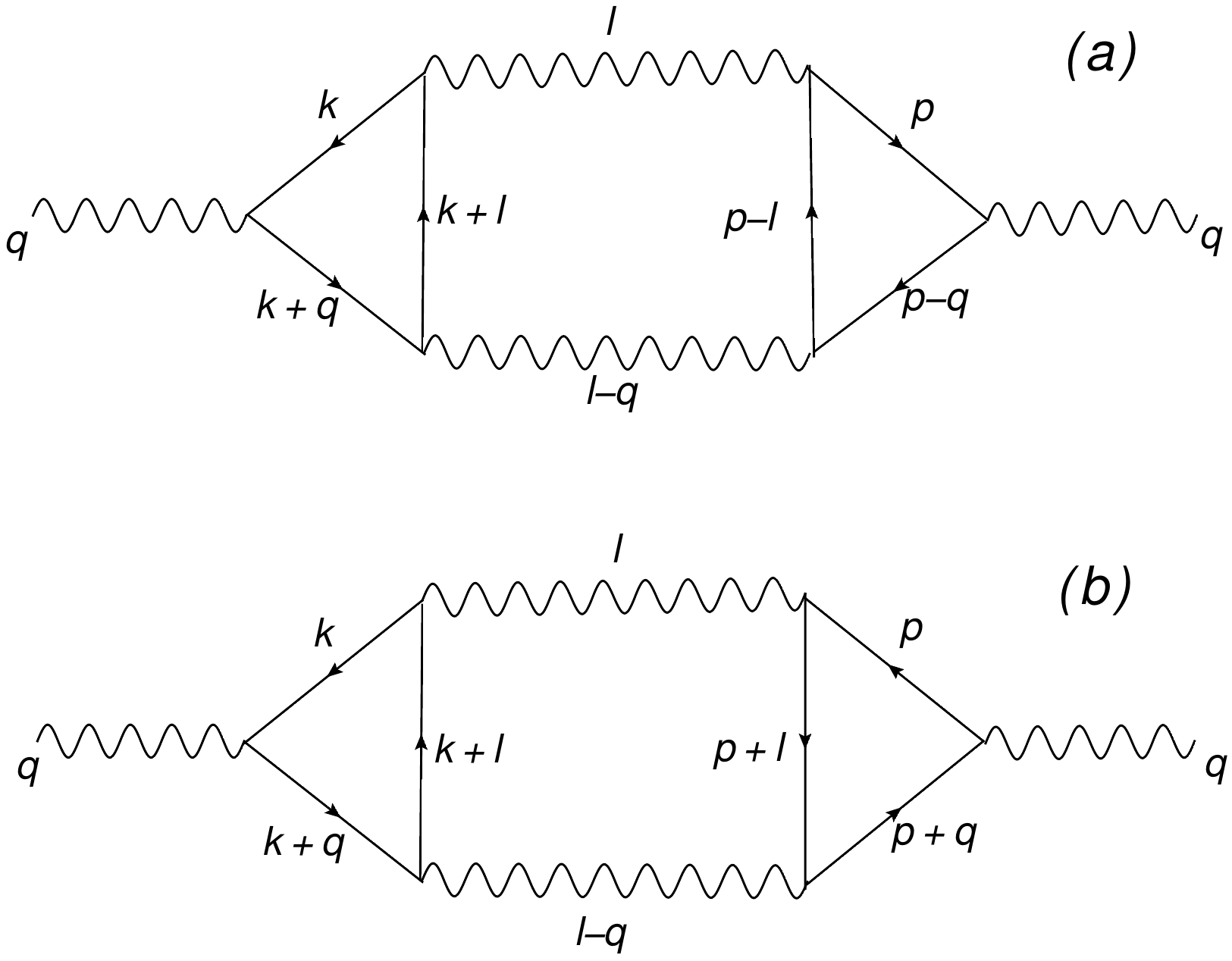}
\caption{Aslamazov-Larkin type contributions to boson self-energy. 
Diagrams (a) and (b) correspond to the particle-particle and 
particle-hole channels respectively.}
\label{fig:ALbos}
\end{figure}

The  Aslamazov-Larkin (AL) type diagrams shown in Fig.~\ref{fig:ALbos} are the lowest order diagrams that can renormalize the boson kinetic term \cite{metlsach1,mross}. These
give a three-loop contribution to boson self-energy as
\beq
\label{AL0}
\Pi_{AL} (q) = \Pi_{pp} (q) + \Pi_{ph} (q) =\int  dl \,  D_1 (l) \, D_1 (l-q) \, f_t(l,q) \,
[ \,f_t(l,q)+ f_t(-l,-q) \, ] \,.
\eeq
We will consider $\vec Q =0$ for simplicity, which is enough to examine the divergences. Also, the coordinate system is oriented such that $ \vec{ L }_{(q)}=  \left( q_{d-m+1},0, \ldots, 0\right) $\,.

%For $m=1$, we use Eq.~(\ref{c2}) to confirm that $\Pi_{AL} (q)$ is independent of $k_F$.

%%%%%%%%%%
For $\lambda_{ \text{cross} } >> 1 $, we have
%%%%%%%%%%%%%%%
 \begin{eqnarray}
  f_t(l,q; \vec Q=0 ) 
 %%%%%%%%%%%%%%
&=&  \frac{e^3 \mu^{3 \, x/2 }} 
{\sqrt{N} } 
\int \frac { d \vec P} { ( 2 \, \pi )^{d-m}} 
\frac { 
\Big [ \vec P \cdot (\vec P + \vec L   ) 
- | \vec P | \, | \vec P + \vec L |
\Big ] \, 
\Big [ \Theta \left (  l_{d-m+1} \right) -\Theta \left(  l_{d-m+1} -  q_{d-m+1} \right ) \Big ] }
{  | \vec P | \, | \vec P + \vec L | } \nn
&& \times \,
%%%%%%%%%%%
\frac{ \delta_l -  \frac{l_{d-m+1}} {  q_{d-m+1} } \, \delta_q
-\frac { 2 \, \vec u_{(l)}^ 2 } {3} } 
 { 2^{m + 3 }  \, \pi \, |\vec{u}_{ (l) } |^2 
 \,  q_{d-m+1}
 }
 \left(  \sqrt{ \frac{ k_F} {3 \pi} } \right)^{m-3} \,.
 \label{fc2}
\end{eqnarray}

%%%%%%%%%%%%%
%%%%%%%%%%%%%%%
For $\lambda_{ \text{cross} } << 1 $, which includes the case $m=1$, 
we have
%%%%%%%%%%%%%%%
 \begin{eqnarray}
  f_t(l,q; \vec Q=0 ) 
 %%%%%%%%%%%%%%
&=& \frac{e^3 \mu^{3 \, x/2 }} 
{\sqrt{N} } 
\int \frac { d \vec P} { ( 2 \, \pi )^{d-m}}
 \frac { 
\Big [ \vec P \cdot (\vec P + \vec L   ) 
- | \vec P | \, | \vec P + \vec L |
\Big ] \, 
\Big [ \Theta \left (  l_{d-m+1} \right) -\Theta \left(  l_{d-m+1} -  q_{d-m+1} \right ) \Big ] }
{  | \vec P | \, | \vec P + \vec L | } \nn
&& \qquad \qquad \times \, 
%%%%%%%%%% 
\frac{  \left(  \sqrt{ \frac{ k_F} {3 \pi} } \right)^{m- 1 } } 
 { 2^{m + 2 } \,
 q_{d-m+1}
 } 
\frac{ \delta_l -  \frac{l_{d-m+1}} {  q_{d-m+1} } \, \delta_q
-\frac { 2 \, \vec u_{(l)}^ 2 } {3} } 
{   
  \left( \delta_l -  \frac{l_{d-m+1}} {   q_{d-m+1} } \, \delta_q
-\frac { 2 \, \vec u_{(l)}^ 2 } {3}  \right)^2   
  +
  %%%%
\Big [
 |\vec P| +  |\vec P +\vec L | 
\Big]^2
} \,. 
\end{eqnarray}
%%%%%%%%%%%%%%%

First, let us focus on this limit of $\lambda_{ \text{cross} } << 1 $ in order to see if $Z_3 $ gets a correction from the AL terms for this range.
For the particle-hole channel containing $f(l,q)\, f(l,q)$,
we redefine variables as
$ y =   \delta_l -  \frac{l_{d-m+1}} {  q_{d-m+1} } \, \delta_q
-\frac { 2 \, \vec u_{(l)}^ 2 } {3} $, and 
integrate over $l_{d-m}$ to obtain
%%%%%%%%%%%%%%
\bqa
\label{pipp}
\Pi_{ph}  ( q;\vec Q=0 )  &= &
\frac{e^6 \mu^{3 \, x}} 
{N \, q_{d-m+1}^2} 
\int \frac{d\vec P \, d\vec K \, d\vec L \, d\vec L_{l} } 
{(2\pi)^{ 3d+1 - 2m  }}
 D_1 (l) D_1 (l-q) 
 \Big [ \Theta \left (  l_{d-m+1} \right) -\Theta \left(  l_{d-m+1} -  q_{d-m+1} \right ) \Big ]^2 \nn
 && \qquad \qquad 
 \times 
\frac{\left( \,  [\vec P \cdot (\vec P +\vec L)] - |\vec P| \, |\vec P +\vec L| \,
\right)
\left( \, [ \vec K \cdot (\vec K +\vec L)] -|\vec K| \, |\vec K +\vec L| \, \right)
}
{8 \, |\vec P| \, |\vec K| \,|\vec P +\vec L|\,|\vec K +\vec L| 
\left[ \,|\vec P|+ |\vec P +\vec L| + |\vec K | +|\vec K +\vec L| \, \right]  }.
\eqa
%%%%%%%%%%%%%%%
To calculate the contribution in the particle-particle channel 
containing $f(l,q) f(-l,-q) $, we define 
$ \tilde y = l_{d-m} -  \frac{l_{d-m+1}} {  q_{d-m+1} }  \, q_{d-m}  $, and
integrate over $l_{d-m }$ to get
%%%%%%%%%%%%%%
\bqa
\label{piph}
\Pi_{pp} ( q;\vec Q=0 ) &=& - \frac{e^6 \mu^{3 \, x}} 
{N \, q_{d-m+1}^2} 
\int \frac{d\vec P \, d\vec K \, d\vec L \, d\vec L_{l} } 
{(2\pi)^{ 3d+1 - 2m  }}
 D_1 (l) \,D_1 (l-q ) 
 \Big [ \Theta \left (  l_{d-m+1} \right) -\Theta \left(  l_{d-m+1} -  q_{d-m+1} \right ) \Big ]^2  \nn
&& \qquad \qquad   \times \, 
\frac{\left( [\vec P \cdot (\vec P +\vec L)] -|\vec P| \, |\vec P +\vec L|
\right)
\left( [\vec K \cdot (\vec K +\vec L)] -|\vec K| \, |\vec K +\vec L| \right)
}
{8 \, |\vec P| \, |\vec K| \,|\vec P +\vec L|\,|\vec K +\vec L|}  \nn
&&  \qquad \qquad   \times \,
\frac{  |\vec P|+ |\vec P +\vec L| + |\vec K| +|\vec K +\vec L|  } 
{\left[ \, |\vec P|+ |\vec P +\vec L| + |\vec K| +|\vec K +\vec L| \, \right]^2 +
4 \,   \left( l_{d-m+1}^2 - l_{d-m+1} \, q_{d-m+1}   + \frac {  \vec u_{(l)}^ 2 } {3}  \right )^2
} \nn
%%%%%%%%%%%%%%
&=& - \frac{ 4 \, e^6 \mu^{3 \, x}} 
{N \, q_{d-m+1}^2} 
\int \frac{d\vec P \, d\vec K \, d\vec L \, d\vec u_{l} } 
{(2\pi)^{ 3d+1 - 2m  }}
\int_0^{ | q_{d-m+1} | } d l_{d-m+1} \,
 D_1 (l) \, D_1 (l-q ; q_{ d-m+1 } \rightarrow | q_{ d-m+1 } |)   \nn
&& \qquad \qquad   \times \, 
\frac{\left( [\vec P \cdot (\vec P +\vec L)] -|\vec P| \, |\vec P +\vec L|
\right)
\left( [\vec K \cdot (\vec K +\vec L)] -|\vec K| \, |\vec K +\vec L| \right)
}
{8 \, |\vec P| \, |\vec K| \,|\vec P +\vec L|\,|\vec K +\vec L|}  \nn
&&  \qquad \qquad   \times \,
\frac{  |\vec P|+ |\vec P +\vec L| + |\vec K| +|\vec K +\vec L|  } 
{\left[ \, |\vec P|+ |\vec P +\vec L| + |\vec K| +|\vec K +\vec L| \, \right]^2 +
4 \, \left ( l_{d-m+1}^2 - l_{d-m+1} \, |q_{d-m+1}|    + \frac {  \vec u_{(l)}^ 2 } {3} \right )^2 } \,.\nn
\eqa
%%%%%%%%%%%%%%%%
Although $\Pi_{pp} (q)$ and $\Pi_{ph} (q)$ 
are  individually UV divergent, their sum results in a UV finite correction.
Rescaling $l_{d-m+1} $ as
\beq
l_{d} \rightarrow l_{d-m+1}\, |q_{d-m+1}|
\eeq      
to make the integral over $l_{d-m+1} $ run from $0$ to $1$, and rescaling
\bqa
&& \vec L \rightarrow   2 \, q_{d-m+1}^2 \, l_{d-m+1} \, (1-l_{d-m+1}) \, \vec L\,, \,
\vec P \rightarrow   2 \,  q_{d-m+1}^2 \, l_{d-m+1}\, (1-l_{d-m+1})\, \vec P \,, \nn
&&
\vec K \rightarrow  2 \, q_{d-m+1}^2 \, l_{d-m+1}\, (1-l_{d-m+1}) \, \vec K \, ,
\eqa
we arrive at the expression:
%%%%%%%%%%%%%%%%%%%
\bqa
\label{ALres}
\Pi_{AL} ( q;\vec Q=0 ) & = &  \frac{e^6\, \mu^{3\,x} \, |\vec L_{q}|^{6\,(d-m-1)-m } }  {N} 
\int \frac{d\vec P d\vec K d\vec L}{(2\pi)^{3 \,(d-m)}}
\frac{{\cal J}_m ( |\vec L| )}
{  \left ( \,|\vec P| + |\vec P +\vec L| + |\vec K| +|\vec K +\vec L| \, \right )^2 +1  } \nn
&\cdot&
\frac{\left( [\vec K \cdot (\vec K +\vec L)] -|\vec K| \, |\vec K +\vec L| \right)\,
\left( [\vec P \cdot (\vec P +\vec L)] - |\vec P | \, |\vec P +\vec L|
\right)}
{2 \, |\vec P |\, |\vec K |\,|\vec P +\vec L|\,|\vec K +\vec L| 
\left[ \, |\vec P | + |\vec P +\vec L| + |\vec K|  +|\vec K +\vec L| \, \right]},
\eqa
where
%%%%%%%%%%%%%%%
\bqa
{\cal J}_m ( |\vec L |)  &\sim& \int_0^1 \frac{dl_{d-m+1} } {2 \, \pi} 
\frac{2^{3 d-6} \, [\, l_{d-m+1} \, ( 1-l_{d-m+1} ) \,]^{2\, (d-m)}}
{ l_{d-m+1}^{3-d+m } + \tilde \alpha 
\, [ \, 2 \,  (1-l_{d-m+1} )\, | \vec L| \,]^{d-m} }
 \frac{1}  {(1-l_{ d-m+1}  )^{3-d+m} 
+\tilde \alpha
\, \left(\, 2 \, l_{d-m+1}\,| \vec L| \, \right )^{d-m}}\,.\nn
\eqa
%%%%%%%%%%%%%%%%
Here $\vec P, \vec K$ and $ \vec L$ have been rescaled to be dimensionless
in the unit of $ q_{d-m+1}^2$. Since ${\cal J}_{ m }  ( |\vec L| )$ decays as $  |\vec L |^{ -2 \,(d-m)} $ 
in the $ |\vec L| \rightarrow \infty$  limit, 
the overall degree of divergence of the $\vec P, \vec L$ and $ \vec K$ integrals is
$-3+d-m $, which is UV-finite.
To estimate the dependence on $\tilde e$ and $k_F $, 
we  note that ${\cal J}_{ m } ( |\vec L| )$ has a non-trivial dependence on $\tilde \alpha $, 
and behaves differently depending on whether $  |\vec L| $ is large or small compared to 
$L_* = \tilde \alpha^{- \frac{1} {d-m} } $ (in the unit of $q_{d-m+1}^2$) : 
\bqa\label{jint1} 
{\cal J}_{ m } (L) \approx \left\{ \begin{array}{cl}
C_1 ,& \quad |\vec L| \ll L_* \\
\displaystyle\frac{C_2}{\tilde \alpha ^2 \, |\vec L|^{2\,(d-m) } }, & \quad |\vec L| \gg L_*  \end{array} \right\} \,,
\eqa
%%%%%%%%%%%%
where $C_1$ and $C_2$ are constants
which are independent of $\tilde e$ and $k_F $. 
Thus the Aslamazov-Larkin diagrams contribute
only a finite renormalization to the boson kinetic term and the $m=2 $ case in the $\lambda_{ \text{cross} } << 1 $ limit still has $Z_3=1$ even at this three-loop order.

For the sake of completeness, let us also enumerate the behaviour of the AL terms in some other specific limits.

For $\frac{ |q_{d-m}| }  { |\vec{L}_{(q)}| \sqrt{2 k_F} } , \frac{  | \vec Q| }  { |\vec{L}_{(q)} | \sqrt{2 k_F}} << 1$:

\begin{enumerate}
\item
For  $\frac{ | l_{d-m}| }  { |\vec{L}_{(l)}| \sqrt{2 k_F} } , \frac{  | \vec L| }  { |\vec{L}_{(l)} | \sqrt{2 k_F}} << 1$ and $m>1$, we use Eq.~(\ref{c1}) to get
\bqa
&& \big [\mbox{Integral for  } |{ \vec{L}}_{(l)}|> \frac{\Lambda} {\sqrt{k_F}} 
\mbox{ contributing to } \Pi_{AL} (q)  \big ] \nn
&\propto& \frac{ e^6 \, \mu^{3  x} \, k_F^{m-3} }  { |\vec{L}_{(q)} |^4 }
 \int_{ |{ \vec{L}}_{(l)}|> \frac{\Lambda} {\sqrt{k_F}}  } 
 \frac{dl}  {(2\pi)^{d+1} } 
\frac{   |{ \vec{L}}_{(l)}| }
{ |\vec{L}_{(l)}|^3 + \tilde{\alpha}    \, |\vec L|^{d-m}  
}  \frac{   |{ \vec{L}}_{(l-q)}| }
{ |\vec{L}_{(l-q)}|^3 + \tilde{\alpha}    \, |\vec L - \vec Q|^{d-m}
} \times fn( q, l) \,.
\nn
\eqa
The positive powers of $k_F$ in the denominator of the boson propagator will further suppress the final expression by overall negative powers of $k_F$. But let us estimate the overall powers by ignoring these. Then the factors go as
\beq
e^2\, k_F^{\frac{m-1} {2}} \times
\frac{  {\tilde{e}} ^{ \frac{6}{(m+1)} }  }  { k_F^{\frac{9-2m+m^2 }{2 \,(m+1) }} } \,.
\label{cc1}
\eeq

\item

In the limit $ \frac{ |q_{d-m}| }  { |\vec{L}_{(q)}| \sqrt{2 k_F} } , 
 \frac{ |\vec P | + |\vec P + \vec Q| }  { |\vec{L}_{(q)} | \sqrt{2 k_F}}, \frac{ |\vec P | + |\vec P + \vec L| }  {  |\vec{L}_{(q)}| \sqrt{2 k_F} } << 1  $ and $ \frac{ |l_{d-m}| }  { |\vec{L}_{(l)}| \sqrt{2 k_F} } ,  \frac{ | \vec L| }  {  |\vec{L}_{(l)}| \sqrt{2 k_F} } >> 1  $,
we have
\begin{eqnarray}
\kappa_1 &\simeq& \frac{1}{2\,|{\vec{L}}_{(q)}| }
\int \frac{dx_1}{2\pi }
\frac{x_1 \, \left ( x_1 + l_{d-m} \right)   
\exp\left( {-\frac{ 3 {\vec{u}}_{(p)}^2 +  {\vec{L}}_{(q)}^2  } 
{k_F }} \right)
}
{ 
\Big \lbrace  x_1^2  +\vec P ^2 \Big \rbrace   
\Big \lbrace  \left( x_1 + l_{d-m} \right)^2  +(\vec P +\vec L)^2 \Big \rbrace } 
\nn && \times \, 
\int \frac{ d x_2 }{2\pi }
\frac{ 
\left ( x_1+x_2 + \delta_q \right)
\exp\left( {-\frac{ 3 x_2^2 \, + \, 4 \, \vec{L}_{(q)}^2  \, x_2 } { 4\, \vec{L}_{(q)}^2  \, k_F }} \right)
}
{  \left( x_1 + x_2  + \delta_q \right)^2  +(\vec P +\vec Q)^2 }
\nn
%%%%%
&=& \frac{1}{2\,|{\vec{L}}_{(q)}| }
\int \frac{dx_1}{2\pi }
\frac{x_1 \, \left ( x_1 + l_{d-m} \right)   
\exp\left( {-\frac{ 3 {\vec{u}}_{(p)}^2 + \frac{2}{3} {\vec{L}}_{(q)}^2  } 
{k_F }} \right)
}
{ 
\Big \lbrace  x_1^2  +\vec P ^2 \Big \rbrace   
\Big \lbrace  \left( x_1 + l_{d-m} \right)^2  +(\vec P +\vec L)^2 \Big \rbrace } \nn
&& \qquad  \quad  \times \,
\int \frac{ d x_2 }{2\pi }
\frac{ 
\left ( x_1+x_2 + \delta_q  - \frac{2}{3} \vec{L}_{(q)}^2 \right)
\exp\left( {-\frac{ 3 x_2^2  } { 4\, \vec{L}_{(q)}^2  \, k_F }} \right)
}
{  \left( x_1 + x_2  + \delta_q - \frac{2}{3} \vec{L}_{(q)}^2 \right)^2  +(\vec P +\vec Q)^2 } \,.
\end{eqnarray}
This implies that
\begin{equation}
f_t(l,q) = \frac{ e^3 \, k_F^{\frac{m-2}{2}} }{2\,|{\vec{L}}_{(q)}| } \times fn( \vec{L},  q, l_{d-m}  )
\label{c3}
\end{equation}
in these limits.

Hence, for  $\frac{ | l_{d-m}| }  { |\vec{L}_{(l)}| \sqrt{2 k_F} } , \frac{  | \vec L| }  { |\vec{L}_{(l)} | \sqrt{2 k_F}} >> 1$ and $m>1$,  Eqs.~(\ref{c3}) and Eq.~(A15) of Ref.~\cite{ips1} give us:
%%%%%%%%%%%%%%%%%%%%%%%%%%%
\bqa
&& \big [\mbox{Integral for  } |{ \vec{L}}_{(l)}|< \frac{\Lambda} {\sqrt{k_F}} 
\mbox{ contributing to } \Pi_{AL} (q) \big ] \nn
&\propto& \frac{ e^6  \, k_F^{m-2} }  { |\vec{L}_{(q)} |^2 }
 \int_{ |{ \vec{L}}_{(l)}|< \frac{\Lambda} {\sqrt{k_F}} } \frac{dl}  {(2\pi)^{d+1} } 
 \frac{   fn( \vec L,  l_{d-m} , q ) }
 { {\vec{L}}_{(l)}^2   + e^2 \, \mu^x \, J^{m-1} \sqrt{ k_F} \, \,
\tilde{f} \left( |\vec L |, l_{d-m} \right)  } 
\frac{  |{ \vec{L}}_{(q)}| }
{ |\vec{L}_{(q)}|^3 + \tilde{\alpha}    \, |\vec L - \vec Q|^{d-m}
}  \nn
%%%
&\propto& \frac{ e^4  \, \Lambda^{m} }  { |\vec{L}_{(q)} |^2 \, k_F^2 }
 \int \frac{d \vec L \, d l_{d-m}}  {(2\pi)^{d-m+1} } 
\frac{  fn( \vec L,  l_{d-m} , q  ) }
{ |\vec{L}_{(q)}|^3 + \tilde{\alpha}    \, |\vec L - \vec Q|^{d-m} 
} \,. \nn
\eqa
%%%%%%%%%%%%%%%%%%%%%%%%%%%%%
Again, ignoring the negative powers of $k_F$ coming from $\tilde \alpha$, we get the factors as
\beq
e^2\, k_F^{\frac{m} {2}} \times
\frac{  {\tilde{e}} ^{ \frac{ 3 }{(m+1)} }  }  { k_F^{\frac{m^2 + m +3 } {(m+1) }} } \,.
\label{cc2}
\eeq
\end{enumerate}

\vspace{2 mm}
%%%%%%%%%%%%%%%%%%%%%%%%%%%%%
%%%%%%%%%%%%%%%%%%%%%%%%%%%%%
For $\frac{ |q_{d-m}| }  { |\vec{L}_{(q)}| \sqrt{2 k_F} } , \frac{  | \vec Q| }  { |\vec{L}_{(q)} | \sqrt{2 k_F}}  >> 1 $:
 
 \begin{enumerate}
\item
In the limit $  \frac{ | l_{d-m } | }  {\vec{L}_{(l)} \sqrt{2 k_F} }  << 1  $, we get
\begin{eqnarray}
\kappa_1  &\simeq&
\exp\left( {-\frac{   {\vec{L}}_{(q)}^2 +   {\vec{L}}_{(l)}^2} { 3\, k_F }} \right) 
\int \frac{dx_1 \, d p_{d-m+1}}{(2\pi)^2 }
\frac{
\exp\left( {-\frac{ 3 \left( {\vec{L}}_{(p)} +\frac{1}{3} {\vec{L}}_{(q)} +\frac{1}{3} {\vec{L}}_{(l)} \right)^2  } { k_F }} \right)
}
{ 
\Big \lbrace  x_1^2  +\vec P ^2 \Big \rbrace   
\Big \lbrace  \left( x_1 + q_{d-m}  \right)^2  +(\vec P +\vec Q)^2 \Big \rbrace  }
 \nn
&& \qquad \qquad \qquad \qquad \qquad \qquad \qquad 
 \times
\frac{
x_1 \, 
(x_1 + q_{d-m} ) \, 
 \Big \lbrace x_1 + {\vec{L}}_{(l)}^2 
+ 2 {\vec{L}}_{(p)} \cdot {\vec{L}}_{(l)} 
\Big \rbrace 
 }
{ 
 \Big \lbrace x_1 + {\vec{L}}_{(l)}^2 
+ 2 {\vec{L}}_{(p)} \cdot {\vec{L}}_{(l)} 
\Big \rbrace ^2  +(\vec P +\vec L)^2 } \nn
%%%%%%%%%%%%%%
\implies && \int \frac{ d \vec{u}_{(p)} }  { (2 \pi)^{m-1}} \, \kappa_1 \nn
%%%%%%%%%%5
&=& \exp\left( {-\frac{   {\vec{L}}_{(q)}^2 +   {\vec{L}}_{(l)}^2} { 3\, k_F }} \right) 
\left ( \frac{k_F} {12 \pi} \right)^{\frac{ m-1 }{2}}
\sqrt{\frac{ 3 } { \pi k_F} }
\int \frac{dx_1  }{ 2\pi }
\frac{   x_1 \, 
(x_1 + q_{d-m} ) \, 
( x_1 + \frac{1}{3}   {\vec{L}}_{(l)}^2 - \frac{2}{3}  {\vec{L}}_{(l)} \cdot  {\vec{L}}_{(q)} )   }
{  \Big \lbrace  x_1^2  +\vec P ^2 \Big \rbrace   
\Big \lbrace  \left( x_1 + q_{d-m}  \right)^2  +(\vec P +\vec Q)^2 \Big \rbrace } \,.\nn
\end{eqnarray}
This implies
\begin{equation}
f_t(l,q) \propto e^3 \,  k_F^{\frac{m-2}{2}}    \label{c5}
\end{equation} in the above limits.

Therefore, for  $\frac{ | l_{d-m}| }  { |\vec{L}_{(l)}| \sqrt{2 k_F} } , \frac{  | \vec L| }  { |\vec{L}_{(l)} | \sqrt{2 k_F}} << 1$ and $m>1$, we use Eq.~(\ref{c5}) to get
\bqa
&& \big [\mbox{Integral for  } |{ \vec{L}}_{(l)}|> \frac{\Lambda} {\sqrt{k_F}} 
\mbox{ contributing to } \Pi_{AL} (q)  \big ] \nn
 &\propto&  e^6  \, k_F^{m-2} 
 \int_{ |{ \vec{L}}_{(l)}|> \frac{\Lambda} {\sqrt{k_F}}  } 
 \frac{dl}  {(2\pi)^{d+1} } 
\frac{   |{ \vec{L}}_{(l)}| }
{ |\vec{L}_{(l)}|^3 + \tilde{\alpha}    \, |\vec L|^{d-m}  
}  \frac{   |{ \vec{L}}_{(l-q)}| }
{ |\vec{L}_{(l-q)}|^3 + \tilde{\alpha}    \, |\vec L - \vec Q|^{d-m}
} \times  fn( \vec L, \vec L_{(l)}, q) \,.
\nn
\eqa
The positive powers of $k_F$ in the denominator of the boson propagator will further suppress the final expression by overall negative powers of $k_F$. Again, let us estimate the overall powers by ignoring these. The factors go as
\beq
e^2\, k_F^{\frac{m} {2}} \times
\frac{  {\tilde{e}} ^{ \frac{6}{(m+1)} }  }  { k_F^{\frac{8-3m+m^2 }{2 \,(m+1) }} } \,.
\label{cc3}
\eeq
%%%%%%%%%%%

\item

In the limit $  \frac{ | l_{d-m } | }  {\vec{L}_{(l)} \sqrt{2 k_F} }  >> 1  $, we get
\begin{eqnarray}
\kappa_1 &\simeq&
\exp\left( {-\frac{ 3 {\vec{u}}_{(p)}^2 +2 {\vec{u}}_{(p)} \cdot {\vec{u}}_{(l)}+  {\vec{L}}_{(q)}^2 +   {\vec{L}}_{(l)}^2} { k_F }} \right) \nn
&& \times \int \frac{dx_1 \, d p_{d-m+1}}{(2\pi)^2 }
\frac{
\exp\left( {-\frac{ 3 p_{d-m+1}^2 +2 ( \,l_{d-m+1}+ |{\vec{L}}_{(q)}| \,) \, p_{d-m+1}} { k_F }} \right)
}
{ 
\Big \lbrace  x_1^2  +\vec P ^2 \Big \rbrace   
\Big \lbrace  \left( x_1 + q_{d-m}  \right)^2  +(\vec P +\vec Q)^2 \Big \rbrace  }
\frac{
x_1 \, 
(x_1 + q_{d-m} ) \, 
(x_1  + l_{d-m} )
 }
{ 
 \Big \lbrace  \left(x_1    + l _{d-m} \right)^2  +(\vec P +\vec L)^2 \Big \rbrace }  \nn
%%%%%%%%%%%%%%%
&=& \exp\left( {-\frac{ 3 {\vec{u}}_{(p)}^2 +2 {\vec{u}}_{(p)} \cdot {\vec{u}}_{(l)}+  {\vec{L}}_{(q)}^2 +   {\vec{L}}_{(l)}^2} { k_F }}\right)
\exp\left( {\frac{ l_{d-m+1}^2 + |{\vec{L}}_{(q)}|^2 } { 3 \, k_F }} \right) 
\sqrt{ \frac{k_F}{ 12 \,  \pi} }  \nn
&& \times
\int \frac{dx_1}{2\pi }
\frac{
x_1 \, 
(x_1 + q_{d-m} ) \, 
 (x_1  + l_{d-m} )
}
{ 
\Big \lbrace  x_1^2  +\vec P ^2 \Big \rbrace   
\Big \lbrace  \left( x_1 + q_{d-m}  \right)^2  +(\vec P +\vec Q)^2 \Big \rbrace 
\Big \lbrace   \left(  x_1    + l _{d-m}
\right )^2  +(\vec P +\vec L)^2 \Big \rbrace
 }  \nn
%%%%%%%%%%%%%%%
\implies &&
\int \frac{ d \vec{u}_{(p)} }  { (2 \pi)^{m-1}} \, \kappa_1
=
\exp\left( {-\frac{ 2\, {\vec{L}}_{(q)}^2 +  2\, {\vec{L}}_{(l)}^2} {3\, k_F }}\right)
\left ( \frac{k_F}{ 12 \,  \pi} \right )^{m/2}  f_{tt} \left (l_{d-m} ,\,q_{d-m} ,   P, |\vec P +\vec Q|, |\vec P +\vec L| \right ) ,\nonumber 
\end{eqnarray}
where the function $f_{tt}$ is of mass dimension $-2$.
This leads to
\begin{equation}
f_t(l,q) \propto e^3 \,  k_F^{\frac{m}{2}}  \,. \label{c4}
\end{equation}
%}

Thus for  $\frac{ | l_{d-m}| }  { |\vec{L}_{(l)}| \sqrt{2 k_F} } , \frac{  | \vec L| }  { |\vec{L}_{(l)} | \sqrt{2 k_F}} >> 1$ and $m>1$,  Eqs.~(\ref{c4}) and Eq.~(A15) of Ref.~\cite{ips1} give us
\bqa
&& \big [ \mbox{Integral for  } |{ \vec{L}}_{(l)}|< \frac{\Lambda} {\sqrt{k_F}} 
\mbox{ contributing to } \Pi_{AL} (q)  \big ]   \nn
&\propto & e^6  \, k_F^{m} 
 \int_{ |{ \vec{L}}_{(l)}|< \frac{\Lambda} {\sqrt{k_F}} } \frac{dl}  {(2\pi)^{d+1} } 
 \frac{   fn( \vec L,  l_{d-m} , q ) }
 { {\vec{L}}_{(l)}^2   + e^2 \, \mu^x \, J^{m-1} \sqrt{ k_F} \, \,
\tilde{f} \left( |\vec L |, l_{d-m} \right)  } \nn
&& \qquad \qquad \qquad \times \,
\frac{ 1 }
{ |\vec{L}_{(l-q)}|^2 +e^2 \, \mu^x \, J^{m-1} \sqrt{ k_F} \, \,
\tilde{f} \left( |\vec L -\vec Q |, l_{d-m} - q_{d-m} \right)
}  \nn
%%%
&\propto& \frac{ e^2  \, \Lambda^{m} }  {  k_F^{m/2} }
 \int \frac{d \vec L \, d l_{d-m}}  {(2\pi)^{d-m+1} } 
 \, fn( \vec L,  l_{d-m} , q  )  \,. \nonumber
\eqa
This results in the factors
\beq
e^2\, k_F^{\frac{m} {2}} \times
\frac{  1  }  { k_F^m } \,.
\label{cc4}
\eeq

\end{enumerate}

\vspace{5 mm}

From the behaviour of the AL terms in all the above limits, we conclude that for $\frac{ |q_{d-m}| }  { |\vec{L}_{(q)}| \sqrt{2 k_F} } , \frac{  | \vec Q| }  { |\vec{L}_{(q)} | \sqrt{2 k_F}} << 1$ as well as$\frac{ |q_{d-m}| }  { |\vec{L}_{(q)}| \sqrt{2 k_F} } , \frac{  | \vec Q| }  { |\vec{L}_{(q)} | \sqrt{2 k_F}} >> 1$, $\Pi_{AL} (q)$ is suppressed by positive powers of $k_F$ compared to the one-loop result.

\bibliography{NFL_3-loop}

\end{document}